\documentclass[11pt]{article}

\usepackage{authblk} 
\usepackage{epsfig}
\usepackage{graphicx}
\usepackage{amsmath,amsfonts}
\usepackage[margin=1in]{geometry}
\usepackage{array}
\usepackage{caption}

\usepackage{tikz}
\usetikzlibrary{snakes}
\usetikzlibrary{plotmarks}
\def\begeq{\begin{equation}}
\def\endeq{\end{equation}}

\def\begeqar{\begin{eqnarray}}
\def\endeqar{\end{eqnarray}}

\def\<{\langle}
\def\>{\rangle}

\newcommand{\q}{\mathfrak{q}}
\newcommand{\be}{\begin{equation}}
\newcommand{\ba}{\begin{eqnarray}}
\newcommand{\ea}{\end{eqnarray}}
\newcommand{\ee}{\end{equation}}

\newcommand{\XXX}[1]{}

\def\begeq{\begin{equation}}
\def\endeq{\end{equation}}

\def\begeqar{\begin{eqnarray}}
\def\endeqar{\end{eqnarray}}

\def\({\left(}		\def\){\right)}
\def\[{\left[}          \def\]{\right]}

\title{Non compact conformal field theory \\ and the $a_2^{(2)}$ (Izergin-Korepin) model in regime III}
\author[1,2]{\'Eric Vernier}
\author[1,3]{Jesper Lykke Jacobsen}
\author[2,4]{Hubert Saleur}
\affil[1]{LPTENS, \'Ecole Normale Sup\'erieure, 24 rue Lhomond, 75231 Paris, France}
\affil[2]{IPhT, CEA Saclay, 91191 Gif-sur-Yvette, France}
\affil[3]{Universit\'e Pierre et Marie Curie, 4 place Jussieu, 75252 Paris, France}
\affil[4]{USC Physics Department, Los Angeles CA 90089, USA}
\date{}
\begin{document}

\maketitle

\begin{abstract}

The so-called regime III of the $a_2^{(2)}$ Izergin-Korepin 19-vertex model has defied understanding for many years. We show in this paper that its continuum limit involves in fact a non compact conformal field theory (the so-called Witten Euclidian black hole CFT), which leads to a continuous spectrum of critical exponents, as well as very strong corrections to scaling. Detailed numerical evidence based on the Bethe ansatz analysis is presented, involving in particular the observation of {\sl discrete states} in the spectrum, in full agreement with the string theory prediction for the black hole CFT. Our results have important consequences for the physics of the $O(n)$ model, which will be discussed elsewhere.

\end{abstract}

\section{Introduction}

The Izergin-Korepin (or $a_2^{(2)}$) model is, together with the 6-vertex model, a central part of  the field of  
integrable statistical mechanics models. It is  directly connected - in particular thanks to the reformulation as an $O(n)$ model 
- with many of the most recent developments in conformal field theory, 2D quantum gravity \cite{OnTwoDGrav},  combinatorics \cite{OnCombinat} probability theory \cite{OnProba},  algebra \cite{OnAlgebra} and has also many physical applications, in particular in polymer theory \cite{OnPoly}. 

More than thirty years after the discovery of the vertex model \cite{IzergKorep}, and more than twenty years after the reformulation as an $O(n)$ model \cite{Nienhuis90,BloteNienhuis89,Nienhuis}, it  seems that everything should be understood about such  basic  aspects as the nature of the continuum limit, or  the spectrum of  critical exponents. Quite surprisingly, this is not the case. The so called regime III (see a detailed definition below) resisted most of the sophisticated attempts started in \cite{Nienhuis}, exhibiting unusually large corrections to scaling, and new, disconcerting behavior of the measured gaps with the twist angle.

This has been considered in the community as  a technical problem more than a fundamental one, but we show here that something very deep and new is happening: the continuum limit of the Izergin Korepin model in regime III is a {\sl non compact conformal field theory}, with a continuous spectrum of critical exponents. This is possible, even though the lattice model is perfectly compact, because the Boltzmann weights are not all positive in the regime III (the Hamiltonian is not hermitian), and the correspondence between the lattice and continuum target spaces is more complicated than in the usual, positive (hermitian) cases. In the latter case, for the simplest statistical mechanics models such as the IK model in the other regimes or the 6 vertex model, the continuum target space is usually a circle, that is, a real variable ``compactified''  by the identification $\Phi\equiv \Phi+2\pi R$. The variable $\Phi$ is a free boson \cite{CoulombGasMap}, and the compactification 
arises because of its angular nature. A non compact target space would correspond, for instance, to a boson with infinite radius of compactification ($R\to \infty$), but in our case we will see that things are a little more complicated.

This is not the first time that such non compact behavior is encountered. In \cite{EsslerFS} and in \cite{IkhlefJS1,IkhlefJS2,IkhlefJS3,CanduIkhlef} two other lattice models were studied that involve similarly finite-dimensional representations on every site or link, but with a non compact continuum limit, and a continuous spectrum of critical exponents. These models however were not the most  natural or useful from a pure statistical mechanics point of view: the one in \cite{EsslerFS} requires  a $sl(2|1)$ super group symmetry, and the one in \cite{IkhlefJS1} involves a delicate staggering of the spectral parameters. In contrast, the IK model in regime III is related with very physical questions, in particular the behavior of a leading candidate for the theta point of polymers. Our results will turn out to have profound consequences on our understanding of this point.

The existence of a continuous spectrum of critical exponents means that, for the corresponding conformal field theory, there will be primary fields with dimensions $(h,\bar{h})$ taking a continuous set of values within an interval. In terms of lattice transfer matrices or hamiltonians, this implies that the rescaled gaps $L(E-E_0)$ --- where $L$ denotes the length of the system, and $E_0$ is the ground state energy --- instead of converging to a discrete set of points in the limit $L\to\infty$, will take continuous values covering some intervals.

The Izergin-Korepin and $O(n)$ models have been studied at great length, and we will not spend much time reminding the reader of their definition and the vast body of knowledge which is available about them. Except for a short reminder of the properties in regime I provided in Appendix \ref{app:B}, and that we will need for purposes of definition and normalization, we will therefore focus entirely on the regime III. We will, however, devote section~\ref{Sec:gen} to a careful review of the conventions and notations in the literature, in order to identify precisely the model we are interested in. Another point worth recalling here is that the IK and $O(n)$ model are not exactly identical: the $O(n)$ model is obtained by combining sectors of the IK vertex model with different ``twisted'' boundary conditions and subtracting states, very much in the same way that the Potts model is obtained from sectors of the 6 vertex model. We focus in this paper on the continuum limit of the IK {\sl vertex} model, where it is 
usually implied that this model has periodic boundary conditions, although we will also consider twisted boundary conditions at various points. The role of these boundary conditions however is mostly to allow us to explore aspects of the IK model itself: the $O(n)$ loop model per se will be considered elsewhere.  

Section~\ref{Sec:betheanalysis} provides a careful, revised analysis of the numerical results from the Bethe ansatz, together with duality arguments that lead to a tentative identification of the continuum limit of the IK model with the Eucidian black hole conformal field theory (CFT) \cite{BlackHoleCFT}, whose main features are recalled at the beginning of section \ref{Sec:BlackHoleIdent}.  This CFT --- which can also be considered as the (GKO) coset $SL(2,\mathbb{R})/U(1)$ --- has the interesting property that it admits both continuous and discrete states.  The rules giving rise to these discrete states are very specific, and result in a pattern that depends crucially on the coupling constant --- which translates here into the anisotropy of the model, or the $n$ variable in the $O(n)$ version. We show in section \ref{Sec:BlackHoleIdent} how the lattice model exactly reproduces the results predicted in \cite{Troost,RibSch}. We then discuss the issue of the density of states, and how the termination of the 
regime is related with the appearance of a marginally relevant operator. This allows us to predict an essential singularity in the free energy (ground state energy) of the model, whose existence we also prove directly from the Bethe ansatz results.  

Two further appendices contain the derivation of an  algebraic RSOS formulation (different from the geometrical one given in \cite{NienhuisRSOS})  used in some of our early arguments about the continuum limit, and a detailed exposition of the numerical method and results.
 
\section{Generalities}\label{Sec:gen}

\subsection{The models}
\label{sec:models}

The Izergin-Korepin model \cite{IzergKorep} is a quantized version of the Bullough-Dodd model \cite{BullDodd}.
It was originally \cite{IzergKorep} formulated as a vertex model, but is also closely related to  an O($n$) type
loop model \cite{Nienhuis90,BloteNienhuis89,Nienhuis,Batchelor} or as an RSOS type height model \cite{NienhuisRSOS} (see also Appendix~\ref{app:A}).

In view of the program we expect to carry out in this  paper and its sequels, it is however most natural to start by giving the standard form of the $\check{R}$ matrix for the $a_2^{(2)}$ algebra \cite{GM2}:
\begin{eqnarray}
\check{R}^{(2)}&\propto &1+{x-1\over x+1}{q^{3}-x\over q^{3}+x}E+{1-x\over 1+x}{1\over q-q^{-1}}\left(B+B^{-1}\right) \,.\label{IKRmat}
\end{eqnarray}
Here the SO($3$) braid and Temperley-Lieb generators satisfy the relationship
\begin{eqnarray}
 B - B^{-1} = (q-q^{-1}) (1 - E) \,,
\end{eqnarray}
and the $\check{R}$ matrices obey the Yang Baxter equation in the form
\begin{eqnarray}
 \check{R}_{12}\check{R}_{23}\check{R}_{12}=\check{R}_{23}\check{R}_{12}\check{R}_{23} \,.
\end{eqnarray}
We denote by $q$ the deformation parameter, and by $x$ the (multiplicative) spectral parameter. We note that a slightly different normalization is often used for the deformation parameter in the $a_2^{(2)}$ case \cite{Takacs}. We will set 
\begin{eqnarray}
q \equiv e^{i\gamma}\nonumber\\
x \equiv e^{2\lambda} \,,
\label{ourparameters}
\end{eqnarray}
in all that follows. 

Explicit calculations show that this matrix matches the more usual (in the statistical mechanics literature) form \cite{Batchelori} in terms of weights for the 19 possible vertices encoded in the $R$ matrix ($R\equiv P\check{R}$) acting on the space
$(\mathbb{C}^3)^{\otimes 2}$
\begin{equation}
R=\left(\begin{array}{ccc|ccc|ccc}
c&0&0&0&0&0&0&0&0\\
0&b&0&e&0&0&0&0&0\\
0&0&d&0&g&0&f&0&0\\ \hline
0&\bar{e}&0&b&0&0&0&0&0\\
0&0&\bar{g}&0&a&0&g&0&0\\
0&0&0&0&0&b&0&e&0\\ \hline
0&0&\bar{f}&0&\bar{g}&0&d&0&0\\
0&0&0&0&0&\bar{e}&0&b&0\\
0&0&0&0&0&0&0&0&c
\end{array}\right) \,,
\label{19vRmatrix}
\end{equation}
where the weights can be written (YB stands for Yung and Batchelor \cite{Batchelori}):
\begin{eqnarray}
a&=&\sinh(u_{\rm YB}-3\lambda_{\rm YB})-\sinh(5\lambda_{\rm YB})+\sinh(3\lambda_{\rm YB})+\sinh(\lambda_{\rm YB})\nonumber\\
b&=&\sinh(u_{\rm YB}-3\lambda_{\rm YB})+\sinh(3\lambda_{\rm YB})\nonumber\\
c&=&\sinh(u_{\rm YB}-5\lambda_{\rm YB})+\sinh\lambda_{\rm YB}\nonumber\\
d&=&\sinh(u_{\rm YB}-\lambda_{\rm YB})+\sinh\lambda_{\rm YB}\nonumber\\
e&=&-2e^{-u_{\rm YB}/2}\sinh(2\lambda_{\rm YB})\cosh({u_{\rm YB}\over2}-3\lambda_{\rm YB})\nonumber\\
\bar{e}&=&-2e^{u_{\rm YB}/2}\sinh(2\lambda_{\rm YB})\cosh({u_{\rm YB}\over2}-3\lambda_{\rm YB})\nonumber\\
f&=&-2e^{-u_{\rm YB}+2\lambda_{\rm YB}}\sinh(\lambda_{\rm YB})\sinh(2\lambda_{\rm YB})-e^{-\lambda_{\rm YB}}\sinh(4\lambda_{\rm YB})\nonumber\\
\bar{f}&=&2e^{u_{\rm YB}-2\lambda_{\rm YB}}\sinh(\lambda_{\rm YB})\sinh(2\lambda_{\rm YB})-e^{\lambda_{\rm YB}}\sinh(4\lambda_{\rm YB})\nonumber\\
g&=&2e^{-{u_{\rm YB}\over 2}+2\lambda_{\rm YB}}\sinh(u_{\rm YB}/2)\sinh(2\lambda_{\rm YB})\nonumber\\
\bar{g}&=&-2e^{{u_{\rm YB}\over 2}-2\lambda_{\rm YB}}\sinh(u_{\rm YB}/2)\sinh(2\lambda_{\rm YB})
\end{eqnarray}
provided one sets
\begin{eqnarray}
\lambda_{\rm YB}=i{\gamma\over 2}\nonumber\\
x=e^{u_{\rm YB}} \,,
\end{eqnarray}
and implements a gauge transformation on the degrees of freedom on every edge, obtained by simple multiplication with
\begin{equation}
\left(\begin{array}{ccc} 
1&0&0\\
0&i&0\\
0&0&1\end{array}\right) \,.
\end{equation}
Each space $\mathbb{C}^3$ is interpreted naturally in terms of spin $S=1$, and the value of $S^z= 0,\pm 1$ can be
represented graphically as an empty edge, or an edge carrying an up/down pointing arrow. The 19 configurations
around a vertex allowed by (\ref{19vRmatrix}) are precisely those compatible with arrow conservation,
generalizing the ice rule (which is well-known from the 6-vertex model).

The model is closely related to  a loop model. The local configurations around a vertex of the square lattice for this model are the following:
\begin{equation}
\begin{tikzpicture}[scale=1.0]
 \draw[thin,black] (0,0)--(1,1);
 \draw[thin,black] (0,1)--(1,0);
 \draw (0.5,0) node[below] {$\rho_1$};
\begin{scope}[xshift=1.5cm]
 \draw[thin,black] (0,0)--(1,1);
 \draw[thin,black] (0,1)--(1,0);
 \draw[blue,ultra thick] (0,0)--(0.36,0.36) arc(-45:45:2mm)--(0,1);
 \draw (0.5,0) node[below] {$\rho_2$};
\end{scope}
\begin{scope}[xshift=3cm]
 \draw[thin,black] (0,0)--(1,1);
 \draw[thin,black] (0,1)--(1,0);
 \draw[blue,ultra thick] (1,0)--(0.64,0.36) arc(225:135:2mm)--(1,1);
 \draw (0.5,0) node[below] {$\rho_3$};
\end{scope}
\begin{scope}[xshift=4.5cm]
 \draw[thin,black] (0,0)--(1,1);
 \draw[thin,black] (0,1)--(1,0);
 \draw[blue,ultra thick] (0,0)--(0.36,0.36) arc(135:45:2mm)--(1,0);
 \draw (0.5,0) node[below] {$\rho_4$};
\end{scope}
\begin{scope}[xshift=6cm]
 \draw[thin,black] (0,0)--(1,1);
 \draw[thin,black] (0,1)--(1,0);
 \draw[blue,ultra thick] (0,1)--(0.36,0.64) arc(-135:-45:2mm)--(1,1);
 \draw (0.5,0) node[below] {$\rho_5$};
\end{scope}
\begin{scope}[xshift=7.5cm]
 \draw[thin,black] (0,0)--(1,1);
 \draw[thin,black] (0,1)--(1,0);
 \draw[blue,ultra thick] (0,0)--(1,1);
 \draw (0.5,0) node[below] {$\rho_6$};
\end{scope}
\begin{scope}[xshift=9cm]
 \draw[thin,black] (0,0)--(1,1);
 \draw[thin,black] (0,1)--(1,0);
 \draw[blue,ultra thick] (0,1)--(1,0);
 \draw (0.5,0) node[below] {$\rho_7$};
\end{scope}
\begin{scope}[xshift=10.5cm]
 \draw[thin,black] (0,0)--(1,1);
 \draw[thin,black] (0,1)--(1,0);
 \draw[blue,ultra thick] (0,0)--(0.36,0.36) arc(-45:45:2mm)--(0,1);
 \draw[blue,ultra thick] (1,0)--(0.64,0.36) arc(225:135:2mm)--(1,1);
 \draw (0.5,0) node[below] {$\rho_8$};
\end{scope}
\begin{scope}[xshift=12cm]
 \draw[thin,black] (0,0)--(1,1);
 \draw[thin,black] (0,1)--(1,0);
 \draw[blue,ultra thick] (0,0)--(0.36,0.36) arc(135:45:2mm)--(1,0);
 \draw[blue,ultra thick] (0,1)--(0.36,0.64) arc(-135:-45:2mm)--(1,1);
 \draw (0.5,0) node[below] {$\rho_9$}; \,,
\end{scope}
\end{tikzpicture}
\label{onweights}
\end{equation}
and there is an additional weight $n$ per closed loop, henceforth referred to as the loop fugacity. To bring the loop model into correspondence with the 19-vertex model one first introduces an extra orientational degree of freedom (clockwise or counterclockwise) for each loop. The fugacity $n$ is then distributed over these orientations and further over the local turns. Summing over the loop connectivities compatible with each choice of arrows (orientation) on the adjacent edges finally produces a vertex model. Choosing correctly the gauge factors allowed by these transformations produces the vertex weights (ZB stands for Zhou and Batchelor \cite{Batchelor})
\begin{eqnarray}
\rho_1&=&1+{\sin u_{\rm ZB} \sin(3\lambda_{\rm ZB}-u_{\rm ZB})
\over \sin2\lambda_{\rm ZB}\sin3\lambda_{\rm ZB}}\nonumber\\
\rho_2&=&\rho_3={\sin(3\lambda_{\rm ZB}-u_{\rm ZB})\over\sin 3\lambda_{\rm ZB}}\nonumber\\
\rho_4&=&\rho_5={\sin u_{\rm ZB} \over\sin3\lambda_{\rm ZB}}\nonumber\\
\rho_6&=&\rho_7={\sin u_{\rm ZB}\sin(3\lambda_{\rm ZB}-u_{\rm ZB})\over \sin 2\lambda_{\rm ZB}\sin3\lambda_{\rm ZB}}\nonumber\\
\rho_8&=&{\sin(2\lambda_{\rm ZB}-u_{\rm ZB})\sin(3\lambda_{\rm ZB}-u_{\rm ZB})\over \sin 2\lambda_{\rm ZB}\sin3\lambda_{\rm ZB}}\nonumber\\
\rho_9&=&-{\sin u_{\rm ZB}\sin(\lambda_{\rm ZB}-u_{\rm ZB})\over \sin 2\lambda_{\rm ZB}\sin3\lambda_{\rm ZB}} \,,
\label{ZBweights}
\end{eqnarray}
and the loop fugacity
\begin{equation}
 n=-2\cos4\lambda_{\rm ZB} \,.
\end{equation}
The correspondence with the earlier parameters is
\begin{eqnarray}
u_{\rm YB}=-2iu_{\rm ZB}\nonumber\\
\lambda_{\rm YB}=-i\lambda_{\rm ZB}+i{\pi\over 2} \,,
\end{eqnarray}
that is as well
\begin{eqnarray}
\gamma=\pi-2\lambda_{\rm ZB}\nonumber\\
x=e^{-2iu_{\rm ZB}} \,.
\end{eqnarray}

Finally, the same  appears in the literature with different conventions. The weights in Warnaar, Batchelor and Nienhuis (WBN) \cite{Nienhuis} are obtained  by yet a further  gauge transformation, where each vertex is multiplied by a factor 
\begin{equation}
\mathrm{e}^{\mathrm{i}{\theta_{\rm WBN} \over 8} \left(a-b+c-d\right)}  \,,
\end{equation}
with $a,b,c,d$ denoting the states on the edges (equal to $\pm 1,0$) and
\begin{equation}
 \gamma = \theta_{\rm WBN} \qquad \lambda = \mathrm{i} {3 \theta_{\rm WBN} + 4 \psi_{\rm WBN} \over 4} \,,
 \label{WBNweights}
\end{equation}

Having clarified the relations between all these different conventions, we shall from now on eschrew the use of the various subscripted parameters and stick exclusively with the multiplicative spectral parameter $x$ and the crossing parameter $q$ (or their additive equivalents $\lambda$ and $\gamma$) defined in (\ref{ourparameters}).
Also, our main focus shall be on the vertex model (although we shall occasionnally need the RSOS version as well), and we shall put the emphasis on the algebraic formulation (\ref{IKRmat}) of the $\check{R}$ matrix.

For a system of length $L$ sites with (twisted) periodic boundary conditions, the row to row transfer matrix in the vertex representation is defined as usual, 
\begin{equation}
 T^{(L)}(\lambda) = \mathrm{Tr}_h \left(\check{R}_{h1}(\lambda)\check{R}_{h2}(\lambda)\ldots \check{R}_{hL}(\lambda) \mathrm{e}^{\mathrm{i} \varphi S_z} \right) \,,
 \label{eq:Tmatrix}
 \end{equation}
 where $h$ denotes for the horizontal, auxilliary space, on which acts the operator $S_z$, and $\varphi$ is the twist angle. 

The corresponding spin chain hamiltonian can be obtained by taking the very anisotropic limit $\lambda \to 0$, namely 
\begin{equation}
 H^{(L)} = -\left.\frac{\mathrm{d}}{\mathrm{d}\lambda}\log T^{(L)}(\lambda) \right|_{\lambda=0} \,,
\end{equation}
It is given by equation (6.1) of \cite{Nienhuis}; a  particularly important aspect is that it is not Hermitian.

\subsection{The Bethe equations}

We consider the $q$-deformed $a_2^{(2)}$ model, which is related to the O($n$) loop model as described in section~\ref{sec:models}.
The twist will play a very important role later in this work, and will be introduced in section~\ref{sec:twist}. But until then we shall consider
exclusively the untwisted case.

Setting $q=e^{i\gamma}$, as in (\ref{ourparameters}), the untwisted Bethe equations have the form (in the homogeneous case)
\begin{equation}
\left( {\sinh\left(\lambda_j-i{\gamma\over 2}\right)\over  
\sinh\left(\lambda_j+i{\gamma\over 2}\right)} \right)^N=
\prod_{i\neq j} {\sinh(\lambda_j-\lambda_i-i\gamma)\over  \sinh
(\lambda_j-\lambda_i+i\gamma)}{\cosh\left(\lambda_j-\lambda_i+i{\gamma\over 2}\right)\over  
\cosh\left(\lambda_j-\lambda_i-i{\gamma\over 2}\right)} \,. \label{basiceqs}
\end{equation}
The eigenenergies of the Hamiltonian  take the form
\begin{equation}
E=-c\sum_i {\sin\gamma\over \cosh 2\lambda_j-\cos\gamma} \,. \label{energy}
\end{equation}
We emphasize that the system of Bethe equations and the energy cannot be mapped onto each other under $\gamma\to \pi-\gamma$. This means that we need to consider $\gamma\in [0,\pi]$ and the two signs of energy in (\ref{energy}). 

\subsection{The regimes}
\label{sec:regimes}

When $\gamma$ runs through the interval $[0,\pi]$, the loop fugacity
\begin{equation}
 n = -2 \cos 2 \gamma \,,
\end{equation}
covers the range $[-2,2]$ twice. In view of the two possible signs of $c$ there are therefore four different sets of integrable weights (\ref{ZBweights}) corresponding to each choice of $n \in (-2,2)$. These were referred to as branches 1,2,3,4 in \cite{BloteNienhuis89}. From the point of view of integrability and conformal properties it is however more important to distinguish between three regimes (called I, II and III in \cite{Nienhuis}) that each correspond to distinct structural properties of the Bethe Ansatz solution and of the CFT universality classes.

The case $c<0$ corresponds to the regime I in \cite{Nienhuis}, where the ground state is obtained with  $\lambda_j$ roots having an imaginary part equal to ${\pi\over 2}$. This covers branches 1 and 2 of \cite{BloteNienhuis89} that can be interpreted physically as the dilute and dense phases of the O($n$) model.
 
The case $c>0$ corresponds to regimes II, III in \cite{Nienhuis}. More precisely we have:
\begin{eqnarray}
 \mbox{Regime II}: & \gamma \in [\frac{\pi}{3},\pi] \,, \nonumber \\
 \mbox{Regime III}: & \gamma \in [0,\frac{\pi}{3}] \,.
 \label{regimes23}
\end{eqnarray}
In the conventions of \cite{BloteNienhuis89}, regime III corresponds to the part of branch 3 with $n \in [-2,1]$, whereas regime II covers the remainder or branch 3 and all of branch 4. The physical behavior on branches 3 and 4 was originally suggested \cite{BloteNienhuis89} to be that of the dilute and dense phases of the O($n$) model with an extra Ising degree of freedom (see also Appendix A of \cite{JacobsenKondev}), but it was noticed \cite{BloteNienhuis89} that for the part of branch 3 that corresponds to regime III the convergence of the finite-size numerical estimates for the central charge and critical exponents was anomalously bad. This issue will be clarified in the present paper.

We also note that the value $n=0$ in regime III can be identified with a model of the theta point of polymers (see below); it corresponds to $\gamma = \frac{\pi}{4}$.
The parameter $\theta$ in \cite{Nienhuis} must be identified with $\theta=-\gamma$. 

A naive analysis would suggest the ground state in regimes II and III is obtained by filling a sea of real $\lambda_j$. This is however not the case. In fact, the ground state is made of complexes with imaginary parts close to $\pm {1\over 4}(\pi-\gamma)$:
\begin{equation}
\lambda_j=\hbox{Re }\lambda_j\pm {i\over 4}(\pi-\gamma) \,.
\end{equation}
Note that these $2$-strings are not the usual ones, in that the gap in imaginary parts is equal to ${1\over 2}(\pi-\gamma)$, not $\gamma$; this is possible because the right-hand side of the Bethe equations contains a ratio of cosine terms, that results from the twisting of $a_2$. We emphasize that here we use  the term `twist' is  in the sense of affine quantum algebras and solutions of the Yang Baxter equation \cite{Braden}. This has nothing to do with the `twist angle' that appears in the Bethe equations. 

The same $2$-strings build the ground state in regime II and regime III. Differences arise however in the corrections to the asymptotic shape of the complexes, as well as the  analytical behavior of the Bethe kernels.

The two signs of energy in (\ref{energy}) can be identified with two particular values of the spectral parameter $\lambda$ corresponding to local maxima of the Bethe eigenvalues:
\begin{equation}
 \lambda_c = \mathrm{i}\left( \frac{3 \gamma}{4} - \frac{\pi}{4} \right) \qquad \lambda_d = \mathrm{i}\left( \frac{3 \gamma}{4} + \frac{\pi}{4} \right) \,.
 \label{lambdacd}
\end{equation}
In other words, the states corresponding to the dominant eigenvalues at $\lambda_c$ (resp.\ $\lambda_d$) are the low-energy states for $c>0$ (resp.\ $c<0$).
Note that in terms of the O($n$) weights (\ref{ZBweights}) the choice $\lambda_c$ (resp.\ $\lambda_d$) corresponds to
$\rho_2 = \rho_3 = -\rho_4 = -\rho_5$ and $\rho_8 = \rho_9$
(resp.\ $\rho_2 = \rho_3 = \rho_4 = \rho_5$ and $\rho_8 = \rho_9$).
Both choices are however isotropic points, since the model is gauge invariant under changing the signs of various weights $\rho_i$ \cite{BloteNienhuis89}.

Two other special choices of the spectral parameter are relevant for bringing the square-lattice O($n$) model in equivalence with the better-known
O($n$) model on the hexagonal lattice \cite{Nienhuis82}. Indeed, when $\lambda = {\rm i} \gamma$  (resp.\ $\lambda = \frac{\rm i}{2} (\gamma - \pi)$) we have $\rho_8 = 0$ (resp.\ $\rho_9 = 0$) and the vertices (\ref{onweights}) can be pulled apart vertically (resp.\ horizontally) so as to produce a pair of vertices on the hexagonal lattice. In both cases this produces an isotropic loop model with monomer fugacity $|K| = (2 \sin \gamma/2)^{-1}$; the sign of $K$ is immaterial since loops on the hexagonal lattice have even length. We note that for the isotropic square-lattice model, all three regimes correspond to the dominant eigenvalues for some range of parameter values, cf.~(\ref{regimes23}). However, for the hexagonal-lattice model regime I always determines the dominant eigenvalues, and so in that case regimes II and III can be considered irrelevant.

\section{The continuum limit in regime III: first features}\label{Sec:betheanalysis}

\subsection{The compact part}

We define  Fourier transforms  via
\begin{equation}
f(\omega)=\int {d\lambda\over 2\pi} e^{i \lambda \omega} f(\alpha) \,,
\end{equation}
and use the basic formulas
\begin{eqnarray}
{d\over d\lambda}\ln {\sinh(\lambda+i\alpha)\over \sinh(\lambda-i\alpha}&=&\int_{-\infty}^\infty d\omega\cos\omega\lambda
{\sinh \omega \left({\pi\over 2}-\alpha\right)\over \sinh{\omega\pi\over 2}} \,, \nonumber\\
{d\over d\lambda}\ln {\cosh(\lambda-i\alpha)\over \cosh(\lambda+i\alpha}&=&\int_{-\infty}^\infty d\omega\cos\omega\lambda
{\sinh \omega \alpha\over \sinh{\omega\pi\over 2}} \,.
\end{eqnarray}

Provided $\gamma\leq {\pi\over 3}$, that is in regime III, the bare equations for the centers of the complexes read 
\begin{equation}
\rho+\rho^h={2\sinh {\omega\gamma\over 2}\cosh \omega\left({\pi+\gamma\over 4}\right)\over \sinh {\omega\pi\over 2}}-{\sinh \omega\left({\pi\over 2}-\gamma\right)-\sinh {3\omega\gamma\over 2}+\sinh {\omega\gamma\over 2}\over \sinh {\omega\pi\over 2}}\rho \,.
\end{equation}
The corresponding physical equations are
\begin{equation}
\rho={1\over 2\cosh {\omega\over 4}(\pi-3\gamma)}-{\sinh {\omega\pi\over 2}\over 
4\sinh {\omega\gamma\over 2}\cosh {\omega\over 4}(\pi+\gamma)\cosh {\omega\over 4}(\pi-3\gamma}\rho^h \,,
\end{equation}
from which we deduce the density of roots and the bulk ground state energy, namely
\begin{equation}
E = \int_{-\infty}^{\infty}\mathrm{d}u \rho(u)
\frac{\cosh(2u)\sin{\gamma\over 2} - \cos\gamma}
{\cos^2 \gamma - 2 \cos\gamma \sin{\gamma\over 2}\cosh(2u) + \frac{1}{2}\left(\cosh(4u) - \cos\gamma \right) } \,,
\label{eq:ethermo}
\end{equation}
where
\begin{equation}
 \rho(u) = \frac{1}{\left|\pi - 3 \gamma\right|} \frac{1}{\cosh\left( \frac{2\pi u}{\pi - 3 \gamma} \right)} \,.
\end{equation}

The model has gapless excitations obtained by making holes in the ground state at large rapidities.  Conformal properties can be obtained by studying corrections to scaling. The central charge of the untwisted model is found, after considerable analytical work, to be \cite{Nienhuis}
\begin{equation}
c=2 \,.
\end{equation}
Excitations obtained by removing complexes from the ground state can be handled analytically \cite{Nienhuis}. The final result is in agreement with the usual formula
\cite{KorepinBook}, based on the kernel $K$ of the general form of the bare Bethe equations $\rho+\rho^h=s+K\rho$. 
We  have here, at vanishing frequency, $1-K=4{\gamma\over\pi}$, so the conformal weights read
\begin{equation}
x=\Delta+\bar{\Delta}={\gamma\over 4\pi}n^2+{\pi\over 16\gamma}w^2 \,,
\label{fssIII}
\end{equation}
where  $n$ is twice the number of holes of complexes (a complex containing two Bethe roots; one has in fact $S^z=n$, with $S^z$ the spin of the excitation, in units where arrow in the vertex model carry $S^z=\pm 1$), and the number $w$ measures global shifts of the Fermi sea (backscattering from left to right or right to left).
This expression for the conformal weights is checked numerically in Appendix~\ref{app:C}.
The conformal weights themselves are also easily extracted from the calculation of the momentum, leading to
\begin{equation}
\Delta (\bar{\Delta})={\gamma\over 8\pi}\left(n\pm {\pi\over 2\gamma}w\right)^2 \,.
\end{equation}
This part is easily interpreted as occuring from a compact bosonic degree of freedom, similar to the one characteristic of regime I, and thus the ``usual'' dilute and dense phases of the $O(n)$ model.

Of course, since the central charge is $c=2$, there must be more degrees of freedom. The possibility of having two Majorana fermions --- each contributing an extra ${1\over 2}$ to the central charge --- is quickly excluded from numerics. There is however evidence for a second bosonic degree of freedom, but of a peculiar nature.

\subsection{Searching for the missing part: A duality argument}

There is an elegant way to see why the continuum limit in regime III should be made of one compact and one non compact boson. It uses a duality argument quite similar to the one presented in \cite{FendleyJacobsen} for a related model. 
 
We recall the fact that there are two $\check{R}$ matrices associated with the fundamental representation of $SO(3)$ \cite{GM2}:
\begin{eqnarray}
\check{R}^{(1)} &\propto& 1+{x-1\over x+1}{q+x\over q-x}E+{1-x\over 1+x}{1\over q-q^{-1}}\left(B+B^{-1}\right) \,, \\
\check{R}^{(2)} &\propto &1+{x-1\over x+1}{q^{3}-x\over q^{3}+x}E+{1-x\over 1+x}{1\over q-q^{-1}}\left(B+B^{-1}\right) \,.
\end{eqnarray}
While $\check{R}^{(1)}$ corresponds to the spin-one version of the XXZ spin chain, $\check{R}^{(2)}$ coresponds to the IK model \cite{IzergKorep}. 

The existence of these two $\check{R}$ matrices is more general, and occurs for all the $SO(N)$ algebras in their  fundamental representation:
\begin{eqnarray}
\check{R}^{(1)}&\propto& 1+{x-1\over x+1}{q^{N-2}+x\over q^{N-2}-x}E+{1-x\over 1+x}{1\over q-q^{-1}}\left(B+B^{-1}\right) \,, \\
\check{R}^{(2)}&\propto &1+{x-1\over x+1}{q^{N}-x\over q^{N}+x}E+{1-x\over 1+x}{1\over q-q^{-1}}\left(B+B^{-1}\right) \,.
\end{eqnarray}

The algebraic relations satisfied by the generators  are
\begin{equation}
B-B^{-1}=(q-q^{-1})(1-E) \,,
\end{equation}
where 
\begin{equation}
E=\left(1+[N-1]\right)P_0,~~~[N-1]={q^{N-1}-q^{1-N}\over q-q^{-1}} \,,
\end{equation}
together with
\begin{eqnarray}
B_iB_{i+1}B_i&=&B_{i+1}B_iB_{i+1} \,, \nonumber\\
E_i^2&=&\left(1+[N-1]\right)E_i \,, \nonumber\\
B_iE_i=E_iB_i&=&q^{1-N}E_i \,, \nonumber\\
B_i^{-1}E_i&=&E_iB_i^{-1}=q^{N-1}E_i \,.
\end{eqnarray}
From this data, it is possible to prove generally that one has the following duality at the algebraic level:
\begin{equation}
 \check{R}^{(1)}, \check{R}^{(2)}, N, q, x \leftrightarrow
 \check{R}^{(2)}, \check{R}^{(1)}, \tilde{N}, q, x^{-1} \,,
\end{equation}
where  $q=e^{i\pi\over N+\tilde{N}-2}$. More precisely, what happens is that, if we have a  pair of $\check{R}$ matrices taking the form $\check{R}^{(1)}$ and $\check{R}^{(2)}$ written earlier, for a given $N,q,x$, these $\check{R}$ matrices can also be interpreted as $\check{R}^{(2)}$ and $\check{R}^{(1)}$ for new values of the parameters $\tilde{N}, q, x^{-1}$. 
The proof is straightforward, and simply based on  matching the relations satisfied by the generators, thanks to  $q^{N-1}=-\left(q^{-1}\right)^{\tilde{N}-1}$.

Whenever  $q=e^{i\pi\over p+1}$, with $p \ge 2$ an integer, that is $\gamma={\pi\over p+1}$, we can thus  relate some of  the properties of the IK chain ($\check{R}^{(2)}$, $N=3$) with those of the $SO(p)$ `ordinary' (i.e., type 1) Yang Baxter chain. Now,  careful study of the domains of variation of spectral parameters shows that the  IK model in regime III maps to the  $SO(p)$ model in the regime where its continuum limit \cite{deVega87} is given by  the conformal cosets
\begin{equation}
{SO(p)_1\times SO(p)_2\over SO(p)_3} 
\end{equation}
and their natural Coulomb gas extensions. Finally, it is easy to recognize (see below) these particular cosets as parafermionic theories. The conclusion of this long chain of arguments is thus that the IK model in regime III should be closely related, when $\gamma={\pi\over p+1}$, $p$ an integer, to parafermionic theories. 

While this general argument is perfectly correct, it requires a bit of care to be applied in detail.  A first  point is, that the $\check{R}$ matrices correspond to a very particular choice of `gauge' in the Yang Baxter equation, and that an integrable  periodic system  built out of them will correspond, in the usual vertex model or spin chain language, to a system with a certain twist angle. Hence the equivalence we discuss, strictly speaking, will  hold only  for specifically twisted versions of the models.  Second, this equivalence cannot hold for all aspects of the model since the size of the Hilbert space is not even the same! 
The duality can however be made totally precise and accurate if we move from the vertex models or spin chains to their quantum group restricted, solid on solid version - the RSOS models. The situation is then similar to what was discussed in the case of $SU(N)$ in  \cite{SalAlt91}. 

We now distinguish between even and odd values of $p$:
\begin{eqnarray}
p=2l : & & {SO(2l)_1\times SO(2l)_2\over SO(2l)_3} \,, \qquad \qquad \  \, c=2-{6\over 4l+2} \,, \\
p=2l+1 : & & {SO(2l+1)_1\times SO(2l+1)_2\over SO(2l+1)_3} \,, \quad c=2-{6\over 4l+4} \,.
\end{eqnarray}
The first theory is known to coincide with $Z_{4l}$ parafermions, and is denoted $D_l^{(2l)}$ in \cite{Fateev}. The second is known to coincide with $Z_{4l+2}$ parafermions, and is denoted $B_l^{(2l+1)}$ in \cite{Fateev}. The conclusion of our analysis is thus that proper RSOS versions of the IK model at these values of $p$ in regime III should be described by  these parafermionic theories in the continuum limit. 

More generally - and although the duality argument does not apply in this case, it is natural to expect that there is nothing special with parafermionic theories with even index, and, therefore, that more generally 
the RSOS versions of the IK model in regime III for $q=e^{2i\pi\over k}$, $k$ an integer (the previous cases correspond only  to $k=2p+2$ even) should be described, at low energy, by the $Z_{k-2}$ conformal field theories. The values of $k$ covers the interval $k\in [6,\infty)$, corresponding to the models $Z_4,Z_5,\ldots,Z_\infty$. 

Before using our results on the RSOS version to deduce properties of the vertex model itself, it is important to present 
some numerical checks of our claim. First, we must warn the reader that there are several ways to associate a RSOS model to the IK model. The best known - and probably most physical - way consists in interpreting the loops in the $O(n)$ model as frontiers of height domains, leading eventually do a `dilute' version of the regular RSOS models associated with the 6 vertex model. This is {\sl not} what we must do here, where instead we with to use the full symmetry algebra of the model at the critical point. Therefore, our RSOS models will be obtained by exploiting the full $U_{\q}sl(2)$ spin one symmetry, interpreting the  quantum spins as heights, and obtaining the weights via 6j calculus. In this sense, our RSOS model will be related with (but different from) the `fused' models \cite{Vincent} of the $U_{\q}sl(2)$ hierarchy. The difference lies in the Boltzmann weights: the usual spin one fused model is based on $\check{R}^{(1)}$, while here we use of course $\check{R}^{(2)}$. 

Details of the model, the  face weights and the construction are provided  in Appendix~\ref{app:A}.
For $k$ integer, the heights on each face take integer or half integer values between $0$ and ${k\over 2}-1$. The heights on neighboring sites vary by $0$, $1$ or $-1$, so the  heights are either all integer or all half integer, which defines two a priori different models. 
Numerical diagonalization of the corresponding transfer matrix however shows that the leading eigenvalues are the same for both models. 

\begin{figure}
\begin{center}
 \includegraphics[scale=0.6]{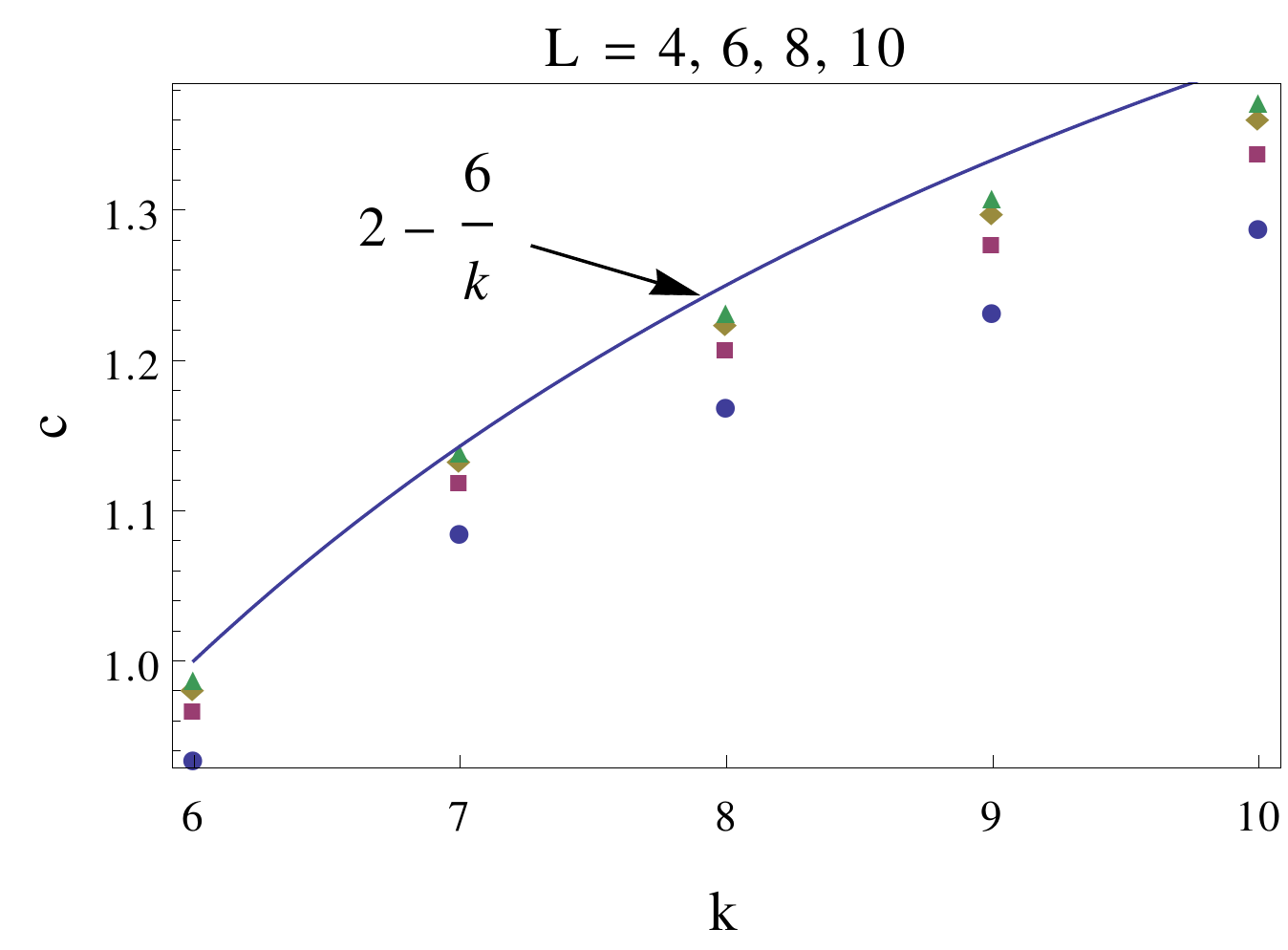}
 \end{center}
\caption{Measures of the central charge of the RSOS model (in any sector) for different values of $k$. 
The blue, purple, yellow and green dots correspond to infinite-size extrapolations from sizes $L=(4,6)$, $(6,8)$, $(8,10)$, $(10,12)$ respectively. 
}                               
\label{fig:c_RSOS}                         
\end{figure}

Measures of the   central charges for $k=6, 7,\ldots,10$ are displayed in figure~\ref{fig:c_RSOS}. They  agree very well with our expectation of a $Z_{k-2}$ parafermionic continuum limit, since for these theories $c=2-{6\over k}$.  
Further evidence regarding the operator content will appear below. 

Now, there is no reason not to assume --- as is usually the case --- that the low energy physics of the IK chain can be described a Coulomb gas  (see e.g.\cite{Gerasimov}), whose field content  (that is, the number and type of free fields) is independent of the coupling $\gamma$ in the regime III. There are, meanwhile, two known Coulomb gas descriptions of the $Z_{k-2}$ parafermions. One of them \cite{Gerasimov} involves a $k$ dependent number of fields, and would lead, via the duality, to a $\gamma$ dependent number of fields in the description of regime III, a fact which is not only unpleasant but for which there is no numerical evidence. The other Coulomb gas description \cite{Narain} involves simply two fields, independently of $k$, and is a natural candidate for us. We will see precisely later how this description fits the IK model. For now, it is enough to recall that it involves two  bosons: one which is compact, and the other which is {\sl non compact}.
The natural conclusion is that the same should apply to the IK model, and that, on top of the compact bosonic degreee of freedom identified earlier, we should have, an extra contribution of $c=1$ to the central charge, corresponding to a non compact degree of freedom. This is what we now establish from a numerical point of view.

\subsection{Numerical evidence for a non compact boson}

\label{section:numerics}

We now discuss the low lying levels in the untwisted (periodic) case in regime III.
We label these levels  by two integers $n$, $j$. The first index is the magnetization ($n=S_z$ in the spin-one language), related to the number of Bethe roots $m_1$ by $m_1 = L - n$. The second index is the level of the excitation within given magnetization sector - we will  make this definition more precise  in terms of the Bethe roots shortly. 
Note that these levels are defined for any value of the twist parameter, even though we first consider the zero-twist case. 
Note also that in this part we restrict to the states of zero momentum, that is, $w=0$ in the notations of (\ref{fssIII}).

For all states we proceed as follows : 
\begin{enumerate}
 \item Obtain eigenvalues and eigenvectors at small sizes $L$, by direct diagonalization of the transfer matrix.
 \item Identify the corresponding Bethe roots by the so-called McCoy method \cite{McCoy1,McCoy2,NepomechieRavanini}: The integrability of the model ensures that the eigenvalues of the properly rescaled transfer matrix are polynomials in $x$, whose coefficients can be expressed in terms of the Bethe roots. Having found an eigenvector, it is then straightforward to identify the corresponding Bethe roots (see Appendix~\ref{app:C} for more details about this method). 
 \item Make a conjecture for the distribution of the roots at large sizes (are they real, forming $2$-strings,\ldots), and write the corresponding real forms of the Bethe equations.
 \item Solve those equations for large sizes $L$ using a Newton-Raphson method.
\end{enumerate}

As announced in section~\ref{sec:regimes}, the ground state in the sector $n=0$ is made of $L \over 2$ 2-strings. 
More generally, the ground state in the sectors of even spin $n$ is made of ${L \over 4} - {n \over 2}$ 2-strings, and the ground state in the sectors of odd spin $n$ is made of ${L \over 4} - {n-1 \over 2}$ 2-strings, plus one root at $\mathrm{i} {\pi \over 2}$.
In all these sectors the $j$'th excitation corresponds to taking $j$ 2-strings off the Fermi sea and replacing them by $j$ antistrings with imaginary part ${\pi \over 2}$.

\begin{figure}
\begin{center}
 \includegraphics[scale=0.6]{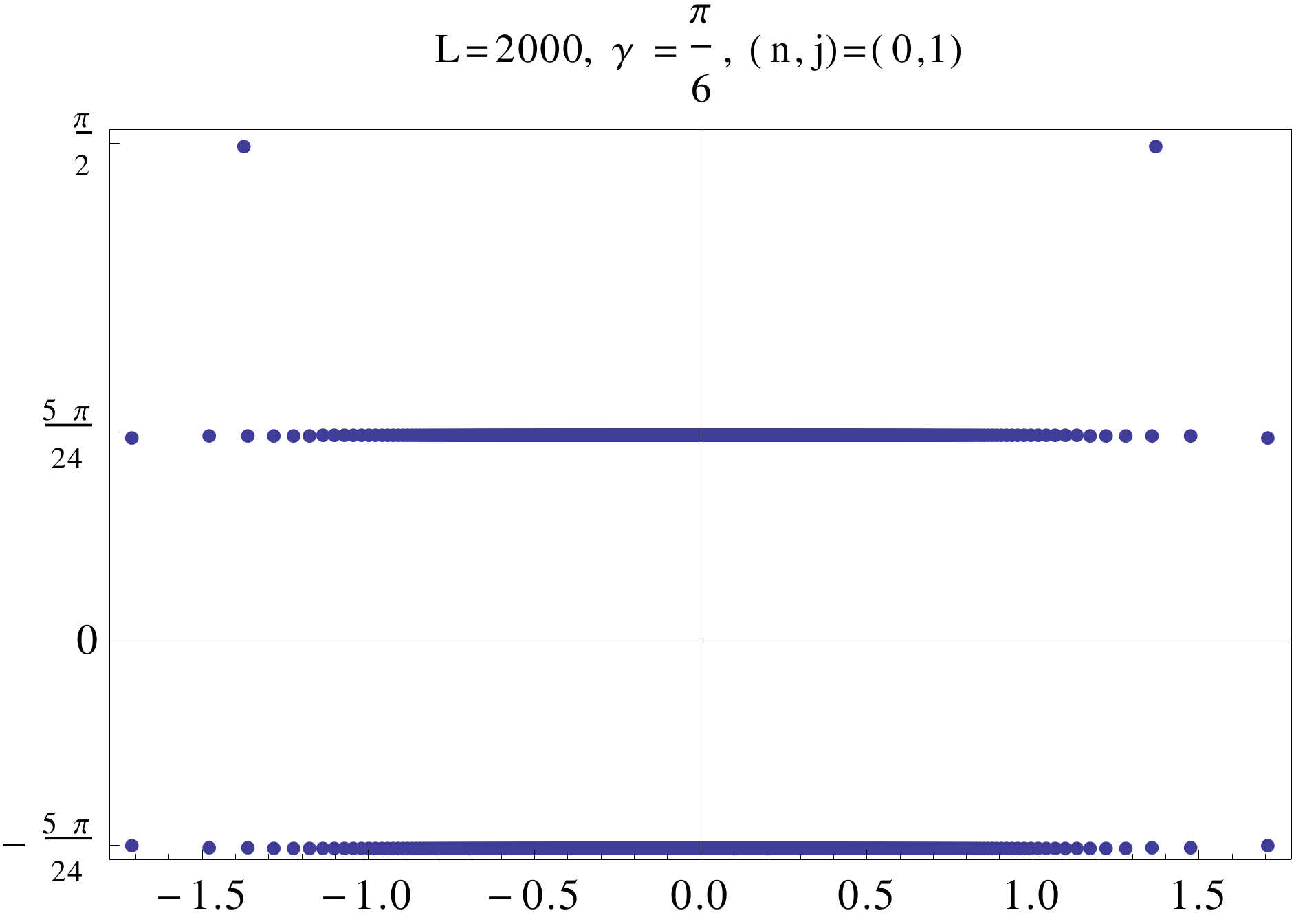}
\end{center}
\caption{Bethe roots $\lambda_j$ for the first excitation in the zero-spin sector in regime III.
The parameters are $\gamma = {\pi \over 6}$ and $L=2000$.
}                           
\label{fig:E1lcpi6_L2000}                          
\end{figure}

For instance, figure~ \ref{fig:E1lcpi6_L2000} shows the roots structure of the excitation $(n,j) = (0,1)$ for $L=2000$.

Let us fix the notations for the study of the conformal weights:
\begin{eqnarray}
 E_{0,0} &=& E_\infty - \frac{\pi v_{\rm F} (\gamma)}{6 L^2}c_{0,0} \,, \\
 E_{n,j} - E_{0,0}  &=&  \frac{2 \pi v_{\rm F} (\gamma)}{L^2} x_{n,j} \,,
\end{eqnarray}
where $x_{n,j} = \Delta_{n,j} + \bar{\Delta}_{n,j}$, and the Fermi velocity is found from the scattering equations to be 
\begin{equation}
 v_{\rm F} ={\pi \over {\pi - 3\gamma}} \,.
\end{equation}

We shall soon see that, unlike the usual case, $c_{0,0}$ contains strong finite-size corrections in $1/\log L$ that tend to mask the asymptotic result $c=2$, hence the distinction between the two quantities in our notations.

To analyze the excitations $x_{n,j}$, we shall find it useful to work with the effective central charges $c_{n,j} \equiv c_{0,0}-12 x_{n,j}$ rather than
with the exponents themselves.

The numerical analysis for sizes ranging up to $L=4800$, and which we present in Appendix~\ref{app:C}, leads us to the following conjecture: 
\begin{equation}
-\frac{c_{n,j}}{12} = n^2\frac{\gamma}{4 \pi} + \left(N_{n,j}\right)^2 \frac{A(\gamma)}{\left[ B_{n,j}(\gamma) + \log L \right]^2}  \,. \label{c/12conj}
\end{equation}
The first term on the right-hand side is in agreement with (\ref{fssIII}) (with $w=0$), whereas the second term is the non compact contribution that we were looking for.
Our numerical results further motivate the conjecture that the $N_{n,j}$ are integers,
\begin{eqnarray}
 N_{n,j} &=& \frac{3 + (-1)^{n+1}}{2} + 2 j  \\
 &=& 1 + \mbox{number of Bethe roots $\lambda_i$ with $\Im \lambda_i = \frac{\pi}{2}$} \,,
\end{eqnarray}
and that the function $A(\gamma)$ is
\begin{equation}
 A(\gamma) = \frac{5}{2} \frac{\gamma \left(\pi - \gamma \right)}{ \left(\pi - 3\gamma \right)^2} \,.
\end{equation}
The functions $B_{n,j}(\gamma)$ are unfortunately not accessible numerically for now. 

To summarize, we have the conjecture:
\begin{equation}
c(L)=2-{12 A(\gamma)\over [B+\log L]^2} \,
\end{equation}
and thus
\begin{equation}
x_{n,j}(L)-{c(L)\over 12}= -\frac{1}{6} + {n^2\gamma\over 4\pi}+(2j+1)^2{A(\gamma)\over [B+\log L]^2} \quad \mbox{for $n$ even} \,.\label{xnone}
\end{equation}
Similarly
\begin{equation}
x_{n,j}(L)-{c(L)\over 12}= -\frac{1}{6} + {n^2\gamma\over 4\pi}+(2j+2)^2{A(\gamma)\over [B+\log L]^2}  \quad \mbox{for $n$ odd} \,.\label{xntwo}
\end{equation}
The difference between the form of corrections for $n$ odd and $n$ even will be discussed below.

\section{The identification of the continuum limit}\label{Sec:BlackHoleIdent}

\subsection{The black hole sigma model and the continuous spectrum}

The main results obtained so far for the IK model in regime III  - that its continuum limit is described by a compact and a non compact boson, and that its RSOS restrictions give  parafermionic theories - are reminiscent 
of those obtained for  the  $Z_2$ staggered 6-vertex model  \cite{IkhlefJS1}. Now, later work \cite{IkhlefJS3} identified the low energy theory of this model with the so called Witten black hole CFT \cite{BlackHoleCFT} (which can also be considered as the coset $SL(2,\mathbb{R})_k/U(1)$).  It is tempting to investigate whether the IK model in regime III is also related with this fascinating CFT. 

Before proceeding in this direction, we find it   useful  to recall some basic features of the black hole CFT. 
The classical action is usually written as
\begin{equation}
 A={k\over 4\pi}\int d^2x \sqrt{h}h^{ij}\left(\partial_i r\partial_j r+\tanh^2 r\partial_i\theta\partial_j\theta\right) \,,
\end{equation}
Here $h$ is the (fixed) world sheet (WS) metric; $r$ and $\theta$ are the fluctuating fields. To understand their meaning and domain of variation, we write the corresponding target space metric 
\begin{equation}
 ds^2={k\over 2} d\sigma^2,~d\sigma^2=(dr)^2+\tanh^2r (d\theta)^2 \,.
\end{equation}
It is associated  to a two dimensional surface in three dimensions with the rough shape of a cigar, hence the familiar name `cigar CFT'. More precisely, the target has rotational invariance around the $z$ axis, while the radius in the $x,y$ plane is given by $\tanh r$, where $r\geq 0$ denotes the geodesic distance from the origin. We have
\begin{eqnarray}
 x=\sinh r\cos\theta \,, \nonumber\\
 y=\sinh r\sin\theta \,,
\end{eqnarray}
with $\theta\in [0,2\pi]$ an angle. The metric reads as well
\begin{equation}
 d\sigma^2={dx^2+dy^2\over 1+x^2+y^2} \,.
\end{equation}
The Gaussian curvature is 
\begin{equation}
 K={2\over 1+x^2+y^2}={2\over\cosh^2 r} \,.
\end{equation}
As discussed by Witten \cite{BlackHoleCFT}, the target space metric is not Ricci flat, so if one thinks of this as an ordinary sigma model, it seems it should flow and not describe a CFT. The point is, that  there is also an extra term making the theory gapless: the `dilaton' field with corresponding action
\begin{equation}
 A={k\over 4\pi}\int d^2x \sqrt{h}h^{ij}\left(\partial_i r\partial_j r+\tanh^2 r\partial_i\theta\partial_j\theta\right)-{1\over 8\pi}\int d^2x \sqrt{h}\Phi(r,\theta)R^{(2)} \,,
\end{equation}
where $R^{(2)}$ is the WS curvature. One has moreover
\begin{equation}
 \Phi(r,\theta)=2\ln\cosh r+\Phi_0 \,.
 \label{dilaton}
\end{equation}
The central charge can be calculated by going to the flat region ($r\to\infty$) where $\Phi\approx 2r$ and $r$ behaves like a non compact free boson. One finds 
\begin{equation}
 c=2+{6\over k} \,.
\end{equation}
This gets corrected in the full quantum theory into 
\begin{equation}
 c=2+{6\over k-2} \,.
\end{equation}
Finally, note that we can introduce the complex field $\Psi=x+iy$ (Kruskal coordinates)  and rewrite the classical action (minus the dilaton term) as
\begin{equation}
 A={k\over 4\pi}\int d^2x \sqrt{h}h^{ij}{\partial_i \Psi\partial_j \Psi^*\over 1+|\Psi|^2} \,.
\end{equation}
 
The best way to understand the physics of this CFT is to study it within the minisuperspace approximation, that is, solve the Laplacian on the target \cite{RibSch}
\begin{equation}
 \Delta=-{2\over k}\left[\partial_r^2+\left(\coth r+ \tanh r \right)\partial_r+\hbox{coth}^2 r\partial_\theta^2\right] \label{LapTar}\,.
\end{equation}
In this limit, there are no $L^2$ normalizable eigenfunctions. The whole spectrum is obtained from $\delta$ function normalizable eigenfunctions, which depend on two parameters: one is $n\in Z$, the angular momentum of rotations around the axis, and the other, $j=-{1\over 2}+is$, is related with the momentum $s$ along the $\rho$ direction of the cigar. The corresponding eigenvalue of the Laplacian reads
\begin{equation}
 x=h+\bar{h}=-{2J(J+1)\over k}+{n^2\over 2k} \,.
\end{equation}
and normalizability restricts the allowed values of $J$ to 
\begin{equation}
 J=-{1\over 2}+is \,, \quad \mbox{with } s \in \mathbb{R} \,.
\end{equation}
These formulas still hold in the quantum theory, after renormalization $k\to k-2$ for the non compact part, and up to the introduction of winding modes:
\begin{equation}
h(\bar{h})=-{J(J+1)\over k-2}+{(n\pm kw)^2\over 4k} \,.
\label{hBHwind}
\end{equation}
Note  that the identity field, which would correspond to $j=0$, does not correspond to a normalizable state, and thus is not present in the spectrum. In fact, the state with the lowest conformal weight in the spectrum (obtained with $j=-{1\over 2}$ ($s=0$), $n=w=0$) has $x={1\over 2(k-2)}$, leading to the effective central charge
\begin{equation}
c_{\rm eff}=2+{6\over k-2}-12\times {1\over 2(k-2)}=2 \,,
\end{equation}
independently of the level $k$. 

Using  the notation $\phi_n^s$ for the eigenfunction of (\ref{LapTar}) with the quantum numbers $n,s$, it is easy to find the asymptotic behavior
at large $r$ 
\begin{equation}
 \phi_n^s\approx e^{-r}e^{-isr+in\theta}+R_{\rm cl}(s,n)e^{-r}e^{isr+in\theta},r\to\infty  \,,  \label{asympWF}
\end{equation}
corresponding to a left and right movers decomposition. The so-called classical reflection amplitude is given by 
\begin{equation}
 R_{\rm cl}(s,n)={\Gamma(2is)\Gamma^2\left({1\over 2}-is+{n\over 2}\right)\over \Gamma(-2is)\Gamma^2\left({1\over 2}+is+{n\over 2}\right)} \,.
\end{equation}
The quantum corrections to the reflection amplitude lead to the expression
\begin{equation}
R_{qu}=R_{\rm cl}\times{\Gamma(1+2isb^2)\over \Gamma(1-2isb^2)}\left({\Gamma(1-b^2)\over \Gamma(1+b^2)}\right)^{2is} \,,
\end{equation} 
where $b^2={1\over k-2}$ \cite{BlackHoleCFT,RibSch}.
 
The consideration of the wave function $\phi_n^s$ is crucial to determine the density of levels $\rho(s)$; this, in turn, is the main quantity allowing one to distinguish one theory with a non compact degree of freedom from another. It is  indeed the direct measurement of $\rho(s)$ in the staggered 6 vertex model - and the successful comparison with known exact formulas for this quantity \cite{Troost} - that led unambiguously to the identification of it continuum limit with the black hole CFT.  Unfortunately, it seems extremely difficult at the present time  to determine analytically or numerically the density of states for the regime III of the O($n$) model. This is due mostly to  the nature of the Bethe roots, which were aligned, in finite size, on real lines for the staggered 6 vertex model, while here their imaginary parts strongly depend on the size. We will, in what follows, use a new and  alternative strategy based on a detailed analysis of the {\sl discrete states}. 

\subsection{Twists and discrete states}
\label{sec:twist}

Another key property of the black hole CFT is that, on top of the continuum of normalizable states discussed in the previous section, it also admits {\sl discrete states}. Although these states cannot be seen in the spectrum of the minisuperspace Laplacian (which is exact only at large $k$), one can understand them qualitatively simply as bound states. Their existence translates into the presence, on top of the continuum of exponents we discussed earlier, of an additional  discrete set of values of exponents which are also allowed in the theory, and should be seen in our model if the identification of the continuum limit is correct. After some discussion \cite{BlackHoleCFT}, the now accepted values of the quantum numbers for these discrete states are \cite{RibSch,Troost}
\begin{equation}
J\in \left[{1-k\over 2},-{1\over 2}\right]\cap \left( \mathbb{N}-{1\over 2}|kw|+{1\over 2}|n| \right) \label{discstat} \,.
\end{equation}
The question for us is now, can we observe these states numerically in our model, and do their properties match the very stringent bounds given in (\ref{discstat})?

It turns out that the analysis is made much easier by moving away from periodic boundary conditions, and introducing a twist
$\varphi$ in the model.  We start from the corresponding  Bethe equations 
\begin{equation}
\left( {\sinh\left(\lambda_j-i{\gamma\over 2}\right)\over  
\sinh\left(\lambda_j+i{\gamma\over 2}\right)} \right)^N=
{\rm e}^{i \varphi}
\prod_{i\neq j} {\sinh(\lambda_j-\lambda_i-i\gamma)\over  \sinh
(\lambda_j-\lambda_i+i\gamma)}{\cosh\left(\lambda_j-\lambda_i+i{\gamma\over 2}\right)\over  
\cosh\left(\lambda_j-\lambda_i-i{\gamma\over 2}\right)} \,. \label{basiceqstwisted}
\end{equation}
In the O($n$) loop model this modifies the weight of non contractible loops from $n$ to $\tilde{n} = 2 \cos \varphi$.
Due to reflection symmetry only the absolute value $|\varphi|$ is physically relevant. We shall however write $\varphi$ and assume $\varphi \ge 0$ throughout.

\begin{figure}
\begin{center}
 \includegraphics[scale=0.6]{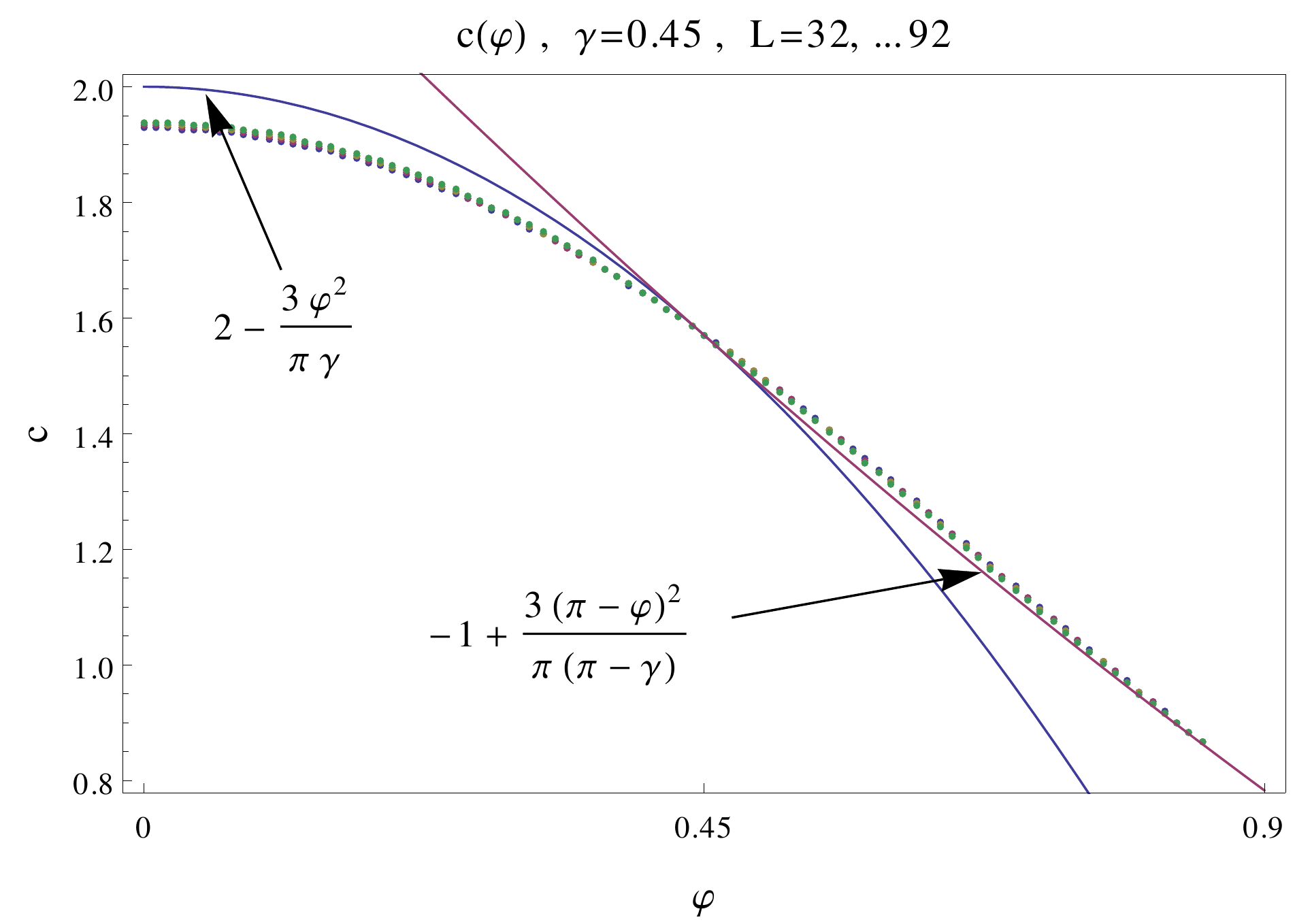}  
 \end{center}
\caption{
Central charge measured as a function of the twist for the particular value $\gamma = 0.45$. 
The two plotted curves are the two branches observed in \cite{Nienhuis}.
}                               
\label{fig:ctwist}                          
\end{figure}

A complete numerical discussion of what happens when the twist $\varphi$ is turned on is provided in Appendix~\ref{app:C}.
In particular, we recovered numerically the (apparently) puzzling observation of \cite{Nienhuis}  that the scaling of the  ground state as a function of the twist angle $\varphi$ has a different analytical form depending on the magnitude of $\varphi$. Recall that, in this reference,  the effective central charge as a function of the twist angle was found to be given by %
\begin{equation}
c = \begin{cases}  2 - {3 \phi^2 \over \pi \gamma } & \mbox{for } \phi\leq\gamma \,,
\\   -1 + {3(\pi - \phi)^2 \over \pi (\pi - \gamma)} & \mbox{for } \gamma \leq \phi \,.
\end{cases}
\label{eq:ctwist}
\end{equation}
where $\phi=\varphi$. We fully  confirmed this by calculations  for sizes up to $L=92$, as shown in figure~\ref{fig:ctwist}.

To proceed, it is convenient to switch to the notation using $k$ (recall that $\gamma={2\pi\over k}$), since this makes the comparison with the black hole theory easier.
The  numerical results  thus give rise to the following two expressions
\begin{eqnarray}
c^* &\equiv&2-6k\left({\varphi\over 2 \pi}\right)^2 \,, \qquad \qquad \quad \ \ \,  \mbox{for } \varphi\leq {2\pi\over k} \,, \nonumber\\
c_1&\equiv&-1+{3k\over k-2}\left({\varphi-\pi\over\pi}\right)^2 \,, \quad \mbox{for } {2\pi\over k}\leq \varphi \,. \label{twocs}.
\end{eqnarray}
We observe that 
\begin{equation}
c_1=c^*+{6\over \pi^2(k-2)}\left(\pi-k{\varphi \over 2}\right)^2 \,.
\label{firstdiscstat}
\end{equation}
This agrees with the fact (clear in figure~\ref{fig:ctwist}) that the two determinations of the central charge are tangent when $\varphi={2\pi\over k}$. Moreover, the two formulas $c_1$ and $c^*$ {\sl never cross}. In other words, the change of behavior does not seem to be related with a level crossing (which is well confirmed numerically), since the `level' that would give rise to $c_1$ would in fact always determine the ground state even for $\varphi\leq {2\pi\over k}$. Rather than level crossing, the most natural explanation for the observed results is that the level giving rise to $c_1$ is simply not there for  $\varphi\leq {2\pi\over k}$. This can be seen also  if we directly put $\varphi=0$ in the formula for $c_1$: we get then the central charge of the black hole sigma model $c_{\rm BH}=2+{6\over k-2}$, which, as mentioned earlier, corresponds to the  non normalizable ground state, that is, a `state' which is not in the spectrum. It is thus natural to expect that the peculiar features observed in this 
regime have to do with normalizability. In fact, what we are 
observing is exactly the appearance of discrete states in the spectrum of the black hole sigma model for large enough twists, in total agreement, as we shall now see, with (\ref{discstat}). 

Let us first analyse the untwisted case. Recall once again that the exponents of the $SL(2,\mathbb{R})/U(1)$ model read
\begin{equation}
h=-{J(J+1)\over k-2}+{(n+kw)^2\over 4k},~~\bar{h}=-{J(J+1)\over k-2}+{(n-kw)^2\over 4k} \,,
\end{equation}
while the central charge is $c_{\rm BH}=2+{6\over k-2}$. For the continuous series, we set $J=-{1\over 2}+is$, so
\begin{equation}
c_{\rm BH}-12(h+\bar{h})=2-24{s^2\over k-2}-6\left({n^2\over k}+kw^2\right) \,.
\end{equation}
For the discrete series, 
\begin{equation}
J\in \left[{1-k\over 2},-{1\over 2}\right]\cap \left( \mathbb{N}-{1\over 2}|kw|+{1\over 2}|n| \right) \label{discstat} \,.
\end{equation}
with 
the corresponding effective central charge 
\begin{equation}
c_{\rm BH}-12(h+\bar{h})=2+{6\over k-2}(2J+1)^2-6\left({n^2\over k}+kw^2\right) \,.
\end{equation}
Note in particular the fact that the $j$ term leads to a positive additional contribution to the value $c=2$, while the $s$ term leads to a negative one. 

\begin{figure}
\centering
\includegraphics[scale=.60]{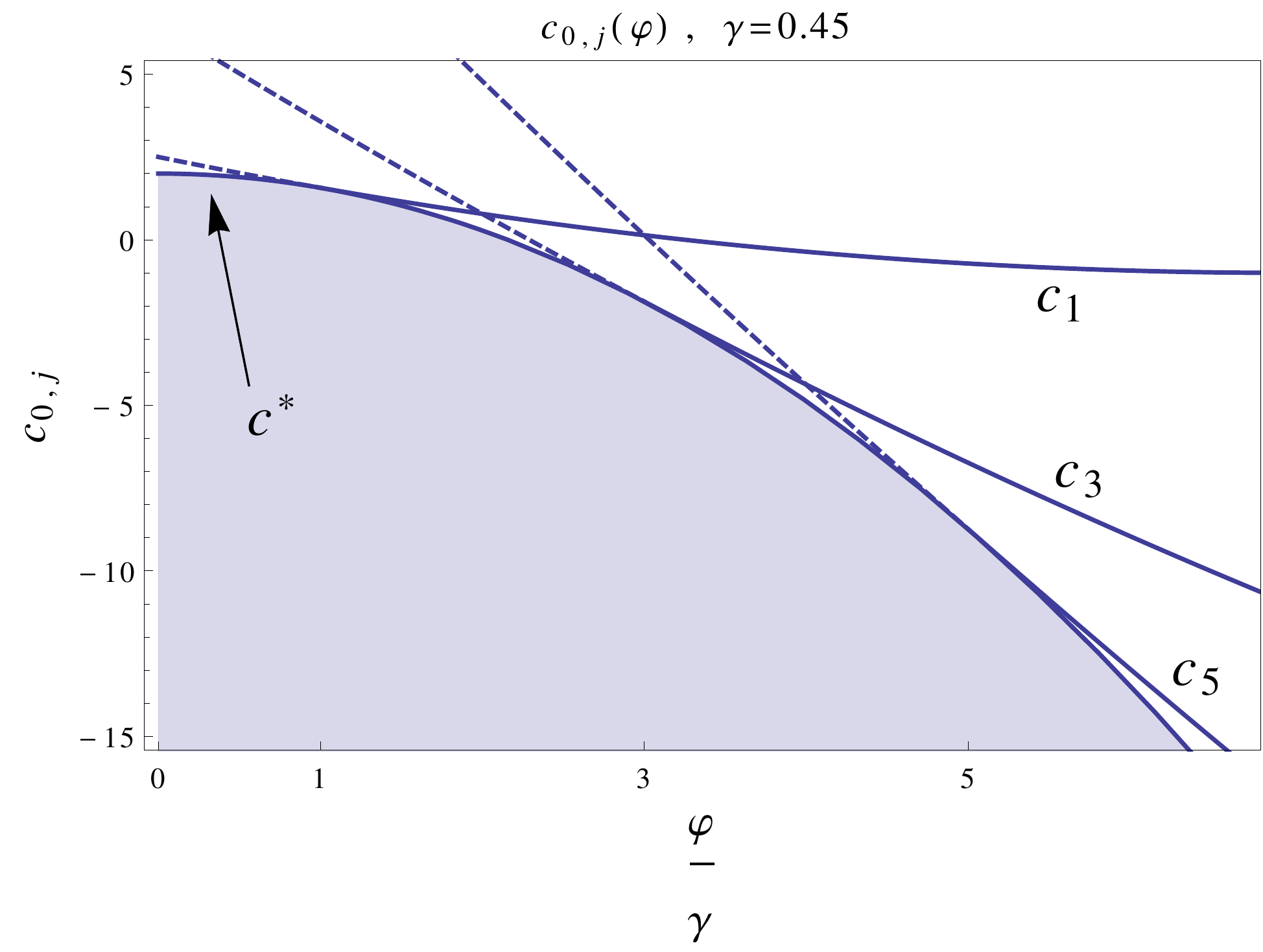}
    \caption{Discrete levels popping out of the continuum in the sector $n=0$ as the twist $\varphi$ is increased (we represent the associated central charges $c_{0,j}$, not the real levels). The shaded zone is the continuum. The discrete states are represented by a dashed line when they are not normalizable}
\label{fig1}
\end{figure}

Now, all these results are for the theory without twist. Meanwhile, for $\varphi\leq {2\pi\over k}$, the effective value of the central charge (generalizing $c^*$ in (\ref{twocs}) to the case $n,w\neq 0$) turns out to be
\begin{equation}
c^*=2-6k\left({\varphi\over 2\pi}+w\right)^2-6{n^2\over k} \,.  \label{cstar}
\end{equation}
This suggests that the twist $\varphi$ provides, on the lattice, a way to adjust, at fixed coupling, the winding $w$ to continuously varying values. This can in fact be justified directly within the black hole sigma model theory, but we will not discuss this further here. Rather, we will now assume that the identification
\begin{equation}
w\equiv {\varphi\over 2\pi} \,,
\end{equation}
holds to identify which possible discrete states might contribute to the spectrum, and see what the consequences are. Choosing first  $n=0$, the set of discrete states (\ref{discstat})  becomes therefore
\begin{equation}
J\in \left[{1-k\over 2},-{1\over 2}\right]\cap \left( \mathbb{N}-{1\over 2}\left|{k\varphi\over 2\pi}\right| \right) \,.\label{discstat1}
\end{equation}
Forgetting for now the bound on the left of the interval, we see  that a new normalizable state should appear whenever there exists an integer $p\in  \mathbb{N}$ such that 
\begin{equation}
p-{1\over 2}\left|{k\varphi\over 2\pi}\right|\leq -{1\over 2} \,.
\end{equation}
The associated conformal weight leads to the effective central charge
\begin{eqnarray}
c_{2p+1}&=&c_{\rm BH}-12(h+\bar{h})=2+{6\over \pi^2(k-2)}\left[(2p+1)\pi-k{\varphi \over 2}\right]^2-6k\left({\varphi\over\pi}\right)^2 \,, \nonumber\\
&=&c^*+{6\over \pi^2(k-2)}\left[(2p+1)\pi -k{\varphi \over 2}\right]^2 \,. \label{c2p+1}
\end{eqnarray}
The behavior of the effective central charge associated with all the exponents is represented in figure~\ref{fig1}. We see that discrete levels should `pop out' of the continuum at regular values of $\varphi$: this is precisely in agreement with the observation (\ref{twocs}) for $p=0$. We also note that, when a level for a given $p$ comes out of the continuum, it does so as the $j=p^{th}$ excited level. 

\begin{figure}
\centering
     \includegraphics[scale=.60]{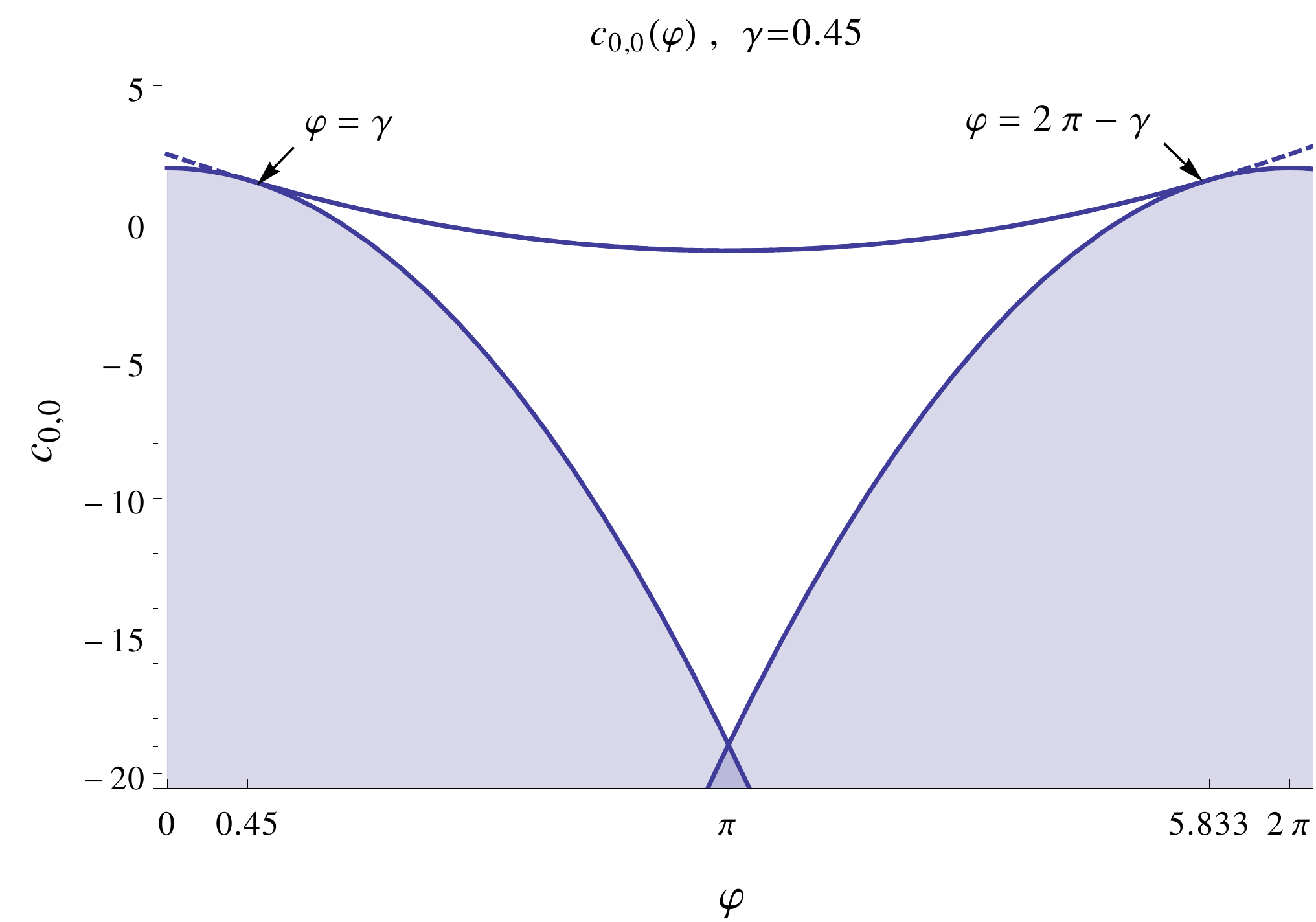}
    \caption{The ground state as a function of $\varphi$. In the domain studied, it is obtained by three sections of parabolas.
 }
\label{fig2}
\end{figure}

Before discussing the other values of $p$, we now go back to the bound on the left of the interval (\ref{discstat1}). It is reached, for $c_{2p+1}$, whenever
\begin{equation}
{1-k\over 2}=p-{1\over 2}\left|{k\varphi\over2\pi}\right| \,,
\end{equation}
or
\begin{equation}
{\varphi\over2\pi}=1+{2p-1\over k} \,.
\end{equation}
If we take the level $p=0$ for instance, we see that it should disappear when ${\varphi\over2\pi}={k-1\over k}$. But this is precisely where the function $c_1$ intersects the `excited winding mode' where in $c^*$, instead of taking just $\varphi$, we can take $\varphi\pm 2 w$, where $w$ is the winding number (usually called electric charge in the context of Coulomb gas analysis). The corresponding central charge is in fact 
\begin{equation}
c_1=-1+{3k\over k-2}\left(1-{2\over k}\right)^2=2-{6\over k} \,,
\end{equation}
again. What happens is then sketched in figure~\ref{fig2}: the discrete level `returns to the continuum'. Of course, $\varphi$ strictly speaking is only defined in the interval $[-2\pi,2\pi]$, but we can always extend this definition precisely by shifting by charges $w$. Alternatively, we can also increase $k$ to put more features in the fundamental $\varphi$ interval. 

We can also investigate what happens when the magnetization is non zero.  In this case, the `normal' value of the effective central charge  is (generalizing what we called $c^*$) earlier
\begin{equation}
c_n^*=2-6k\left({\varphi\over2\pi}\right)^2-{6n^2\over k} \,.
\end{equation}
(note that with $n,\varphi\neq 0$,  $h\neq\bar{h}$, but we focus on the observed ground state, determined by $h+\bar{h}$). 
Moreover, the magnetization  $n$ on the lattice corresponds exactly   to the number $n$ in the general formula for discrete states (\ref{discstat}.
This means, for instance, that the first discrete level now appears for 
\begin{equation}
-{1\over 2}\left|{k\varphi\over2\pi}\right|+{|n|\over 2}\leq -{1\over 2} \,.
\end{equation}
or 
\begin{equation}
\varphi\geq {2\pi\over k}(1+|n|) \,.
\end{equation}
This should  lead to the following measure of the central charge:
\begin{equation}
c_{\rm mea}=c^*_n+{6\over \pi^2(k-2)}\left(1+|n|-k{\varphi \over 2}\right)^2 \,,
\end{equation}
something also observed by Nienhuis {\em et al.} \cite{Nienhuis}. 

A detailed  analysis (summarized   in Appendix~\ref{app:C}), leads to  the main numerical  result:

{\centering
\fbox{
\begin{minipage}{6in}
As long as we do not intersect the excited winding modes, the effective central charge of the excitation $E_{n}^{(j)}$ is given by 
\begin{equation}
  c_{n,j} = \begin{cases}  c_n^{*}  = 2 - 3 {\varphi^2 \over \pi \gamma} -\frac{3 \gamma n^2}{\pi}  & \mbox{for } \varphi\leq\ (|n|+2 j+1) \gamma 
\\  c_n^{*}  + {3 \over \gamma(\pi - \gamma)}\left[\varphi - (|n|+2 j+1)\gamma\right]^2 & \mbox{for } \varphi \geq (|n|+2 j+1)  \gamma  \,.
\end{cases}
\label{eq:cm}
\end{equation}
\end{minipage}
} }

This is still in perfect agreement with the observations of Nienhuis {\em et al.} \cite{Nienhuis} for the effective central charges $c_{n,0}$; see eqs.~(5.4) and (5.5) in the given reference. We conclude therefore that the existence and precise bounds for the discrete states in the black hole CFT are exactly reproduced by the low energy properties of the IK model in regime III, fully confirming the identification of the continuum limit. We also note that similar phenomena associated with normalizable states appearing and disappearing from the spectrum of the black hole CFT have been observed in \cite{Giveon}; see figure~1 in that reference.

We note that, as a function of $\varphi$, the ground state of the chain exhibits a discontinuity of the second order derivative.

\subsection{The parafermions}

It is fair to ask at this stage why the model we are studying displays, on the one hand, all the required properties to be identified with the $SL(2,\mathbb{R})/U(1)$ theory at low energy, and, on the other hand, is related with parafermionic theories, whose coset contruction is well known to be of the $SU(2)/U(1)$ type. The answer to this puzzle lies in the discrete states: the $Z_{k-2}$ parafermionic theory can be obtained, for $\gamma={2\pi\over k}$, $k$ integer, by selecting the discrete states for a set of discrete values of the twist.

To explain this further, we first  recall more details about parafermions.  The central charge of the parafermions is obtained with the twist ${\varphi\over\pi}={1\over k}$, and it reads
\begin{equation}
c_{\rm PF}=2-{6\over k} \,. \label{cpf}
\end{equation}

Let us put ourselves exactly at the twist where the new discrete state appears that corresponds to $c_{2p+1}$, that is 
\begin{equation}
{\varphi\over2\pi}={2p+1\over k} \,.
\end{equation}
At this value, the discrete level that pops out of the continuum coincides with the `normal' level determining $c^*$. This leads to a gap over (\ref{cpf}) given by
\begin{equation}
h_{\rm PF}={1\over 24}\left[c^*\left({\varphi\over2\pi}={2p+1\over k}\right)-c^*\left({\varphi\over2\pi}={1\over 4k}\right)\right]={(2p+1)^2\over 4k}-{1\over 4k}={p(p+1)\over k} \,.
\end{equation}
This is exactly the dimension of the thermal fields \cite{FZparafermions}, which in the usual notations (see, e.g., \cite{GepnerQiu}) we will denote by $\Phi_0^{2p}$ (and their conformal weights by $h^{2p}_0$). Meanwhile, we can consider the conformal weights associated with the discrete levels already present when the twist approaches this value: they are encoded in $c_{2p+1-2r}$, with $0\leq r\leq p$. A short calculation leads to
\begin{equation}
c_{2p+1-2r}-c_{2p+1}=24 {r^2\over k-2} \,,
\end{equation}
from which it follows that 
\begin{equation}
h_{\rm PF}=h^{2p}_0-{(2r)^2\over 4(k-2)}=h^{2p}_{2r} \,,
\end{equation}
the dimension of the parafermionic descendent of $\Phi^{2p}_0$. In particular, the weight associated with $r=p$ comes from the first discrete level, $c_1$, and it is $h_{2p}^{2p}$, the highest weight of the parafermionic module:
\begin{equation}
h_{2p}^{2p}={2p(k-2-2p)\over 2k(k-2)} \,.
\end{equation}
We can also check that the bounds on the discrete states correspond to the bounds for the parafermionic theory. For instance, the discrete state for $c_1$ is always associated with 
\begin{equation}
h_q^q={q(k-2-q)\over 2k(k-2)} \,,
\end{equation}
where here $q$ denotes the $Z_{k-2}$ charge (and obviously not the quantum group deformation parameter).
Moreover, the angle being ${\varphi\over2\pi}={q+1\over k}$ can be used in the formulas only for  $q\leq k-2$, as expected. The levels for $q$ an odd number of course give the corresponding parafermionic modules (with no thermal field). This suggests that the $Z_{k-2}$ theory is entirely described by the discrete levels, for twist angles ${\varphi\over2\pi}={q+1\over k}$. Of course, to really establish this result would require a lot more work on this side of the problem; in particular, one would have to perform the analysis of all the descendents in each parafermionic module etc. The precise relationship between parafermions and the $SL(2,\mathbb{R})/U(1)$ theory will be discussed elsewhere.

\subsection{Massive deformations}

As most often in  integrable lattice models, a gapless spin chain/vertex model, whose continuum limit is a certain CFT, is only one particular case of a more general family of models, whose low energy description corresponds to integrable perturbations of that CFT. This family of models is most conveniently obtained by {\sl imaginary}  staggering of the spectral parameters \cite{ResSal}.%
\footnote{The $Z_2$ staggered 6-vertex model is obtained by a real staggering, with a fixed magnitude. The massive deformation, in contrast, corresponds to staggering with a purely imaginary shift, of varying magnitude.}
A well-known example of this is the antiferromagnetic XXZ chain, whose continuum limit is a $c=1$ boson, and whose staggering provides a regularization of the sine-Gordon model. In the RSOS versions, this gives rise to minimal models perturbed by $\Phi_{13}$ (here, the labels refer to the labels in the Kac formula)\cite{ResSal}. The natural `cousin' of this model is the IK spin chain, whose continuum limit is also a $c=1$ boson in the regime I, and whose staggering provides now a regularization 
of the (imaginary) Bullough-Dodd theory \cite{BullDodd}, or, in the RSOS version, minimal models perturbed by $\Phi_{21}$\cite{Birgit2}. 

In earlier studies \cite{JS_AFPotts} of the $Z_2$ staggered 6-vertex model at its parafermionic points $Z_{k-2}$, it was found that (imaginary) staggering produced a deformation by the first energy operator $\epsilon_1$, of dimension $h=\bar{h}={2\over k}$. A similar analysis of the staggering of the IK model in regime III suggests a pertrubation by the {\sl second} energy operator, of dimension $h=\bar{h}={6\over k}$. This is rather natural, following the duality argument. Indeed, 
it is well known that staggering in  spin chains based on a group $G$, whose continuum limit is a diagonal coset  $G\times G/G$, corresponds  `adjoint' perturbations \cite{ABL}.
In our case, $G=SO(p)$, the diagonal cosets are 
\begin{equation}
{SO(p)_1\times SO(p)_2\over SO(p)_3} \,,
\end{equation}
and the adjoint perturbation has conformal weight
\begin{equation}
h=1- {C^*\over 1+2+C^*}=1-{p-2\over p+1}={3\over p+1}={6\over k} \,,
\end{equation}
where $C^*$ is the dual Coxeter number of $SO(p)$. 

Of course, staggering corresponds to well defined perturbations in the black hole sigma model as well. It is not so clear, however, how one can go from these to the parafermionic perturbations. Within the black hole theory, there are definitely two known integrable perturbations\cite{BD85,Schiff,Fateevperso}. One is the standard complex sinh-Gordon model with action
\begin{equation}
A_{\rm CSG_0}={k\over 4\pi}\int \left[{\partial_\mu\Psi\partial\mu\bar{\Psi}\over 1+|\Psi|^2}+m^2|\Psi|^2\right]d^2x \,,
\label{facti}
\end{equation}
and the other is a variant with a more complicated perturbation
\begin{equation}
A_{\rm CSG_1}={k\over 4\pi}\int\left[{\partial_\mu\Psi\partial\mu\bar{\Psi}\over 1+|\Psi|^2}+m^2\left(|\Psi|^2+|\Psi|^4\right)\right]d^2x 
\label{factii}
\end{equation}
(the notation $CSG_0$ and $CSG_1$ is borrowed from \cite{BD85}).

What these have to do with the $Z_{k-2}$ theories is not entirely clear, and requires a better understanding of the issue of discrete states. This will be discussed elsewhere. 

The important point for us is that the existence of the two versions of the theory matches closely the existence of two lattice models. Moreover, one can go further and analyze, for instance, conserved quantities in both models. They behave very differently. The $CSG_0$ model has local conserved quantities at all grades, both odd and even, a fact deeply related with the underlying symmetry of the $Z_2$ staggered 6-vertex model \cite{IkhlefJS1,IkhlefJS2,IkhlefJS3,CanduIkhlef}. Meanwhile, the variant (\ref{factii}) has only local conserved quantities at odd grades,  in agreement with the fact that there is no extra symmetry to distinguish regime III from the other regimes. 

Let us now go back to (imaginary) staggering of the spectral parameter in the IK model. The corresponding Bethe equations are 
\begin{equation}
\left( {\sinh\left(\lambda_j-\Lambda-i{\gamma\over 2}\right)\over  
\sinh\left(\lambda_j-\Lambda+i{\gamma\over 2}\right)} \right)^{L/2}
\left( {\sinh\left(\lambda_j+\Lambda-i{\gamma\over 2}\right)\over  
\sinh\left(\lambda_j+\Lambda+i{\gamma\over 2}\right)} \right)^{L/2}
=
\prod_{i\neq j} {\sinh(\lambda_j-\lambda_i-i\gamma)\over  \sinh
(\lambda_j-\lambda_i+i\gamma)}{\cosh\left(\lambda_j-\lambda_i+i{\gamma\over 2}\right)\over  
\cosh\left(\lambda_j-\lambda_i-i{\gamma\over 2}\right)} \,,
\label{basicstageqs}
\end{equation}
and the new  physical equations
\begin{equation}
\rho={\cos\Lambda\omega\over 2\cosh {\omega\over 4}(\pi-3\gamma)}-{\sinh {\omega\pi\over 2}\over 
4\sinh {\omega\gamma\over 2}\cosh {\omega\over 4}(\pi+\gamma)\cosh {\omega\over 4}(\pi-3\gamma}\rho^h \,.
\end{equation}
The modification of the source term now leads to massive excitations a finite rapidities. The corresponding mass scale is obtained by taking the leading pole at 
\begin{equation}
\omega={2i\pi\over \pi-3\gamma} \,,
\end{equation}
giving a mass scale 
\begin{equation}
M\propto [\hbox{length}]^{-1}\propto \exp\left[-{2\Lambda\pi\over \pi-3\gamma}\right] \,.
\end{equation}
Meanwhile,  staggering corresponds to perturbing the microscopic RSOS model by a certain operator $\Phi_{\rm pert}$ with a bare coupling constant $g$. The relationship between $g_\Lambda$ and $\Lambda$ is independent of the regime, and can be found from the analysis of regime I (see Appendix~\ref{app:B}). One has 
\begin{equation}
g_\Lambda\propto e^{-4\Lambda} \,.
\end{equation}
It follows that 
\begin{equation}
g_\Lambda\propto [\hbox{length}]^{{6\gamma\over\pi}-2}=[\hbox{length}]^{{12\over k}-2} \,.
\end{equation}
This corresponds to coupling to an operator, in the $Z_{k-2}$ theory, of dimension
\begin{equation}
h={6\over k} \,.
\end{equation}
This agrees with the general form of the $l$'th energy operator, of dimension $h={l(l+1)\over k}$ --- and here $l=2$. 

A particularly important fact for us is that, for the perturbation by the second energy operator, there exists a set of non local conserved quantities generated by the second parafermionic operator $\Psi_2$, of dimension $h=2{(k-4)\over k-2}=2-{4\over k-2}$. Knowing this operator allows us to write down the general form of corrections to the low energy limit in the lattice model. Indeed, the model with and without staggering have the same integrable structure, and irrelevant  operators determining the deviation from the fixed point in the lattice model must be compatible with this structure \cite{Lukyanov}. In other words, for our model we must have
\begin{equation}
A^{\rm lattice}_{IK}=A^{\rm CFT}+g_{\rm irr}\int \Psi_2\bar{\Psi}_2d^2x+\ldots \,,
\end{equation}
while for the staggered 6-vertex model we have 
\begin{equation}
A^{\rm lattice}_{6V}=A^{\rm CFT}+g'_{\rm irr}\int T\bar{T} d^2x\ldots \,.
\end{equation}
The coupling constants $g_{\rm irr}$ and $g_{\rm irr}'$ measure the distance to the fixed points. While the dimension of $g'_{\rm irr}$ is independent of $k$, $[g'_{\rm irr}]=[\hbox{length}]^{-2}$, the dimenson of $g_{\rm irr}$ varies:
\begin{equation}
[g_{\rm irr}]=[\hbox{length}]^{2-{8\over k-2}} \,. \label{girrscal}
\end{equation}
We see therefore that the IK model, which at very low energy is described by the black hole sigma model - or, in the RSOS versions, the $Z_{k-2}$ parafermions - differ from these theories at intermediate energies by corrections to scaling which become increasingly important as $k\to 6$. The coupling of these corrections  is  of course irrelevant throughout regime III (i.e., for $k > 6$), and becomes marginal precisely at the transition to regime II (i.e., for $k=6$), corresponding to $\gamma={2 \pi \over k}={\pi \over 3}$. This is in total agreement with the lattice results. 
Moreover, the fact that the end of the regime is associated with an operator becoming marginally relevant suggests that the associated singularity is essential. This is something we briefly discuss now.

\subsection{Transition between regimes II and III}

We now investigate the possible singularity of the free energy and ground state energy at the transition between regimes II and III, namely at $\gamma = {\pi \over 3}$. 

\begin{figure}
\begin{center}
 \includegraphics[scale=0.6]{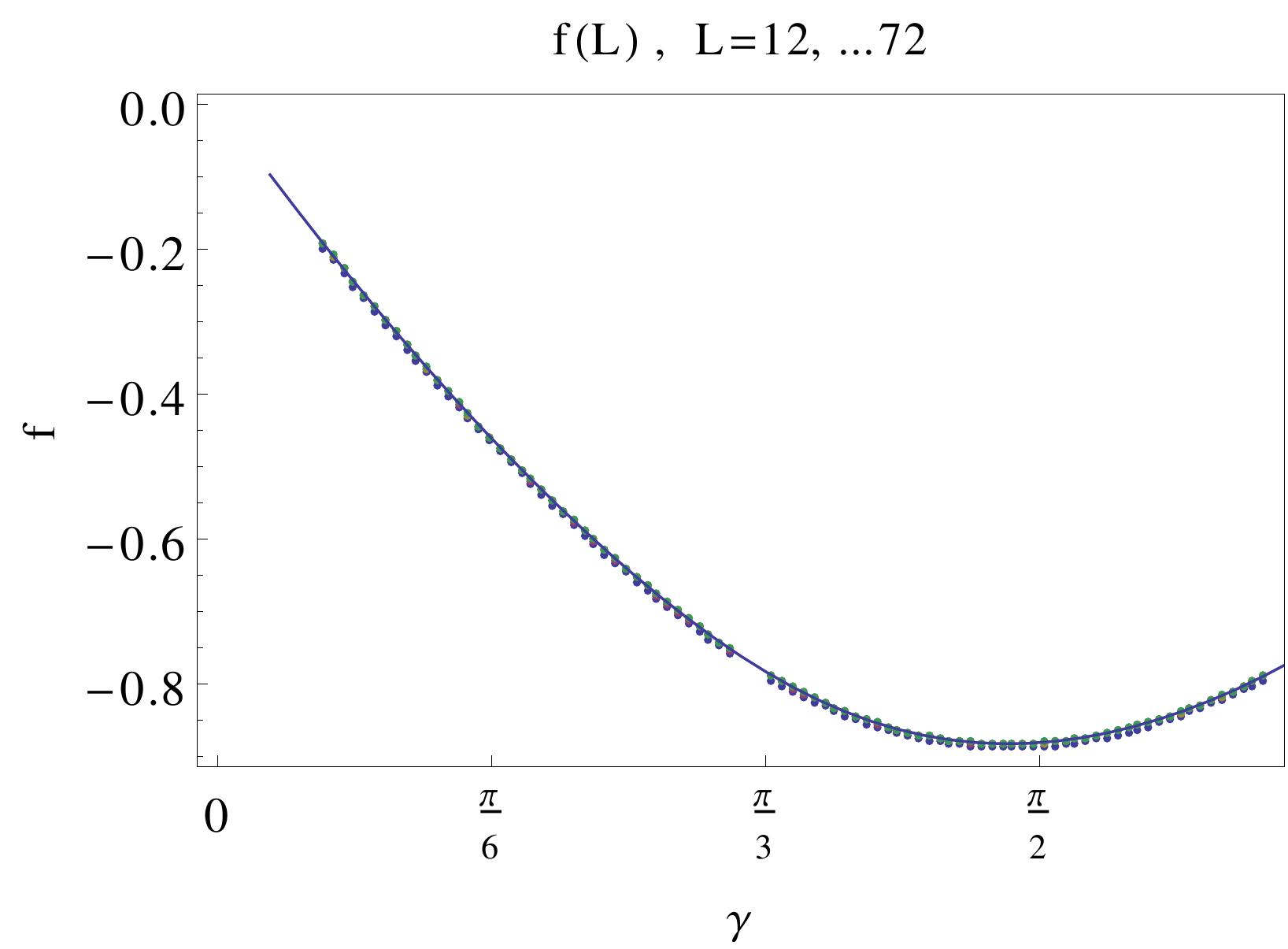}
 \end{center}
\caption{Free energy $f$ across the transition between regimes II and III for various sizes. We plotted in comparison the analytical expression (\ref{fetrans23}) expected in the thermodynamic limit.}                            
\label{fig:Ftransition}                          
\end{figure}

We consider first the free energy. 
The Bethe eigenvalues are written in \cite{GM2} as 
\begin{equation}
 \Lambda_L (\lambda)= a^L \frac{Q\left(\lambda+{ \mathrm{i}\gamma\over 2}\right)}{Q\left( \lambda-{ \mathrm{i}\gamma\over 2} \right)} 
 + d^L \frac{Q\left(\lambda-2\mathrm{i}\gamma+{ \mathrm{i}\pi\over 2}\right)}{Q\left(\lambda-\mathrm{i}\gamma+{ \mathrm{i}\pi\over 2} \right)} 
 + b^L  \frac{Q\left(\lambda-{3 \mathrm{i}\gamma\over 2}\right)}{Q\left( \lambda-{ \mathrm{i}\gamma\over 2} \right)} 
 \frac{Q\left(\lambda+{ \mathrm{i}\pi\over 2}\right)}{Q\left(\lambda-\mathrm{i}\gamma+{ \mathrm{i}\pi\over 2} \right)} \,,
\label{eq:LambdaL}
\end{equation}
where $Q(\lambda)=\prod_j \sinh(\lambda - \lambda_j)$.
At $\lambda = \lambda_c =\mathrm{i}\left({3\gamma \over 4}-{\pi \over 4}\right)$, $a=d={b \over 1-2\sin{\gamma \over 2}}$. 
Moreover for roots configurations as that of the ground state in the thermodynamic limit (2-strings of imaginary part $\pm \left({\pi \over 4}-{\gamma \over 4} \right)$),
the three terms in (\ref{eq:LambdaL}) happen to be exactly equal.
We define the free energy as 
\begin{equation}
 f = -\frac{1}{L}\log \frac{\Lambda_L(\lambda_c)}{a^L} \,.
\end{equation}
In the thermodynamic limit we therefore have the following analytic expression
\begin{equation}
 f = -\int_{-\infty}^{\infty}\mathrm{d}u\rho(u) \log \frac{\left(\cosh 2 u - \cos 2\gamma \right)\left(\cosh 2 u + \cos 3\gamma \right)}{2\sinh^2 u \left( \cos\gamma+\cosh 2u \right)} \,. \label{fetrans23}
\end{equation}
It coincides very well with the finite size results of figure~\ref{fig:Ftransition}.

From numerics $f$ appears to be continuous at $\gamma = {\pi \over 3}$, and so do its derivatives.
Possible essential singularities at ${\pi \over 3}$ can be studied by removing the obviously regular part of the integral and rescaling the roots, yielding
\begin{equation}
 f = f_{\rm reg} - \frac{1}{2} \int_{-\infty}^{\infty}\mathrm{d}v \frac{1}{\cosh \pi v} \log \frac{\cosh (\pi-3\gamma)v - \cos(\pi - 3\gamma)}{\cosh (\pi-3\gamma)v -1} \,.
\end{equation}
Setting $2 \mu = \pi-3\gamma$, the integral is exactly that considered in appendix A of \cite{Abraham}, with $\eta = 1$, that is $\phi_0 = 0$, namely
\begin{equation}
 f = f_{\rm reg} - 2 \Phi(\mu) \,.
\end{equation}

It corresponds to the high temperature free energy of the F model, which is shown to allow for an analytic continuation in the whole complex plane, except from the line $\Im(\mu) = 0, \Re(\mu) \leq 0$.
There it is shown that the low temperature free energy has a series expansion in terms of $\mu$ which coincides with the former analytic continuation on the real axis. 
The conclusion is that the free energy in both low and high temperature phases is described by one single function, infinitely differentiable, but nonanalytic at $\mu = 0$. This corresponds to an infinite order phase transition between the two phases.

In the $a_2^{(2)}$ case the free energy has the same analytical expression in both regimes $\mu > 0$ and $\mu <0$. It is in particular even, and we consider the former case only. 
$\Phi(\mu)$ allows for a series expansion in $\mu$ in the domain $\mathbb{C} \setminus \mathbb{R}_-$, given by eq.~(A41) of \cite{Abraham} with $s=0$.
The series has zero radius of convergence, which ends the demonstration of the non analyticity of $\Phi$ at $\mu = 0$ even though all its derivatives exist and are continuous there.
The explicit form of the singular part is given in equation (8.11.14) of \cite{BaxterBook}, where one needs to take $v=0$. 
One has more precisely 
\begin{equation}
 f_{\rm sing} \propto \mathrm{e}^{- \frac{\pi^2}{\mu}} \,,
 \label{eq:fsing}
\end{equation}
where the proportionality coefficient is finite and nonzero as $\mu \to 0$.  A similar result holds for the energy itself.

\subsection{The amplitude $A(\gamma)$}

We now come back to the question of the amplitude $A(\gamma)$. The black hole sigma model has a non compact target, and therefore the naive density of states is infinite. Of course, in a physical theory, the effective size of the target space is always cut off. This leads in turn to a regularization of the divergence in the density of states, and the existence, on top of the divergence proportional to the size of the target, to a finite part - the latter being the quantity of interest. The phenomenon is easier to study if one works in the semi classical approximation. The cutoff for the target space implies that the wave function vanishes at $r=r_M$, which leads, from (\ref{asympWF}) to  the quantization condition 
\begin{equation}
2isr_M+\ln R_{cl} =i\pi (2N+1) \,, \quad \mbox{with } N \in \mathbb{Z} \,.
\end{equation}
Note that we have added here an extra phase shift $e^{i\pi}$ at the extremity $r_M$ since the wave function is required to vanish there. The $R_{cl}$ is the phase shift at the tip of the cigar, and $e^{2isr_M}$ the phase gathered by the plane waves due to motion along the axis of the cigar.

From this one gets a quantization of $s$, and thus a  density of states
\begin{equation}
\rho(s)={\delta N\over \delta s}={1\over\pi} r_M+{1\over 2i\pi}{d\ln R_{cl}\over ds} \,. \label{densnaive}
\end{equation}
The first term is the part that becomes divergent when the cutoff is removed $r_M\to\infty$. The second term in this expression is the finite contribution to the density of states. As discussed in \cite{IkhlefJS3,CanduIkhlef}, it  is in fact related with the amplitude $B$ in equation (\ref{c/12conj}). But while this amplitude could be determined by a mixture of analytical and numerical techniques in the case of the $Z_2$ staggered 6-vertex model, the presence of strings in the ground state of the IK  model in regime III prevents us from doing this --- for now at least. We will thus focus on the first term in (\ref{densnaive}), which, as we will see shortly, is closely related with the  amplitude $A$.

Indeed, we will make the natural assumption that the cutoff in target space is  related to the amplitude of the irrelevant operator determining the corrections to scaling in the lattice model. In general, this operator can be written in the Liouville form $e^{\alpha r}$, up to derivatives. A term  $g_{irr}\times e^{\alpha r}\times \hbox{derivatives}$  in the action will give rise to an effective cutoff 
\begin{equation}
r_M\approx {\ln g_{irr}\over\alpha} \,.
\end{equation}
Assuming that $\alpha$ does not depend on the coupling $k$, we then find, from the scaling dimension of $g_{\rm irr}$ determined earlier in (\ref{girrscal})
\begin{equation}
r_M\propto \left(2-{8\over k-2}\right)\ln L \,,
\end{equation}
and thus, at leading order, we expect that 
\begin{equation}
\rho(s)={1\over\pi}\ln r_M\propto {1\over\pi}\left(2-{8\over k-2}\right)\ln L\label{rhosi} \,.
\end{equation}
To connect this result with the amplitude $A$, it is now convenient to think of the generating function of levels, which takes the form
\begin{equation}
Z=\sum_{h,\bar{h}} y^{h-c/24}\bar{y}^{\bar{h}-c/24}=\left(y\bar{y}\right)^{-1/6} \sum_{n,w} y^{(n+kw)^2/4k}\bar{y}^{(n-kw)^2/4k}\sum_{N=-\infty}^\infty (y\bar{y})^{A(\gamma) (2N+1)^2/(\ln L+B(\gamma))^2} \,.
\end{equation}

We can, at large $L$, transform the discrete sum into an integral, and match the exponents with the expected form ${s^2\over 4(k-2)}$. Hence 
\begin{equation}
\sum_{j=-\infty}^\infty (y\bar{y})^{A(\gamma) (2j+1)^2/(\ln L+B(\gamma))^2}\rightarrow
{\ln L+B(\gamma)\over 4\sqrt{(k-2) A(\gamma)}}\int_{-\infty}^\infty ds (y\bar{y})^{s^2/4(k-2)} \,.
\end{equation}
This shows that the density of states obtained from the lattice model is, at leading order
\begin{equation}
\rho(s)={\ln L\over 4\sqrt{(k-2)A(\gamma)}}\label{rhosii} \,.
\end{equation}
Comparing the expressions (\ref{rhosi}) and (\ref{rhosii}) shows that
\begin{equation}
A(\gamma)\propto {\gamma(\pi-\gamma)\over (\pi-3\gamma)^2} \,,
\end{equation}
in agreement with the numerical results%
\footnote{We note that in the $Z_2$ staggered 6-vertex model the amplitude $A(\gamma)$ takes a different form \cite{IkhlefJS3,CanduIkhlef}. This is related with the fact that the operator determining the corrections to scaling in this model has a coupling-independent dimension --- which is expected, since the model admits only local conserved quantities.}
(see eq.~\ref{numresforA} in Appendix~\ref{app:C}).

We note that, while the argument determined $A(\gamma)$ unambiguously (up to a proportionality constant), it would work as well if the quantization condition was of the form 
\begin{equation}
2isr_M+\ln R_{cl} =i(2\pi N+\delta) \,, \quad \mbox{with } N \in \mathbb{Z} \,.
\end{equation}
with $\delta$ an arbitrary additional phase shift. Such a phase shift might arise from a modification of the reflection amplitude, or a more detailed analysis of the effect and nature of the cutoff at $r_M$. To our knowledge, this has never been discussed before in the literature. It does play a role of course in the finite size corrections for levels of the spin chain. We see for instance in equations (\ref{xnone}) and (\ref{xntwo}) that the phase shift $\delta$ seems to depend on the parity of the quantum number $n$ with 
\begin{eqnarray}
\delta=\pi, ~~n\hbox{ even}\nonumber\\
\delta=2\pi,~~n\hbox{ odd} \,.
\label{phashi}
\end{eqnarray}
While this may seem a minor detail, no such effect was observed in the staggered 6 vertex model, which suggests that the precise mapping of the lattice models to the black hole CFT is a bit different in the two cases. The result (\ref{phashi}) has an important consequence however: the scaling of the ground state corresponds to $n=0,\delta=\pi$ in the IK case, and thus is affected by strong ${1\over( \ln L)^2}$ corrections. This did not occur in the staggered 6 vertex model at all, where the corresponding phase shift was $\delta=0$, and the ground state did not exhibit any unusual corrections to scaling.

\section{Conclusion}

The fact that regime III of the IK model has a non compact continuum limit seems rather unexpected and exciting to us. It is particularly remarkable that, while the staggered 6 vertex model may have appeared a bit artificial in its construction, the $O(n)$ model, which is deeply related with the IK model, and, as will be discussed elsewhere, exhibits the same non compact features, is perfectly natural and `physical'. Our observation of a non compact continuum limit will have especially striking consequences in this case \cite{VJS2}. 

On the other hand, the IK model is technically much more challenging than the staggered 6 vertex model. In the latter case, it was easy to establish analytically the existence of the continuum spectrum by using singularities in the Bethe ansatz kernels. There is no such obvious possibility here. While the roots have imaginary parts that depend on the size of the system, naive equations written for the complexes in the thermodynamic limit do not exhibit any special feature suggesting the existence of a non compact component. Indeed, 
adding roots with imaginary part ${\pi\over 2}$ and density $\sigma$ leads to the following naive physical equations
\begin{equation}
\rho={1\over 2\cosh {\omega\over 4}(\pi-3\gamma)}-{\sinh {\omega\pi\over 2}\over 
4\sinh {\omega\gamma\over 2}\cosh {\omega\over 4}(\pi+\gamma)\cosh {\omega\over 4}(\pi-3\gamma)}\rho^h+
{1\over 2\cosh{\omega\over 4}(\pi+\gamma)}\sigma \,,
\end{equation}
and 
\begin{equation}
\sigma+\sigma^h= {1\over 2\cosh{\omega\over 4}(\pi+\gamma)}\rho^h+{\cosh {\omega\over 4}(\pi-3\gamma)\over \cosh{\omega\over4}(\pi+\gamma)}\sigma \,.
\end{equation}
which, at least at first sight, seem perfectly innocuous. This means also that the world of spin chains with non compact continuum limit might be much bigger than initially suspected. This will be discussed elsewhere. Our future work will in fact include a discussion of the continuous spectrum of exponents in the $O(n)$ and in particular the polymer models \cite{VJS2}, a discussion of $a_3^{(2)}$ with applications to a two colour loop model \cite{VJS3}, and a general discussion of $a_n^{(2)}$ \cite{VJS4}.

\bigskip

\noindent {\bf Acknowledgments}: we thank Y.~Ikhlef, B.~Nienhuis, V.~Pasquier and V.~Schomerus for useful discussions. This work was supported in part by the French Agence Nationale pour la Recherche (ANR Projet 2010 Blanc SIMI4: DIME).

\appendix

\section{RSOS version}
\label{app:A}

In this appendix we define an RSOS model related to the $a_{2}^{(2)}$ model in regime III. 

There exists a geometrical way of assigning an RSOS model to the O($n$) model, such that the loops are the frontiers of height domains \cite{NienhuisRSOS}.
In its simplest form, this procedure requires $n = 2 \cos(\pi/p)$, where $p=3,4,\ldots$ is an integer and the heights take the values $h=1,2,\ldots,p-1$.
In particular one must have $n \ge 1$. Since regime III is defined by $\gamma \in [0,\frac{\pi}{3})$ we have $n = -2 \cos 2 \gamma < 1$, so the geometrical
procedure is not applicable. It is true that one can extend the working of \cite{NienhuisRSOS} to values $n = 2 \cos(\pi p'/p)$, where $p' < p$ and ${\rm gcd}(p,p')=1$,
but this is at the price of introducing negative weights.
We shall therefore use instead an algebraic RSOS construction exploiting the $U_\q sl(2)$ symmetry, and follow the lines of \cite{Vincent}. 

We introduce the parameter 
\begin{equation}
q\equiv \q^2 \,, \quad \q=e^{i\gamma/2}=e^{i\pi/k} \,.
\end{equation}
One can then rewrite the $\check{R}^{(2)}$ matrix (\ref{IKRmat}) as 
\begin{equation}
\check{R}^{(2)}\propto P_2+{\q^4 x-1\over \q^4-x}P_1+{\q^6x+1\over \q^6+x}P_0 \,,
\label{eq:R2appendix}
\end{equation}
where $P_2,P_1,P_0$ are projectors of $U_{\q}sl(2)$ in the product of two spin one representations. Here the quantum group is defined in the usual way
\begin{equation}
[S^z,S^\pm]=\pm S^\pm,~~~[S^+,S^-]=[2S^z]_{\q} \,,
\end{equation}
with the notation for the quantum numbers
\begin{equation}
[x]_\q\equiv {\q^x-\q^{-x}\over \q-\q^{-1}} \,.
\end{equation}

The isotropic point $\lambda_c$ given by (\ref{lambdacd}) and corresponding to regime III is obtained for $x = - \mathrm{i} \mathfrak{q}^3$. We have then
\begin{eqnarray}
\check{R}^{(2)} &\propto& P_2 + \frac{\sin\left( \frac{7\gamma}{4}- \frac{\pi }{4} \right)}{\sin\left( \frac{\gamma}{4}+ \frac{\pi }{4} \right)} P_1 + \frac{\cos\left( \frac{9\gamma}{4} - \frac{\pi}{4} \right)}{\cos\left( \frac{3\gamma}{4} + \frac{\pi}{4} \right)} P_0 \\
&=& P_2 + \frac{\sin\left( \frac{7\pi}{2 k}- \frac{\pi }{4} \right)}{\sin\left( \frac{\pi}{2k}+ \frac{\pi }{4} \right)} P_1 + \left( 1 + 2 \sin \left( \frac{3 \pi}{k} \right) \right) P_0 \,,
\end{eqnarray}
where we recall that $\gamma = \frac{2\pi}{k}$. Note that although the first expression is singular at $\gamma = \frac{\pi}{3}$, the second rewriting is well-defined at $k=6$ and yields $\check{R}^{(2)} \propto P_2 + P_1 + 3 P_0$.

The projectors can be reformulated in their (R)SOS version, namely in terms of height variables. 
One way to do this relies on quantum 6j calculus and follows closely Pasquier \cite{Vincent}. We review this procedure in the first section of this appendix, and correct in passing a few sign mistakes in the reference cited. 
In a second section we provide an independent check using the RSOS version of the matrix $R^{(1)}$, which corresponds to the usual fused RSOS(2,2) model. 

\subsection{6j derivation}
\label{RSOS_R2}

We temporarily generalize to the case where the vertex model variables lie in some representation $\lambda$ (here $1$), and consider the action of some projector $P_J$, defined in its vertex representation as
\begin{equation}
\left(P_J \right)_{\alpha \beta}^{\alpha' \beta'} = \sum_{M \in J} \langle \lambda \alpha' \lambda \beta' | J M \rangle \langle J M  | \lambda \alpha \lambda \beta \rangle  \,.
\end{equation}
To obtain the path representation of $P_J$ we start from a representation $j_1$ on the left corner of the face, and use  
\begin{eqnarray}
 (j_{i-1})\otimes (\lambda) &=& \oplus_{j_i} (j_i) , \qquad  (j_i) \otimes (\lambda) = \oplus_{j_{i+1}}(j_{i+1}) \, \\   
  (j_{i-1})\otimes (\lambda) &=& \oplus_{j'_i} (j'_i) , \qquad (j'_i) \otimes (\lambda) = \oplus_{j_{i+1}}(j_{i+1}) \,.
\end{eqnarray}
This is nothing but the two decompositions associated to the fusion of representations going around the face from below and above respectively
(cf.\ the diagram in (\ref{face6j}) below).
Using (4.7) in \cite{SaleurZuber} on both sides, and keeping in mind that $P_J$ projects on only one term in $\sum_J$, we end up with
\begin{eqnarray}
 & & \sum_{m_{i-1}} |j_{i-1}m_{i-1} \rangle \langle  j_{i-1}m_{i-1} | \otimes P_J \nonumber \\
 & = & \sum_{j_i, j_i '} \sum_{j_{i+1}, m_{i+1}}
 \left\{  \begin{array}{ccc}
  j_{i-1} & \lambda & j_i \\ \lambda & j_{i+1} & J  \\ \end{array}   \right\}_q
   \left\{  \begin{array}{ccc}
  j_{i-1} & \lambda & j'_i \\ \lambda & j_{i+1} & J  \\ \end{array}   \right\}_q
|j_{i+1}m_{i+1} \rangle \langle  j_{i+1}m_{i+1} | \,.
\end{eqnarray}
We can hence deduce the IRF transfer matrix element
\begin{equation}
\begin{tikzpicture}[scale=0.5]
 \draw[thick] (0,1) -- (1,0) -- (2,1) -- (1,2) -- cycle;
 \draw (1,0) node[below] {$j_i$};
 \draw (0,1) node[left] {$\sigma(P_J)^{(j_i j'_i)}_{j_{i-1} j_{i+1}} = j_{i-1}$};
 \draw (2,1) node[right] {$j_{i+1} = \left\{  \begin{array}{ccc}
  j_{i-1} & \lambda & j_i \\ \lambda & j_{i+1} & J  \\ \end{array}   \right\}_q
   \left\{  \begin{array}{ccc}
  j_{i-1} & \lambda & j'_i \\ \lambda & j_{i+1} & J  \\ \end{array}   \right\}_q$ \,.};
 \draw (1,2) node[above] {$j'_i$};
 \draw (1,1) node {$P_J$};  \,.
\end{tikzpicture}
  \label{face6j}
\end{equation}
 
Note that what we defined as the 6j symbols do not have the same normalization as the conventional ones. They correspond rather to the Racah coefficients, related to the conventional quantum 6j coefficients by 
\begin{equation}
 \left\{  \begin{array}{ccc}
  j_1 & j_2 & j_{12} \\ j_3 & j & j_{23}  \\ \end{array}   \right\}_q =  (-1)^{j_1 + j_2 + j_3 + j} \left( [2 j_{12} + 1][2 j_{23} +1] \right)^{1 \over 2}\left\{  \begin{array}{ccc}
  j_1 & j_2 & j_{12} \\ j_3 & j & j_{23}  \\ \end{array}   \right\}_{q, \text{conventional}} \,.
 \end{equation}
We use for the latter the book \cite{qgroups}, section 3.5. There a generic formula for the 6j is given in eq.~(88).
From this formula we extract:
\begin{eqnarray}
  \left\{  \begin{array}{ccc}
  j & 1 & j \\ 1 & j & 0  \\ \end{array}   \right\}_q  &=& (-1)^{2j+1}\sqrt{\frac{1}{[3]}} \\
   \left\{  \begin{array}{ccc}
  j & 1 & j+1 \\ 1 & j & 0  \\ \end{array}   \right\}_q  &=& (-1)^{2j}\sqrt{\frac{[2j+3]}{[3][2j+1]}} \\
   \left\{  \begin{array}{ccc}
  j & 1 & j-1 \\ 1 & j & 0  \\ \end{array}   \right\}_q  &=& (-1)^{2j}\sqrt{\frac{[2j-1]}{[3][2j+1]}} \\
    \left\{  \begin{array}{ccc}
  j & 1 & j \\ 1 & j & 1  \\ \end{array}   \right\}_q  &=& (-1)^{2j}\sqrt{\frac{[2]}{[4]}} \frac{\mathfrak{q}^{2j+1}+\mathfrak{q}^{-2j-1}}{\sqrt{[2j][2j+2]}} \\
    \left\{  \begin{array}{ccc}
  j & 1 & j+1 \\ 1 & j & 1  \\ \end{array}   \right\}_q  &=& (-1)^{2j}\sqrt{\frac{[2]}{[4]}} \sqrt{\frac{[2j][2j+3]}{[2j+1][2j+2]}} \\
    \left\{  \begin{array}{ccc}
  j & 1 & j-1 \\ 1 & j & 1  \\ \end{array}   \right\}_q  &=& (-1)^{2j+1}\sqrt{\frac{[2]}{[4]}} \sqrt{\frac{[2j-1][2j+2]}{[2j][2j+1]}} \\
     \left\{  \begin{array}{ccc}
  j & 1 & j \\ 1 & j+1 & 1  \\ \end{array}   \right\}_q  &=& (-1)^{2j+1}\sqrt{\frac{[2]}{[4]}} \sqrt{\frac{[2j]}{[2j+2]}} \\
       \left\{  \begin{array}{ccc}
  j & 1 & j+1 \\ 1 & j+1 & 1  \\ \end{array}   \right\}_q  &=& (-1)^{2j}\sqrt{\frac{[2]}{[4]}} \sqrt{\frac{[2j+4]}{[2j+2]}}  \\
       \left\{  \begin{array}{ccc}
  j & 1 & j \\ 1 & j-1 & 1  \\ \end{array}   \right\}_q  &=& (-1)^{2j}\sqrt{\frac{[2]}{[4]}} \sqrt{\frac{[2j+2]}{[2j]}} 
   \\
       \left\{  \begin{array}{ccc}
  j & 1 & j-1 \\ 1 & j-1 & 1  \\ \end{array}   \right\}_q  &=& (-1)^{2j-1}\sqrt{\frac{[2]}{[4]}} \sqrt{\frac{[2j-2]}{[2j]}} 
     \\
       \left\{  \begin{array}{ccc}
  j & 1 & j\pm 1 \\ 1 & j\pm 2 & 2  \\ \end{array}   \right\}_q  &=& 1 \,.
  \end{eqnarray}
  
Some of the above formulae are redundant due to the symmetries of the original 6j symbols (namely invariance under the permutation of any two columns or under the simultaneous vertical permutation of elements in any two columns) which in our case translate into  
\begin{equation}
    \left\{  \begin{array}{ccc}
  j_{1} & j_{23} & j \\ j_3 & j_{12} & j_{2}  \\ \end{array}   \right\}_q
  =(-1)^{j_2+j-j_{12}-j_{23}}   \sqrt{\frac{[2 j +1][2j_2 +1]}{[2 j_{12} +1][2 j_{23} + 1]}}    \left\{  \begin{array}{ccc}
  j_{1} & j_2 & j_{12}  \\ j_3 &  j & j_{23}  \\ \end{array}   \right\}_q \,,
\end{equation}
and
\begin{equation}
     \left\{  \begin{array}{ccc}
  j & j_2 & j_{12}  \\ j_3 &  j_1 & j_{23}  \\ \end{array}   \right\}_q
  =  \left\{  \begin{array}{ccc}
  j_{1} & j_2 & j_{12}  \\ j_3 &  j & j_{23}  \\ \end{array}   \right\}_q \,.
\end{equation}
{}From there we get the face matrix elements of the projectors. 

Compared to what is given in eq.~(13) of \cite{Vincent} the projectors are block diagonal, and we can consider the different subspaces. 

\begin{paragraph}{Sector $j_{i-1} = j_{i+1} \equiv j$.}

We write the matrices in the basis $(j-1 , j , j+1)$.
As in \cite{Vincent} we write the elements above the diagonal as $*$; they are equal to those placed symetrically below the diagonal.
\begin{eqnarray}
 P_0 &=& \frac{1}{[3]}\frac{1}{[2j+1]}\left(
 \begin{array}{ccc}
[2j-1] &  * & * \\
  -\sqrt{[2j-1][2j+1]} & [2j+1] & *\\
  \sqrt{[2j-1][2j+3]} &-\sqrt{[2j+1][2j+3]} &  [2j+3] \\
 \end{array}
 \right)
 \\ 
  P_1 &=& \frac{[2]}{[4]}\left(
 \begin{array}{ccc}
1- \frac{[2]}{[2j][2j+1]} &  * & * \\
  -\sqrt{[2j-1]\over [2j+1]}\frac{[4j+2]}{[2j+1]}\frac{1}{[2j]} & \frac{1}{[2j][2j+2]}\left(\frac{[4j+2]}{[2j+1]}\right)^2  & *\\
  -\sqrt{[2j-1][2j+3] \over [2j+1][2j+1]} & \sqrt{[2j+3] \over [2j+1]}\frac{[4j+2]}{[2j+1]}\frac{1}{[2j]} &  1- \frac{[2]}{[2j+1][2j+2]} \\
 \end{array}
 \right) \,.
 \label{P0P1matrix}
\end{eqnarray}
$P_1$ is exactly the same as in \cite{Vincent}. $P_0$, however, differs by some signs. 
It can be checked that our $P_0$ and $P_1$ in this sector are orthogonal (as they should be), whereas those of \cite{Vincent} are not. 
Note that $P_2$ can be inferred from the identity $P_2 = 1 - P_0 - P_1$.
\end{paragraph}

\begin{paragraph}{Sectors $j_{i+1} = j_{i-1}\pm 1$.}
 
In the sector $j_{i+1} = j_{i-1}+1 \equiv j+1$ we have, in the basis $(j,j+1)$, 
\begin{equation}
 P_1 = \frac{[2]}{[4]} \left( \begin{array}{cc} 
                               {[2j] \over [2j+2]} & * \\
                                 -{\sqrt{[2j][2j+4]} \over [2j+2]} &   {[2j+4] \over [2j+2]} \\
                              \end{array}
 \right) \,.
\end{equation}
In the sector $j_{i+1} = j_{i-1}-1 \equiv j-1$ we have, in the basis $(j,j-1)$, 
\begin{equation}
 P_1 = \frac{[2]}{[4]} \left( \begin{array}{cc} 
                               {[2j+2] \over [2j]} & * \\
                                 -{\sqrt{[2j-2][2j+2]} \over [2j]} &   {[2j-2] \over [2j]} \\
                              \end{array}
 \right) \,.
\end{equation}
In the two last formulas there are some off-diagonal sign differences with the expression given by Pasquier \cite{Vincent}.

\end{paragraph}


\medskip
  
We now recast the resulting weights for the $A_2^{(2)}$ model in a more convenient notation.
We rewrite the weight of a face in a way that matches the diagram of (\ref{face6j}) [the time evolution is now in the standard
North-East direction]:
\begin{equation}
  \left. \left( \begin{array}{cc} 2 j_{i-1}+1 & 2 j'_i +1 \\ 2 j_i +1  & 2 j_{i+1} +1 \end{array} \right| u  \right) \,,
 \end{equation}
and make a slight change of gauge, 
\begin{equation}
  \left. \left( \begin{array}{cc} 2 j +1 & 2 j'_i +1 \\ 2 j_i +1  & 2 j +1 \end{array} \right| u  \right)  \to (-1)^{j'_i - j_i} 
    \left. \left( \begin{array}{cc} 2 j +1 & 2 j'_i +1 \\ 2 j_i +1  & 2 j +1 \end{array} \right| u  \right) \,.
\end{equation}
Setting $x = \mathrm{e}^{2 \lambda} = \mathrm{e}^{2 \mathrm{i} u}$, the decomposition (\ref{eq:R2appendix}) yields the following face weights 
\begin{eqnarray}
   \left. \left( \begin{array}{cc} 2 j +1 & 2 j +1 \\ 2 j +1  & 2 j +1 \end{array} \right| u  \right) &=&  
   1 
   + 2\frac{\sin u \cos{2\pi \over k}}{\sin\left({2\pi \over k}-\lambda\right)}   \frac{[2]}{[4]}\frac{1}{[2j][2j+2]}\left( {[4j+2] \over [2j+1]} \right)^2  
  - 2\frac{\sin u \sin{3\pi \over k}}{\cos\left({3\pi \over k}-\lambda\right)} \frac{1}{[3]}
   \nonumber \\
      \left. \left( \begin{array}{cc} 2 j +1 & 2 j + 3 \\ 2 j + 3  & 2 j +1 \end{array} \right| u  \right) &=&  
   1 
   + 2\frac{\sin u \cos{2\pi \over k}}{\sin\left({2\pi \over k}-\lambda\right)}   \frac{[2]}{[4]}\left(1-\frac{[2]}{[2j+1][2j+2]}\right) 
  - 2\frac{\sin u \sin{3\pi \over k}}{\cos\left({3\pi \over k}-\lambda\right)} \frac{1}{[3]}\frac{[2j+3]}{[2j+1]}
   \nonumber \\
      \left. \left( \begin{array}{cc} 2 j +1 & 2 j - 1 \\ 2 j -1  & 2 j +1 \end{array} \right| u  \right) &=&  
   1 
   + 2\frac{\sin u \cos{2\pi \over k}}{\sin{\left({2\pi \over k}-\lambda\right)}}   \frac{[2]}{[4]}\left(1-\frac{[2]}{[2j][2j+1]}\right) 
  - 2\frac{\sin u \sin{3\pi \over k}}{\cos\left({3\pi \over k}-\lambda\right)} \frac{1}{[3]}\frac{[2j-1]}{[2j+1]}
   \nonumber \\
      \left. \left( \begin{array}{cc} 2 j +1 & 2 j - 1 \\ 2 j +3  & 2 j +1 \end{array} \right| u  \right) &=&  
   1 
   + 2\frac{\sin u \cos{2\pi \over k}}{\sin{\left({2\pi \over k}-\lambda\right)}} \left(- \frac{[2]}{[4]} \sqrt{[2j-1][2j+3] \over [2j+1][2j+1]}\right) \nonumber \\ & &
  - 2\frac{\sin u \sin{3\pi \over k}}{\cos\left({3\pi \over k}-\lambda\right)}  \frac{1}{[3]}\sqrt{[2j-1][2j+3] \over [2j+1][2j+1]}
   \nonumber \\
    -  \left. \left( \begin{array}{cc} 2 j +1 & 2 j + 1 \\ 2 j +3  & 2 j +1 \end{array} \right| u  \right) &=&  
    2\frac{\sin u \cos{2\pi \over k}}{\sin{\left({2\pi \over k}-\lambda\right)}} \frac{[2]}{[4]}\frac{1}{[2j]}\frac{[4j+2]}{[2j+1]}\sqrt{[2j+3] \over[2j+1]} \nonumber \\ & &  
  - 2\frac{\sin u \sin{3\pi \over k}}{\cos\left({3\pi \over k}-\lambda\right)} \left(- \frac{1}{[3]}\sqrt{[2j+3] \over[2j+1]}\right)
     \nonumber \\
    -  \left. \left( \begin{array}{cc} 2 j +1 & 2 j + 1 \\ 2 j -1  & 2 j +1 \end{array} \right| u  \right) &=&  
    2\frac{\sin u \cos{2\pi \over k}}{\sin{\left({2\pi \over k}-\lambda\right)}} \frac{[2]}{[4]}\frac{1}{[2j]}\frac{[4j+2]}{[2j+1]}\left(-\sqrt{[2j-1] \over[2j+1]}\right)   \nonumber \\ & &
  - 2\frac{\sin u \sin{3\pi \over k}}{\cos\left({3\pi \over k}-\lambda\right)} \left(- \frac{1}{[3]}\sqrt{[2j-1] \over[2j+1]}\right)
     \nonumber \\
    \left. \left( \begin{array}{cc} 2 j +1 & 2 j + 1 \\ 2 j +1  & 2 j +3 \end{array} \right| u  \right) &=&  
    1+
    2\frac{\sin u \cos{2\pi \over k}}{\sin{\left({2\pi \over k}-\lambda\right)}} \frac{[2]}{[4]} \frac{[2j]}{[2j+2]}  
         \nonumber \\
    \left. \left( \begin{array}{cc} 2 j +1 & 2 j + 3 \\ 2 j + 3  & 2 j +3 \end{array} \right| u  \right) &=&  
    1+
    2\frac{\sin u \cos{2\pi \over k}}{\sin{\left({2\pi \over k}-\lambda\right)}} \frac{[2]}{[4]} \frac{[2j+4]}{[2j+2]}
             \nonumber \\
    \left. \left( \begin{array}{cc} 2 j +1 & 2 j + 3 \\ 2 j + 1  & 2 j +3 \end{array} \right| u  \right) &=&  
      2\frac{\sin u \cos{2\pi \over k}}{\sin{\left({2\pi \over k}-\lambda\right)}} \frac{[2]}{[4]} \left(-\sqrt{\frac{[2j][2j+4]}{[2j+2][2j+2]}}\right)
    \nonumber \\
    \left. \left( \begin{array}{cc} 2 j +1 & 2 j + 1 \\ 2 j +1  & 2 j -1 \end{array} \right| u  \right) &=&  
    1+
    2\frac{\sin u \cos{2\pi \over k}}{\sin{\left({2\pi \over k}-\lambda\right)}} \frac{[2]}{[4]} \frac{[2j+2]}{[2j]}  
         \nonumber \\
    \left. \left( \begin{array}{cc} 2 j +1 & 2 j -1 \\ 2 j -1  & 2 j -1 \end{array} \right| u  \right) &=&  
    1+
    2\frac{\sin u \cos{2\pi \over k}}{\sin{\left({2\pi \over k}-\lambda\right)}} \frac{[2]}{[4]} \frac{[2j-2]}{[2j]}
             \nonumber \\
    \left. \left( \begin{array}{cc} 2 j +1 & 2 j -1 \\ 2 j + 1  & 2 j -1 \end{array} \right| u  \right) &=&  
    2\frac{\sin u \cos{2\pi \over k}}{\sin{\left({2\pi \over k}-\lambda\right)}} \frac{[2]}{[4]} \left(-\sqrt{\frac{[2j-2][2j+2]}{[2j][2j]}}\right) \,.
  \end{eqnarray}
We checked explicitly for $k=6,7,\ldots,10$ that these weights are all positive at the isotropic point $u = {3 \gamma  \over 4} - {\pi \over 4} = {3 \pi  \over 2 k} - {\pi \over 4} $.
 
In the next section, we shall check our results by an alternative derivation based on the RSOS fusion procedure.
 
Before closing this section we look more carefully at what happens when the heights reach their extremal values, namely $0$ and ${k \over 2}-1$ (with $k$ integer).
Notice first that the decomposition
\begin{equation}
 (j)\otimes (1) = \oplus_{j'}(j') 
\end{equation}
contains, for generic values of $j$, three terms on the right-hand side. However, when $j$ is equal to one of the extremal values there is only one term.
In other words, all faces involving two neighbouring $(0)$ or $j={k \over 2}-1$ are actually forbidden, which can be seen in the expression (\ref{face6j}).

It is easily seen from the expressions of the 6j coefficients that starting from an admissible configuration $\left| j_{i-1} j_i j_{i+1}\right\rangle$, no such forbidden configuration $\left| j_{i-1} j'_i j_{i+1}\right\rangle$ can be produced, in the sense that the corresponding face matrix element vanishes. 
In this way the model, for either even or odd values of $j$, is well-defined.


\subsection{RSOS(2,2) fused model}

We recall the two integrable $\check{R}$ matrices that can be built from the SO(3) BWM algebra:
\begin{eqnarray}
\check{R}^{(1)} & \propto & P_2  + \frac{\mathfrak{q}^4 x -1}{\mathfrak{q}^4 -x} P_1 +  \frac{\mathfrak{q}^4 x -1}{\mathfrak{q}^4 -x} \frac{\mathfrak{q}^2 x -1}{\mathfrak{q}^2 -x}P_0  \,, \\
\check{R}^{(2)} &\propto & P_2  + \frac{\mathfrak{q}^4 x -1}{\mathfrak{q}^4 -x} P_1 +  \frac{\mathfrak{q}^6 x +1}{\mathfrak{q}^6 +x}P_0 \,,
\end{eqnarray}
where $\mathfrak{q}^2 = q = \mathrm{e}^{\mathrm{i} \gamma}$.

These $\check{R}$ matrices are associated with the spin-one representation of $U_{\mathfrak{q}}\left( sl_2 \right)$. 
The RSOS representation of $\check{R}^{(2)}$ has already been discussed in section~\ref{RSOS_R2}.
The RSOS model associated to $R^{(1)}$ can be obtained from that associated with the spin-$1 \over 2$ (6V) model by the fusion procedure. 
This is what we shall make explicit in this section.
We stick closely to \cite{Date} to build the face weights $W_{2,2}$ from the given $W_{1,1}$. 
We take $\xi = 0$, and take the trigonometric limit, such that $[x]$ is just the usual $\q$-deformed number 
\begin{equation}
 [y] = \frac{\mathfrak{q}^{y}-\mathfrak{q}^{-y}}{\mathfrak{q}-\mathfrak{q}^{-1}} = \frac{\sin{\pi y \over k}}{\sin {\pi \over k} } \,.
\end{equation}
Therefore the $W_{1,1}$ weights of eq.~(3) in \cite{Date} are just the gauge-deformed usual `RSOS spin-$1 \over 2$' model: 
\begin{equation}
 \left. \left( \begin{array}{cc} d & c \\ a & b \end{array} \right| u  \right)_{1,1} = [1+u]\delta_{a,c} + 1\sqrt{\frac{[a][c]}{[b][d]}}\frac{g_a}{g_c}\delta_{b,d} \,,
\end{equation}
where $g_a = \frac{(-1)^{a\over 2}}{\sqrt{[a]}}$. This is similar to \cite{Pearce}, except for a typo in the definition of $g_a$ in the latter. 

The weights $W_{2,1}$ are given explicitly in \cite{Date}, eq.~(5). From there we can compute the $W_{2,2}$, which we rescale by $[2]$ for later convenience: 
\begin{eqnarray}
  \left. \left( \begin{array}{cc} a & a \\ a & a \end{array} \right| u  \right)_{2,2} &=& 
  \frac{[a-1-u][a+u]}{[a-1][a]} + \frac{[a-1][a+2][u][1+u]}{[2] [a] [1+a]} \\
  \
   \left. \left( \begin{array}{cc} a & a+2 \\ a & a \end{array} \right| u  \right)_{2,2} &=& 
  \frac{[a-1-u][a+2][u-1]}{[a-1][a]} + \frac{[a+2][1+a-u][1+u]}{ [a] [1+a]} \\
    \
   \left. \left( \begin{array}{cc} a & a-2 \\ a & a \end{array} \right| u  \right)_{2,2} &=& 
   \frac{[a-1+u][a-2][u+1]}{[a-1][a]} + \frac{[a-2][1+a+u][-1+u]}{ [a] [1+a]} \\
     \
   \left. \left( \begin{array}{cc} a & a \\ a+2 & a \end{array} \right| u  \right)_{2,2} &=& 
   \frac{[a-1][a-u][u]}{[2][a+1][a]}  \\
     \
   \left. \left( \begin{array}{cc} a & a \\ a-2 & a \end{array} \right| u  \right)_{2,2} &=& 
     \frac{[a+1][a+u][u]}{[2][a-1][a]}  \\
       \
   \left. \left( \begin{array}{cc} a & a+2 \\ a+2 & a \end{array} \right| u  \right)_{2,2} &=& 
   \frac{[a-u][1+a-u]}{[a][1+a]} \\
         \
   \left. \left( \begin{array}{cc} a & a-2 \\ a-2 & a \end{array} \right| u  \right)_{2,2} &=& 
   \frac{[a+u][-1+a+u]}{[a][-1+a]} \\
         \
   \left. \left( \begin{array}{cc} a & a+2 \\ a-2 & a \end{array} \right| u  \right)_{2,2} &=& 
   \frac{[a+1][a+2][-1+u][u]}{[2][a][-1+a]}  \\
         \
   \left. \left( \begin{array}{cc} a & a-2 \\ a+2 & a \end{array} \right| u  \right)_{2,2} &=& 
   \frac{[a-1][a-2][-1+u][u]}{[2][a][1+a]}  \\
      \
   \left. \left( \begin{array}{cc} a & a \\ a & a+2 \end{array} \right| u  \right)_{2,2} &=& 
   \frac{[a+u][u+2]}{[2][a]} + \frac{[a-1][a+2+u][u]}{[a][a+1][2]}  \\
    \
   \left. \left( \begin{array}{cc} a & a+2 \\ a & a+2 \end{array} \right| u  \right)_{2,2} &=& 
   \frac{[a+3][u][u+1]}{[2][a+1]}  \\
    \
   \left. \left( \begin{array}{cc} a & a \\ a+2 & a+2 \end{array} \right| u  \right)_{2,2} &=& 
   \frac{[a-1][u][u+1]}{[2][a+1]}  \\
       \
   \left. \left( \begin{array}{cc} a & a+2 \\ a+2 & a+2 \end{array} \right| u  \right)_{2,2} &=& 
   \frac{[a+1-u][u+1]}{[a+1]} \,.
  \end{eqnarray}
Note that the weights are symmetric under a NW-SE exchange, and all other weights can be obtained from this symmetry.
   
We perform a change of gauge on these weights, so as to make them symmetric under NE-SW exchange. 
The change of gauge is defined as 
\begin{equation}
  \left. \left( \begin{array}{cc} d & c \\ a & b \end{array} \right| u  \right)_{2,2} \to  \frac{s_{c,b,d}}{s_{a,b,d}}  \left. \left( \begin{array}{cc} d & c \\ a & b \end{array} \right| u  \right)_{2,2} \,,
\label{eq:gaugeRSOS}
  \end{equation}
where:
\begin{itemize}
 \item If $d=b$,
 \begin{eqnarray}
  s_{b\pm2, b, b} &=& \frac{1}{[2]} \sqrt{\frac{[b\mp 1]}{[b\pm 2][b]}} \,, \\ 
  s_{b,b,b} &=& 1 \,.
 \end{eqnarray}
 \item If $d=b-2$,
 \begin{eqnarray}
  s_{b,b,b-2} &=& 1 \,, \\ 
  s_{b-2,b,b-2} &=& \sqrt{\frac{[b+1]}{[b-3]}} \,.
 \end{eqnarray}
  \item If $d=b+2$
 \begin{eqnarray}
  s_{b,b,b+2} &=& 1 \,, \\ 
  s_{b+2,b,b+2} &=& \sqrt{\frac{[b-1]}{[b+3]}} \,.
 \end{eqnarray}
\end{itemize}

The following decomposition holds
\begin{eqnarray} 
 \left. \left( \begin{array}{cc} d & c \\ a & b \end{array} \right| u  \right)_{2,2} &=& 
 \frac{[1+u][2+u]}{[2]} P^{(2)}\left( \begin{array}{cc} d & c \\ a & b \end{array} \right) \nonumber \\
 &+&  \frac{[1+u][2-u]}{[2]} P^{(1)}\left( \begin{array}{cc} d & c \\ a & b \end{array} \right) 
 +  \frac{[1-u][2-u]}{[2]} P^{(0)}\left( \begin{array}{cc} d & c \\ a & b \end{array} \right) \,,
\end{eqnarray}
which is just the decomposition of $R^{(1)}$ over (spin 1 $\times$ spin 1) projectors, once we identify $x = \mathrm{e}^{\mathrm{i} u \gamma}$.

This allows to make explicit the face weights of the projectors. 
Following the language of \cite{Vincent} and previous section, we rotate the faces so that the NW-SE ($b-d$) direction becomes the horizontal one, such that
$$ \left.\left( \begin{array}{cc} 2 j_{i-1} +1 & 2 j'_i +1 \\ 2 j_i +1 &  2 j_{i+1}+1 \end{array} \right| u  \right)_{2,2}$$
takes the state $\left| j_{i-1} j_i j_{i+1}\right\rangle$ to $\left| j_{i-1} j'_i j_{i+1}\right\rangle$.
It is then easily checked that the face matrix elements of the projectors coincide exactly with those obtained previously.

%

\section{Reminders about regime I}
\label{app:B}

This corresponds to $c<0$ in (\ref{energy}), where the ground state is obtained by taking the roots $\lambda_i$ to have imaginary part $\Im \lambda_i = {\pi\over 2}$. 
The Bethe equations read, in Fourier form
\begin{equation}
\rho+\rho^h={\sinh {\omega\gamma\over 2}\over\sinh{\omega\pi\over 2}}+{2\sinh{\omega\over 4}(\pi-\gamma)\cosh{
\omega\over 4}(\pi-3\gamma)\over \sinh{\omega\pi\over 2}}\rho \,,
\end{equation}
so the physical equations obtained by putting the density of excitations over the physical ground state on the right  read
\begin{equation}
\rho+\rho^h={\cosh{\omega\over 4}(\pi-\gamma)\over \cosh{3\omega\over 4}(\pi-\gamma)}-{\sinh{\omega\over 2}(\pi-\gamma)\cosh{\omega\over 2}(\pi-3\gamma)\over\sinh{\omega\gamma\over 2}\cosh {3\omega\over 4}(\pi-\gamma)}\rho^h \,. \label{BethephysI}
\end{equation}
The matrix $K$ obtained by writing 
\begin{equation}
\rho+\rho^h=s+K\rho \,,
\end{equation}
and taking $K$ at zero frequency is simply $K=1-{\gamma\over\pi}$, so $1-K={\gamma\over\pi}$, and thus the gaps associated with holes and shifts of the sea read
\begin{equation}
\Delta+\bar{\Delta}={1-K\over 4}n^2+{1\over 1-K}p^2={\gamma\over 4\pi}n^2+{\pi\over\gamma}p^2 \,,
\end{equation}
where $n$ is the number of holes. The central charge is $c=1$. 

To understand the model in more details, we imagine staggering the bare spectral parameter  as discussed in the text, 
by amounts $\pm\Lambda$. This modifies the physical equation for densities slightly, as now the source term becomes, in Fourier space
\begin{equation}
s(\omega)={\cos\Lambda\omega \cosh{\omega\over 4}(\pi-\gamma)\over \cosh{3\omega\over 4}(\pi-\gamma)} \,.
\end{equation}
When going back to real space, this becomes a complicated expression in terms of the rapidity $u$ of the holes. The field theoretic limit is obtained close to vanishing energy/momentum. This requires taking $\Lambda$ large, and focusing on  a region where the source term is dominated by the poles nearest the origin, here $\omega=\pm {2i\over 3}{\pi\over (\pi-\gamma)}$. In this limit, the source term  is proportional to
$$\exp\left[-{2\Lambda\over 3}{\pi\over \pi-\gamma}\right]\cosh {2\over 3}{\pi\over \pi-\gamma}\lambda \,.$$
This leads to the mass scale
\begin{equation}
M\propto \exp\left(-{2\Lambda\over 3}{\pi\over \pi-\gamma}\right) \,,
\label{massscale}
\end{equation}
and the physical rapidity is $\theta={2\over 3}{\pi\over \pi-\gamma}\lambda$, so the source term reads $s(\lambda)=M\cosh \theta$. 
 
Meanwhile, the kernel in the Bethe equation (\ref{BethephysI}) corresponds to the 
known S-matrix \cite{Takacs,Birgit1} for the Bullough-Dodd model \cite{BullDodd} with action
\begin{equation}
S={1\over 2} \int dx_1dx_2\left[(\partial_{x_1}\Phi)^2+(\partial_{x_2}\Phi)^2+g_{\rm BD}(e^{-2i\beta\Phi}+e^{i\beta\Phi})\right] \,, \label{BDact}
\end{equation}
where one should set 
\begin{equation}
{\beta^2\over 8\pi}={\gamma\over 2\pi} \,.
\end{equation}
In our units, this is the conformal weight of $e^{i\beta\Phi}$. Knowing the  action in the continuum limit allows us  to obtain the relationship between the bare coupling $g_{\rm BD}$ in (\ref{BDact})  and the staggering $e^{-\Lambda}$ on the lattice. Imagine indeed computing perturbatively the ground state energy of the model with action (\ref{BDact}). This will expand in powers of $g_{\rm BD}^3$ since only three point functions involving one insertion of the first exponential and two insertions of the  second one will contribute. These insertions are then integrated over two-dimensional space. By dimensional analysis, it follows that
$$[g_{\rm BD}]=[\hbox{length}]^{-2+{\beta^2\over 2\pi}}=[\hbox{length}]^{-{2(1-{\gamma\over\pi})}} \,.$$
Comparing with (\ref{massscale}), we get thus that
\begin{equation}
g_{\rm BD}\propto e^{-{4\over 3}\Lambda} \,. \label{couprel}
\end{equation}
Twisting the theory such that $e^{-2i\beta\Phi}$ becomes a screening operator leads to a central charge
\begin{equation}
c=1-{6\over x(x+1)} \,,
\end{equation}
where $x={\pi\over 2\gamma-\pi}$. It is well-known \cite{Takacs, Efthimiou} that this staggering corresponds to  
perturbing by the operator $\Phi_{21}$, with  $=h=\bar{h}=h_{21}$ exponent 
\begin{equation}
h_{21}={x+3\over 4x}={3\gamma\over 2\pi}-{1\over 2} \,,
\end{equation}
so the perturbed action reads
\begin{equation}
S=S_{\rm CFT}+g_{21}\int dx_1dx_2~ \Phi_{21} \,.
\end{equation}
Now imagine calculating again the ground state energy of this theory. This will now involve an expansion in powers of 
$g_{21}^2$, with now, by dimensional analysis again, 
$$[g_{21}]=[\hbox{length}]^{-2+2h_{21}}=[\hbox{length}]^{-3+{3\gamma\over\pi}} \,.$$
Hence, $g_{21}^2\propto g_{\rm BD}^3$ and 
\begin{equation}
g_{21}\propto e^{-2\Lambda} \,.
\end{equation}
If we were to perturb some other twisted version by an operator whose odd point functions would be non zero --- such as the second energy operator in parafermions ---  with a coupling  $g_{\rm PF}$, we would have a similar relation, with now $g_{\rm PF}\propto g_{21}^2$, hence 
\begin{equation}
g_{PF}\propto e^{-4\Lambda} \,,
\end{equation}
which is the relation we use in the text. 
 
\section{Numerical methods and results}
\label{app:C}

In this appendix we present numerical results for both the untwisted and twisted models, relying on conjectures on the Bethe roots configurations describing the low-energy levels of the model. These conjectures were made from observations at small system sizes, for which direct diagonalization of the model's transfer matrix gives us access to the eigenvalues and eigenvectors. This information in turn provides knowledge about the root configurations via the so-called McCoy method \cite{McCoy1,McCoy2,NepomechieRavanini}, which we shall describe in some detail before going any further.  
 
\subsection{McCoy method for finding the Bethe roots corresponding to a known eigenstate}
 
Let us assume that we can diagonalize the transfer matrix for a (say) periodic system of small size $L$ in the sector of magnetization $n$. 
We look for the $L-n$ roots associated with one given eigenstate.

Up to some global rescaling, the coefficients of the $a_{2}^{(2)}$ $\check{R}$ matrix are all polynomials of degree 2 in $x$, which we recall is related to the spectral parameter $\lambda$ by $x=\mathrm{e}^{2\lambda}$. 
Integrability of the system ensures that the considered eigenvector can be chosen independently of $x$. 
Once given its coordinates $(v)_i$, with $i=1,\ldots,S$, in a basis of the space of states for the $L$-site chain, chosen such that $(v)_1 \neq 0$, we can therefore write the corresponding eigenvalue as 
\begin{equation}
 \Lambda(\lambda) = \frac{(T(\lambda) \cdot v)_1}{(v)_1} \,,
\end{equation}
Since $T(\lambda)$ is a product of $L$ $\check{R}$-matrices, the element $(T(\lambda) \cdot v)_1$ is then a polynomial of degree $2 L$ in $x$.  
This proves that $\Lambda(\lambda)$, properly scaled, is a polynomial of degree $2 L$ in $x$.  
%

Now turn to the Bethe expression of $\Lambda(L)$ in terms of the set of roots $\{ \lambda_j\}_{j=1..L-n}$ we are looking for \cite{GM2}:
\begin{equation}
 \Lambda_L (\lambda)= a^L \frac{Q\left(\lambda+{ \mathrm{i}\gamma\over 2}\right)}{Q\left( \lambda-{ \mathrm{i}\gamma\over 2} \right)} 
 + d^L \frac{Q\left(\lambda-2\mathrm{i}\gamma+{ \mathrm{i}\pi\over 2}\right)}{Q\left(\lambda-\mathrm{i}\gamma+{ \mathrm{i}\pi\over 2} \right)} 
 + b^L  \frac{Q\left(\lambda-{3 \mathrm{i}\gamma\over 2}\right)}{Q\left( \lambda-{ \mathrm{i}\gamma\over 2} \right)} 
 \frac{Q\left(\lambda+{ \mathrm{i}\pi\over 2}\right)}{Q\left(\lambda-\mathrm{i}\gamma+{ \mathrm{i}\pi\over 2} \right)}  \,,
\label{eq:LambdaLappendix}
\end{equation}
where $Q(\lambda)=\prod_{j=1}^{L-n} \sinh(\lambda - \lambda_j)$, and $a , b, d$ are polynomials of degree $2$ in the variable $x= \mathrm{e}^{2\lambda}$.

Multiplying both sides of (\ref{eq:LambdaLappendix}) by $Q\left(\lambda-\mathrm{i}\gamma+{ \mathrm{i}\pi\over 2} \right)$, we see that we can rewrite it as an identity between two polynomials of degree $2L + 2(L-n)$. 
Closer inspection reveals that, once the $(v)_i$ evaluated (from direct diagonalization), this identity simply consists in a triangular system of linear equations in terms of the $x_j \equiv \mathrm{e}^{2 \lambda_j}$, from which the determination of the $\lambda_j$ is straightforward.

\subsection{Excitations at zero twist}
 
\subsubsection{Structure of the excited states} 
 
We use the notations of section \ref{section:numerics} for the low-lying excitations in regime III, which we briefly recall here:
\begin{itemize}
 \item The ground state is made of a sea of 2-strings with imaginary parts close to ${\pi \over 4} - {\gamma \over 4}$.
 \item Excitations consisting in making holes in the Fermi-sea and/or adding antistrings with imaginary part $\pi \over 2$ can be labeled by two indices $(n,j)$.
 \item All the excitations mentioned above have a symmetric distribution of roots with respect to the imaginary axis, or in other terms a symmetric distribution of the Bethe integers. Left- or right-backscattered states can be created by shifting all the Bethe integers by some fixed amount, hence creating states of nonzero momenta $w$. Only if nonzero will the latter be specified in the labeling of states. 
\end{itemize}

\subsubsection{Study of the conformal weights}

\begin{figure}
\begin{center}
 \includegraphics[scale=0.6]{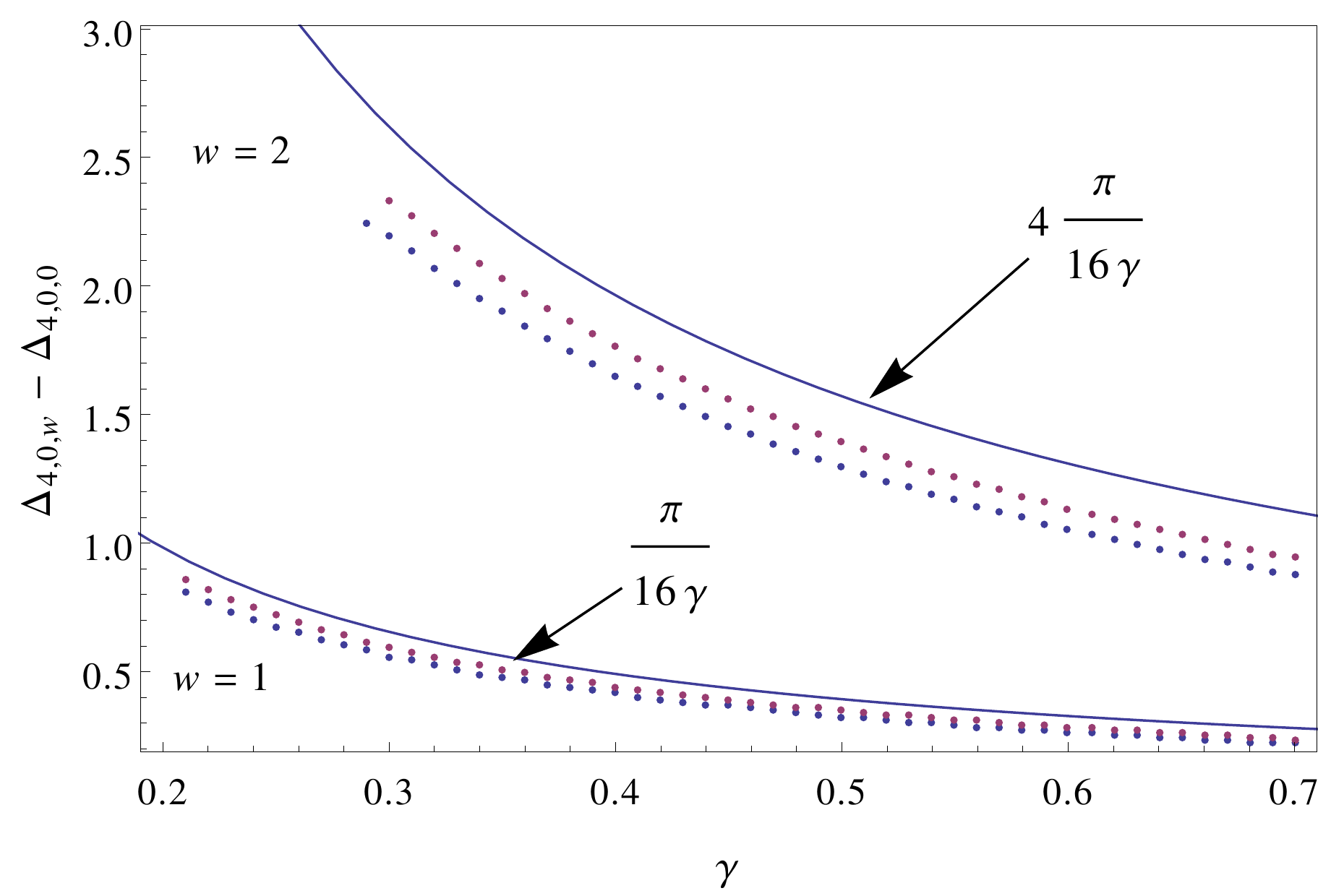}     
 \end{center}
\caption{Conformal dimensions of the $w=1$ and $w=2$ backscattered ground states in the sector $n=4$, measured relatively to that of the $w=0$ state at sizes $L=24$ and $L=32$ (blue and red dots respectively). The conjectures corresponding to (\ref{confweightso1}) are also plotted.}
\label{fig:DBackscattered}                          
\end{figure}

Finite-size scaling of the different conformal weights lead us to the following result 
\begin{equation}
 x_{n,j,w} = \Delta + \bar{\Delta} = {\gamma \over 4 \pi}n^2 + {\pi \over 16 \gamma} w^2 + o(1) \,,
\label{confweightso1}
\end{equation}
where the last term denotes corrections that go to $0$ in the limit $L\to \infty$, and whose study will be the object of the following paragraphs.
This is in full agreement with (\ref{fssIII}).
For instance we show in figure~\ref{fig:DBackscattered} the conformal dimensions of the backscattered states $w=1$ and $w=2$ measured in the sector $n=4$ with respect to the $w=0$ ground state. The data agree well with (\ref{confweightso1}).

\subsubsection{Form of the non compact corrections} 
 
We now focus on the corrections in (\ref{confweightso1}), which will provide indications of a non compact degree of freedom. 
Let us focus on the states with $w=0$ (the $w$ label is hence omitted) and set
\begin{equation}
 -{c_{n,j} \over 12} = n^2 {\gamma \over 4 \pi} + f_{n,j}(\gamma, L) \,.
\end{equation}

\begin{figure}
\begin{center}
 \includegraphics[scale=0.5]{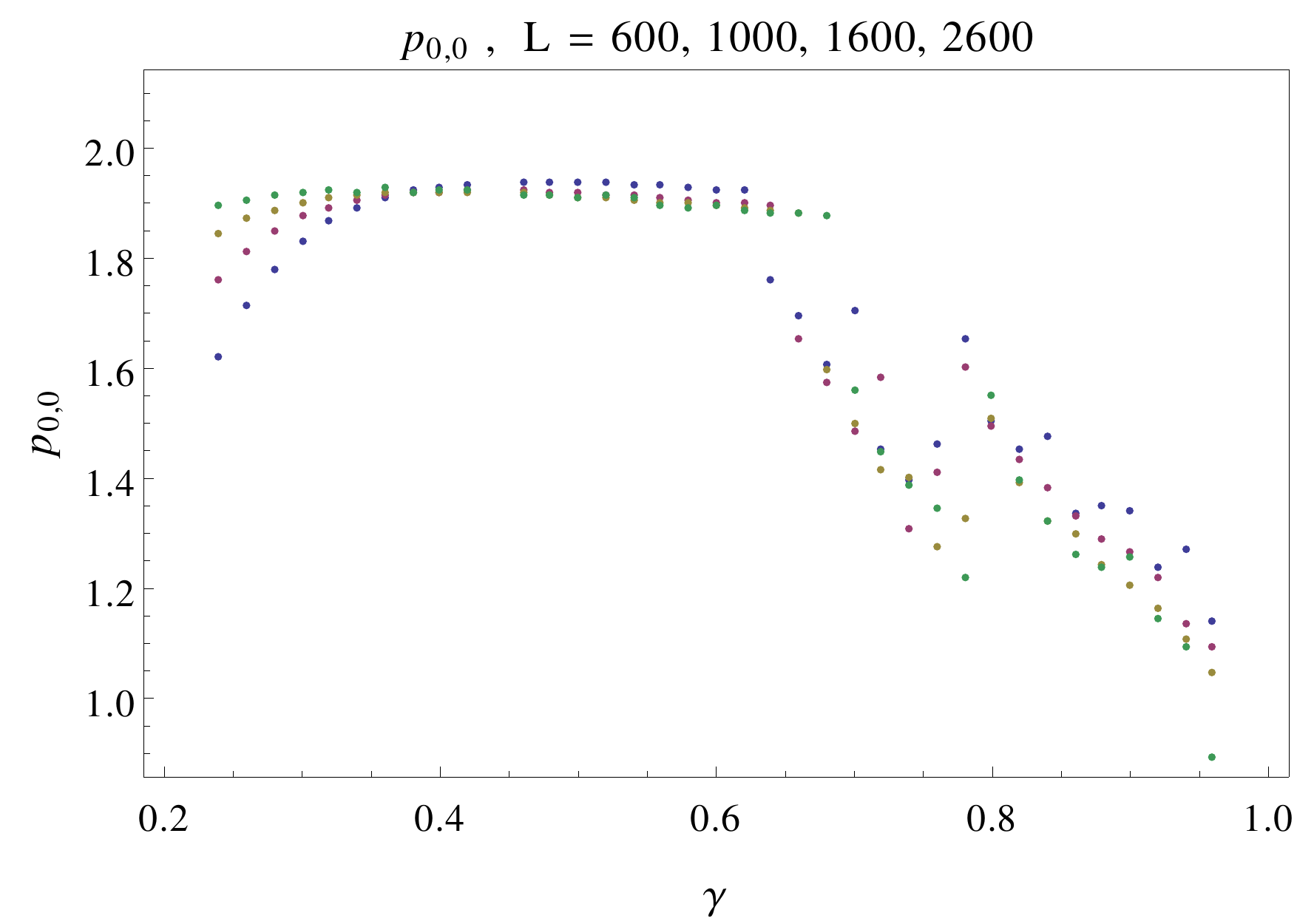}
 \includegraphics[scale=0.5]{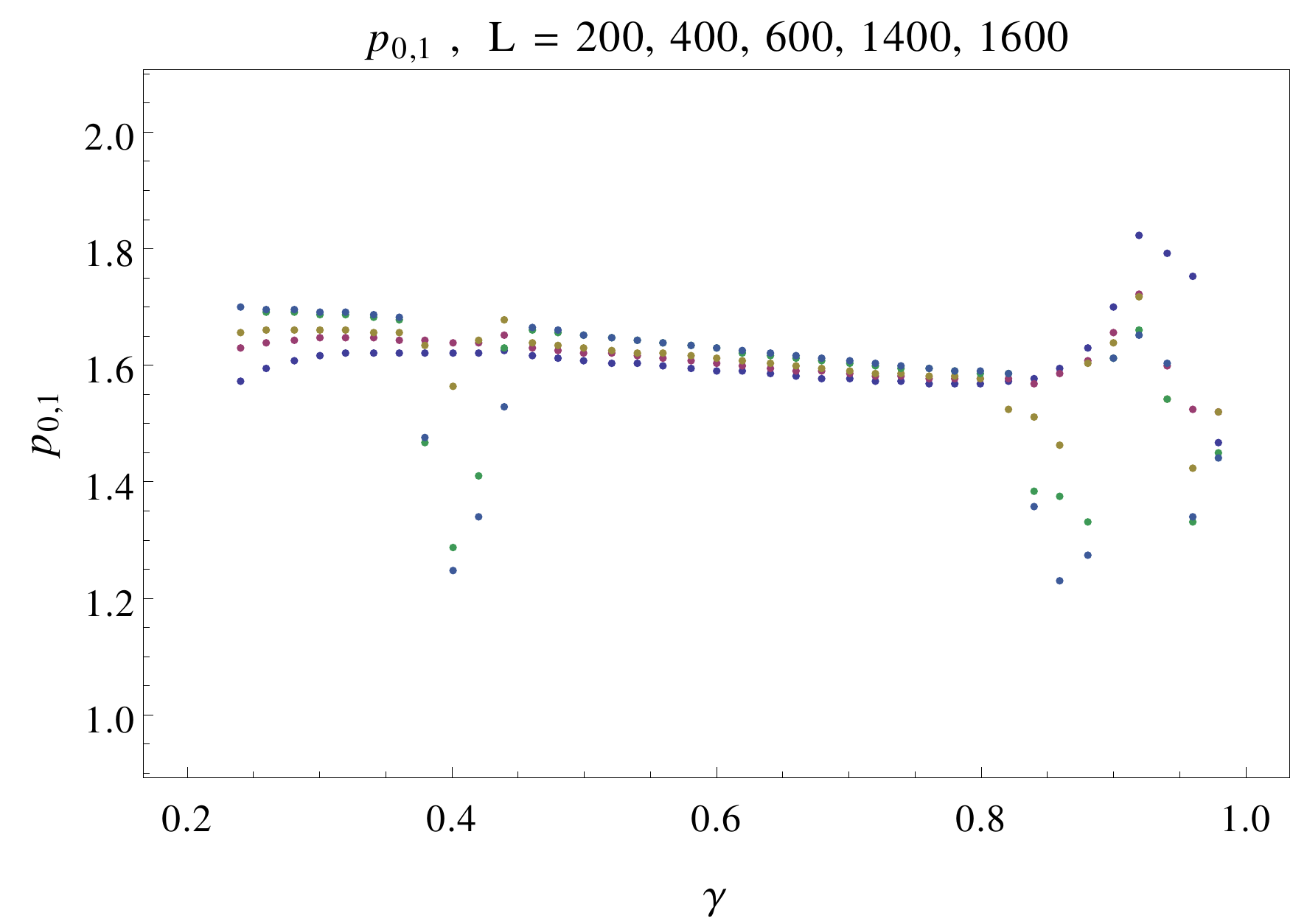}        
 \end{center}
\caption{Estimations of the exponents $p_{0,0}$ and $p_{0,1}$ from resolution of the BAE at different finite sizes.}                                 
\label{fig:pc_pDelta}                          
\end{figure}

In analogy with obvervations for other models with non compact continuum limit \cite{EsslerFS,IkhlefJS3,CanduIkhlef}, and inspired by our discussion of the continuum limit, we look for finite-$L$ corrections of the form 
\begin{equation}
f_{n,j}(\gamma, L) = \frac{A_{n,j}(\gamma)}{\left[ B_{n,j}(\gamma) + \log L \right]^p_{n,j}} \,,
\label{fconjecture}
\end{equation}
and expect $p_{n,j}=p=2$. 
This is confirmed by figure~\ref{fig:pc_pDelta}, where we measured the exponents $p_{0,0}$ and $p_{0,1}$.%

\begin{figure}
\begin{center}
 \includegraphics[scale=0.5]{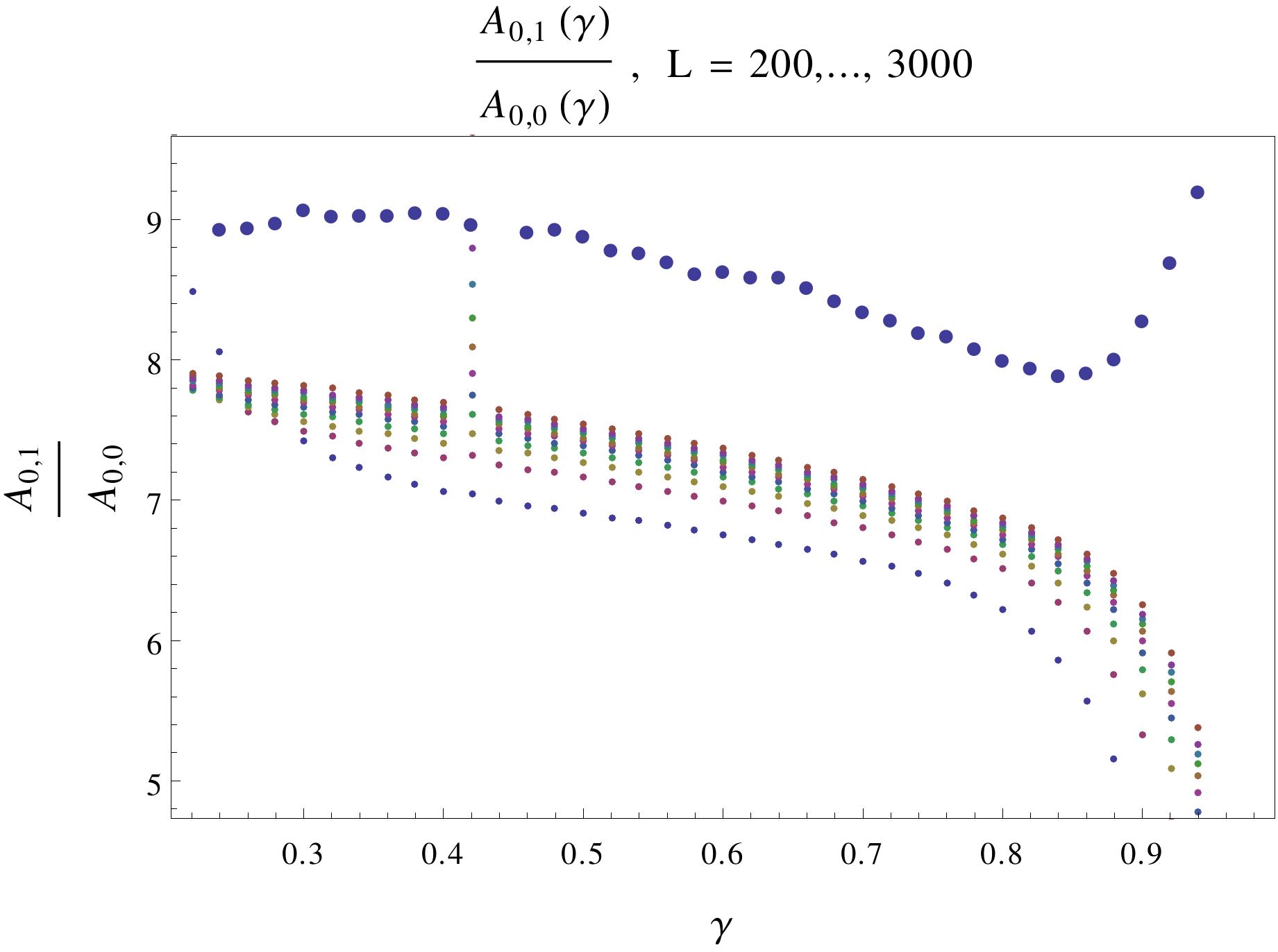}
 \end{center}
\caption{Estimations of the ratio ${A_{0,1} \over A_{0,0}}$ from resolution of the BAE at different finite sizes. 
Thick blue dots show an extrapolation to $L\to \infty$ obtained from fitting with an affine function of $1\over \log L$.
The ratio seem quite constant with $\gamma$, and its apparent value ${A_{0,1} \over A_{0,0}} = 9$ will be confirmed in the following. 
}                                 
\label{fig:ADeltasAc_BDeltasBc}                          
\end{figure}

Setting $p_{n,j}=2$ in (\ref{fconjecture}), we can now focus on the $(n,j)$-dependence of $A_{n,j}$.
Examination of figure \ref{fig:ADeltasAc_BDeltasBc}, where the ratio ${A_{0,1} \over A_{0,0}}$ is plotted as a function of $\gamma$, provides evidence that this ratio may be independent of $\gamma$, and our data indicates that the same holds for ${A_{n,j} \over A_{0,0}}$ with general values of $(n,j)$. As to $B_{n,j}$ we expect a more complex $n$ and $j$ dependence for $n$ and/or $j$ large, which is out of the reach of our numerics.

Hence the conjecture
\begin{equation}
f_{n,j}(\gamma, L) = \left(N_{n,j}\right)^2 \frac{A(\gamma)}{\left[ B_{n,j}(\gamma) + \log L \right]^p} \,.
\label{fconjecture2}
\end{equation}

\subsubsection{Determination of the numbers $N_{n,j}$}

We now fix $N_{0,0} = 1$ up to some rescaling of $A(\gamma)$, and study the low lying excitations in the first three spin sectors at some specific point $\gamma=0.45$.
The quantities $\left(N_{n,j}\right)^2 A(\gamma)$ were extrapolated from two successive sizes $L$ and $L+20$ with $L$ ranging to $\simeq 1000$ (depending on the level).
From there one obtains the ratios ${N_{n,j}\over{N_{0,0}}} = N_{n,j}$.

The finite-size behavior of some of these ratios are estimated in figure \ref{fig:n_{s,i}}. In each case several plots are given: 
\begin{itemize}
\item The blue curves show the estimations of $N_{n,j}$ from the measure of $(N_{n,j})^2 A$  as a function of ${1 \over (\log L)^2}$.
\item The yellow curves show the estimations of $N_{n,j}$ from the measure of ${(N_{n,j})^2 A \over \left[B_{n,j} + \log L\right]^2}$ as a function of ${1 \over \log L}$.
\item The purple and green curves are the respective $L\to \infty$ extrapolations of the two latter between pairs of successive points $L$ and $L+20$.
\end{itemize}
\begin{figure}
\begin{center}
 \includegraphics[scale=0.5]{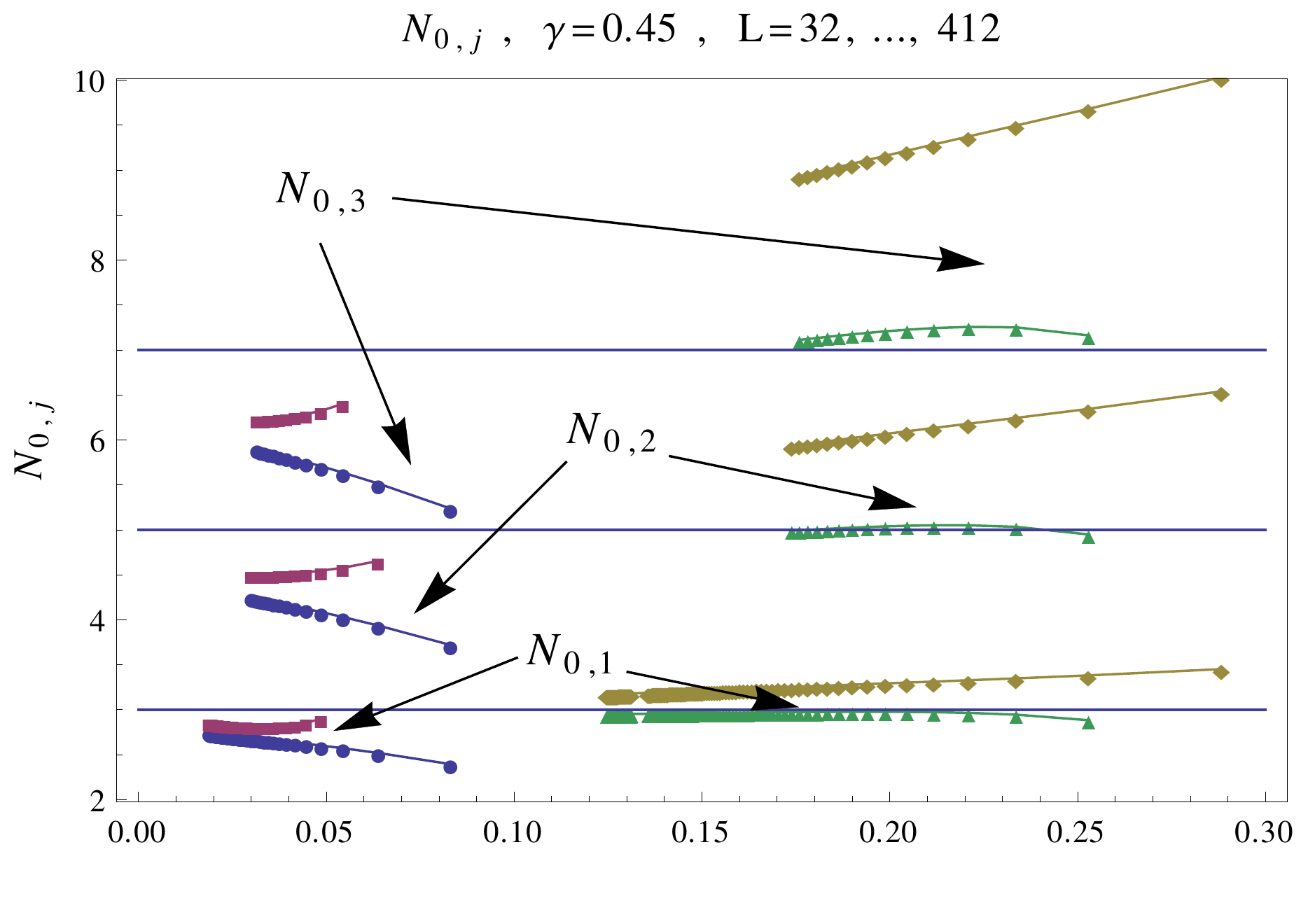}
 \vspace{0.1cm}
 \includegraphics[scale=0.5]{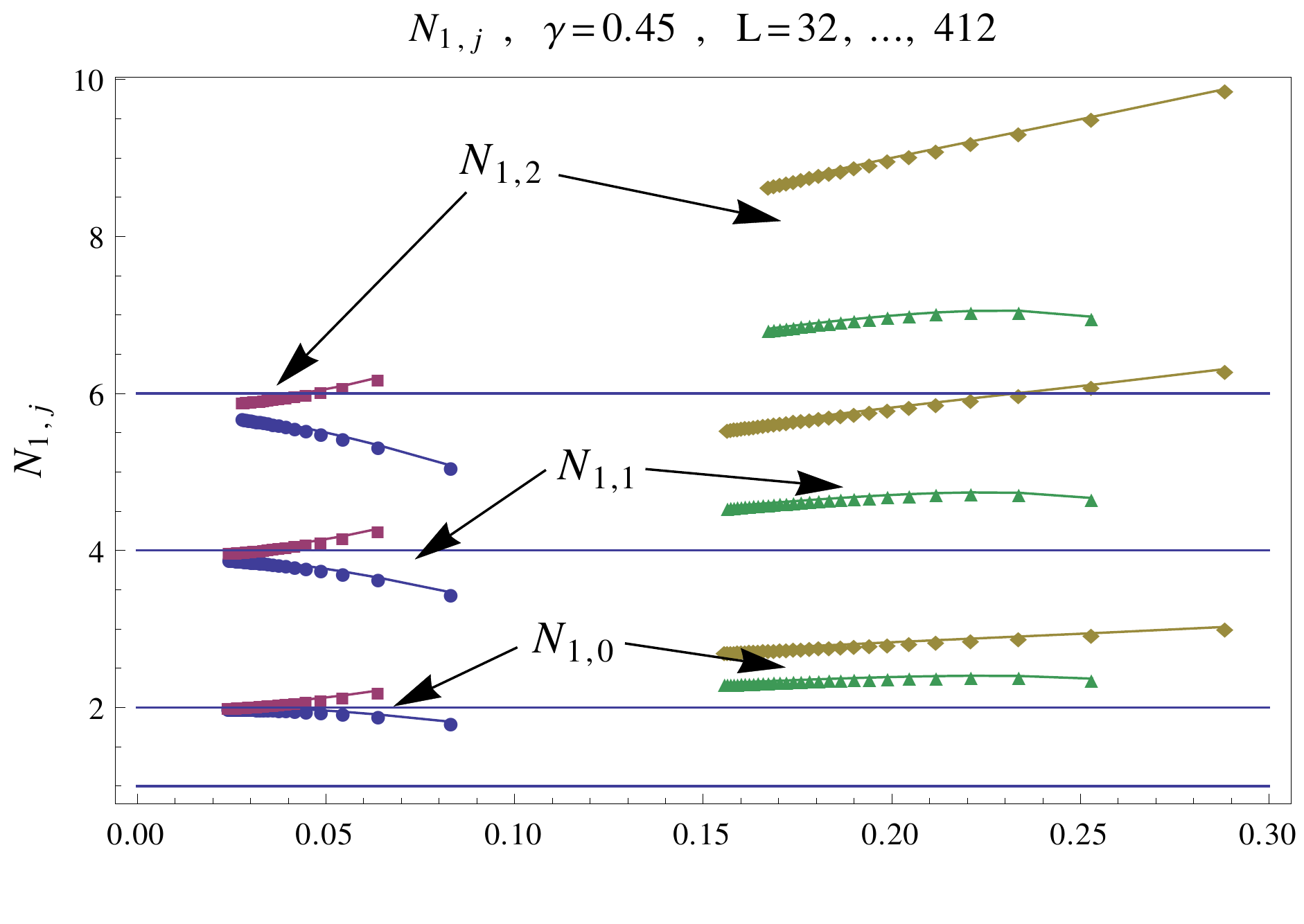}
  \vspace{0.1cm}
\includegraphics[scale=0.5]{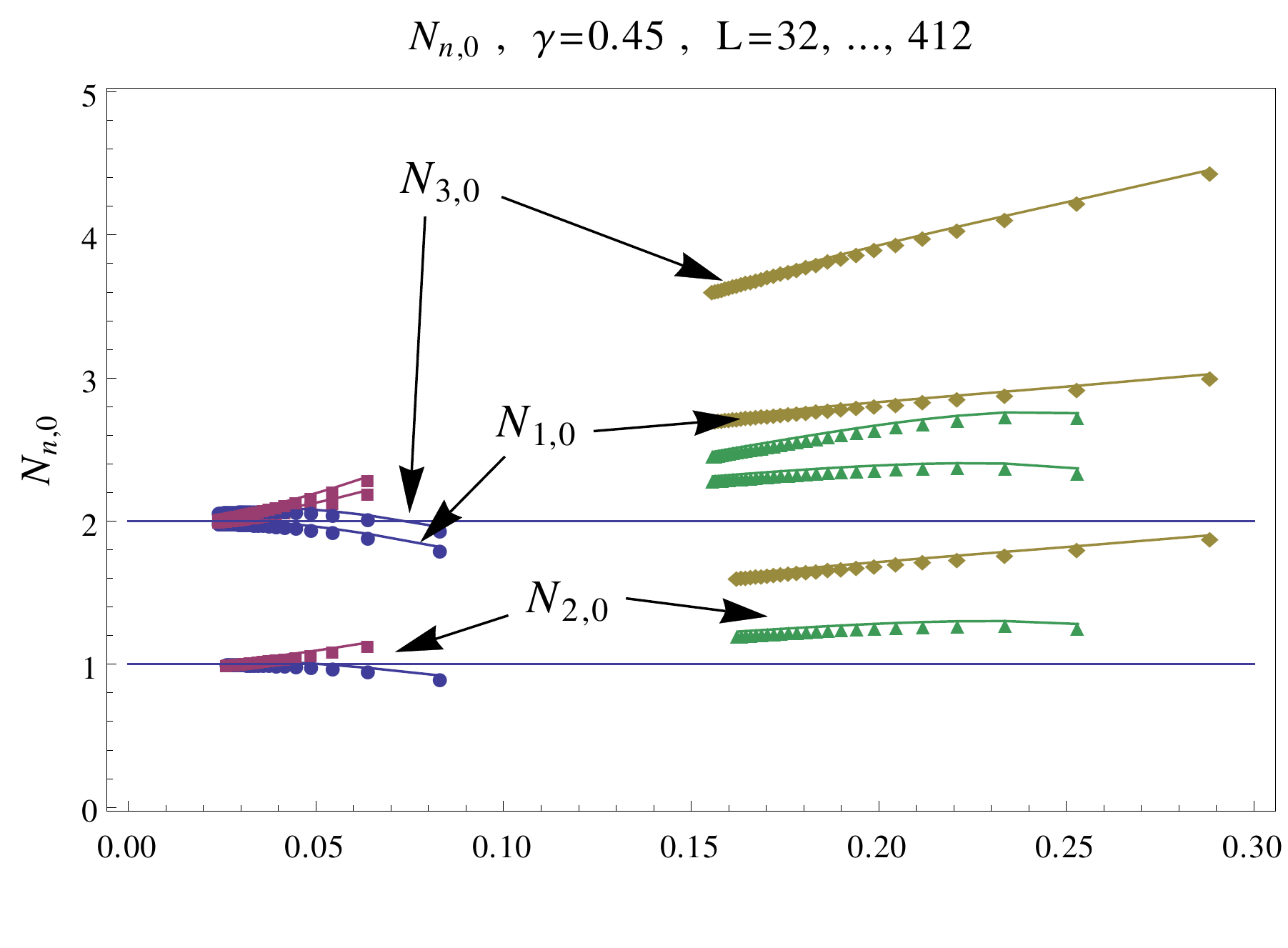}
 \end{center}
 \captionsetup{singlelinecheck=off}
 \caption[f f f]{Determinations of the charges $N_{n,j}$ of the first excitations in the sectors $n=0$ and $n=1$, as well as for the ground states of the first three spin sectors. All results were obtained for $\gamma=0.45$. }
\label{fig:n_{s,i}}                          
\end{figure}
The blue lines in the figure are the conjectures that we made from there.
In some cases the agreement between data obtained from $(N_{n,j})^2 A$ and ${(N_{n,j})^2 A \over \left[B_{n,j} + \log L\right]^2}$ is not really good. 
As will be detailed in section \ref{section:Bni} the function $B_{n,j}(\gamma)$ depends on $j$ through the ratio $\frac{j}{\log L}$, and therefore varies slowly with small values of $j$. This allows to trust estimations from  ${(N_{0,i})^2 A \over \left[B_{0,j} + \log L\right]^2}$ to conjecture $N_{0,j}$. 
From this we deduce in particular $N_{0,1}=3$, which seems confirmed by the very large size estimations of figure \ref{fig:n_{0,1}}. 

Since however $B_{n,j}(\gamma)$ can be different from one value of $n$ to another, we prefered to use $(N_{n,j})^2 A$ for estimations of $N_{n,j}$ with $n\neq 0$.
 
 \begin{figure}
\begin{center}
 \includegraphics[scale=0.7]{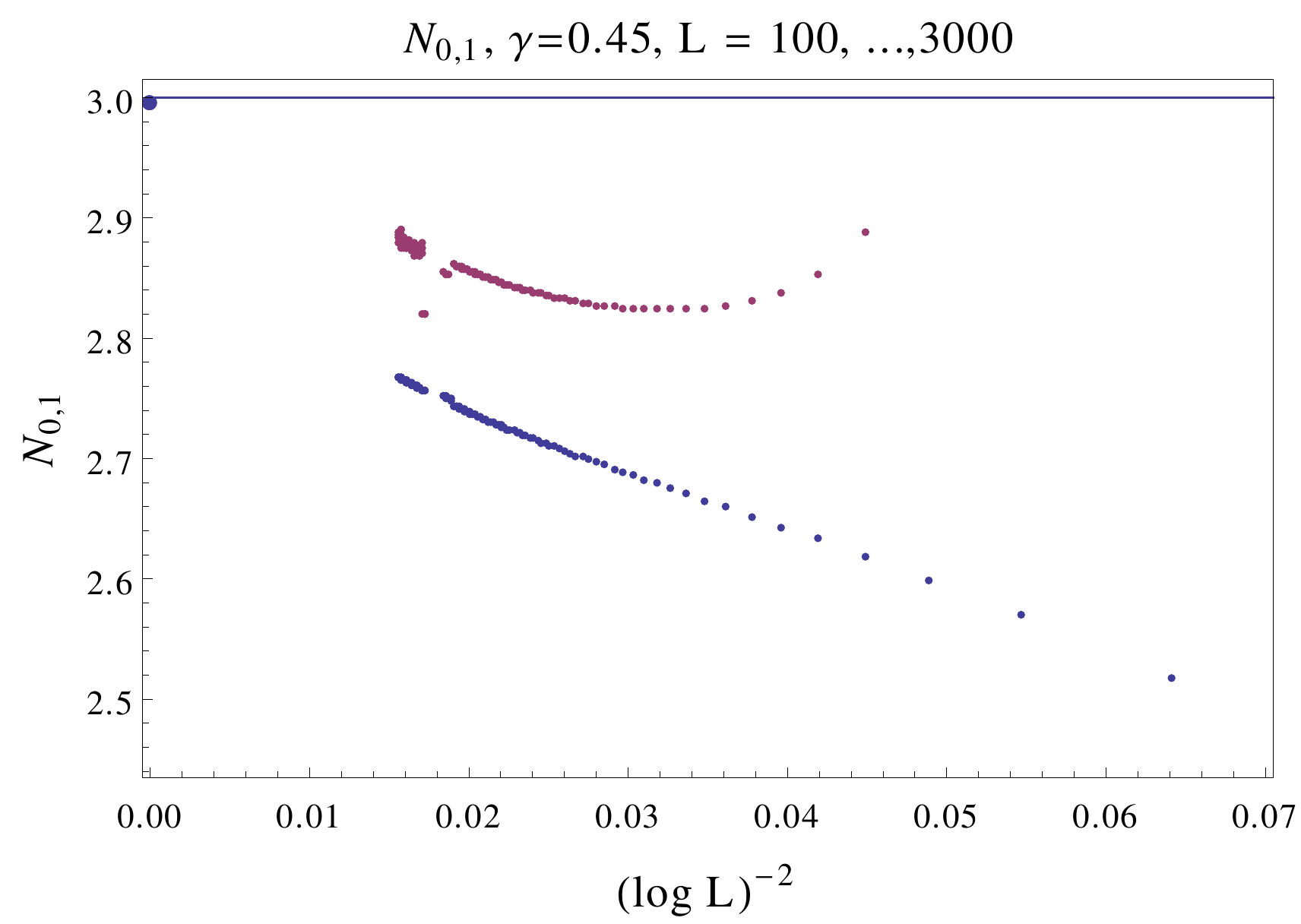}
 \end{center}
 \caption{Estimation of $n_{0,1}$ for $\gamma=0.45$ for sizes up to $L=3000$. The curves have the same meaning as in figure~\ref{fig:n_{s,i}}.}
\label{fig:n_{0,1}}                          
\end{figure}

The (sometimes very arguable) conclusions are summarized in the following table: 
\begin{center}
\begin{tabular}{|l|c|c|c|}
  \hline
  $(n,j)$ & number of 2-strings & number of roots with $\Im \lambda_i = {\pi \over 2}$  & $N_{n,j}$  \\ \hline 
  (0,0) & $L/4$ & 0 & 1 \\ 
  (0,1) & $L/4-1$ & 2& 3 \\ 
  (0,2) & $L/4-2$ & 4 & 5 \\ 
  (0,3) & $L/4-3$ & 6 &  7 \\ 
  \hline
  (1,0) & $L/4-1$ & 1& 2 \\ 
    (1,1) & $L/4-2$ & 3& 4 \\ 
      (1,2) & $L/4-3$ & 5& 6 \\ 
   \hline
  (2,0) & $L/4-2$ & 0& 1 \\ 
     \hline
  (3,0) & $L/4-3$ & 1& 2 \\ 
  \hline
\end{tabular}
\end{center}
These results all support the following conjecture:
\begin{eqnarray}
 N_{n,j} &=& \frac{3 + (-1)^{n+1}}{2} + 2 j  \\
 &=& 1 + \mbox{number of Bethe roots $\lambda_i$ with $\Im \lambda_i = \frac{\pi}{2}$} \,,
\end{eqnarray}

\subsubsection{Determination of the function $A(\gamma)$}

\begin{figure}
\begin{center}
 \includegraphics[scale=0.7]{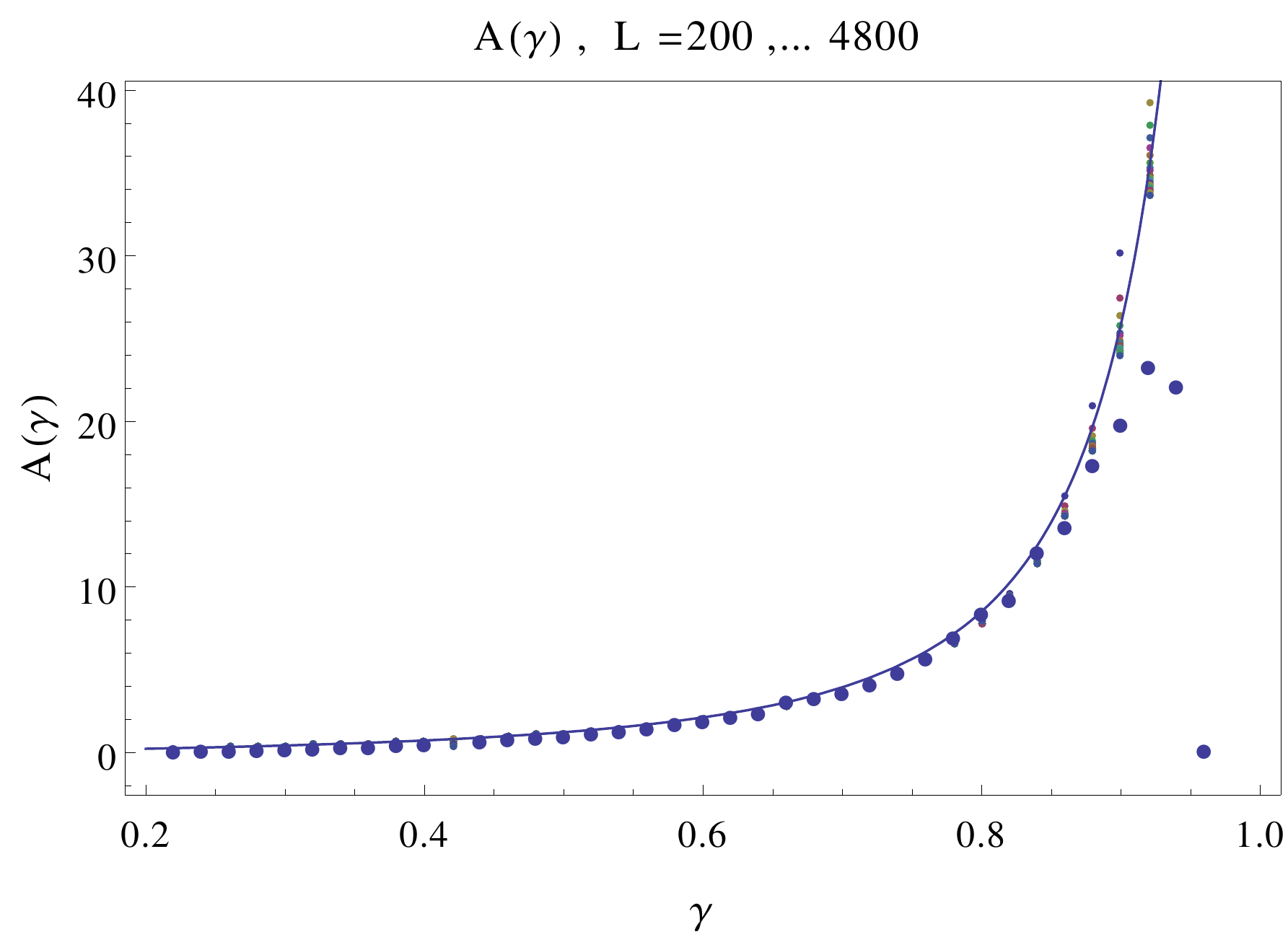}     
 \end{center}
\caption{Estimation of $A(\gamma)$ for different sizes. The thick blue dots are an extrapolation to $L\to \infty$. We plotted in comparison the conjecture (\ref{numresforA}).
}                                 
\label{fig:AList}                          
\end{figure}

We determined $A(\gamma)$ from the ground state energies, which are easier to obtain numerically. 
See figure \ref{fig:AList}. 
 
\begin{figure}
\begin{center}
 \includegraphics[scale=0.7]{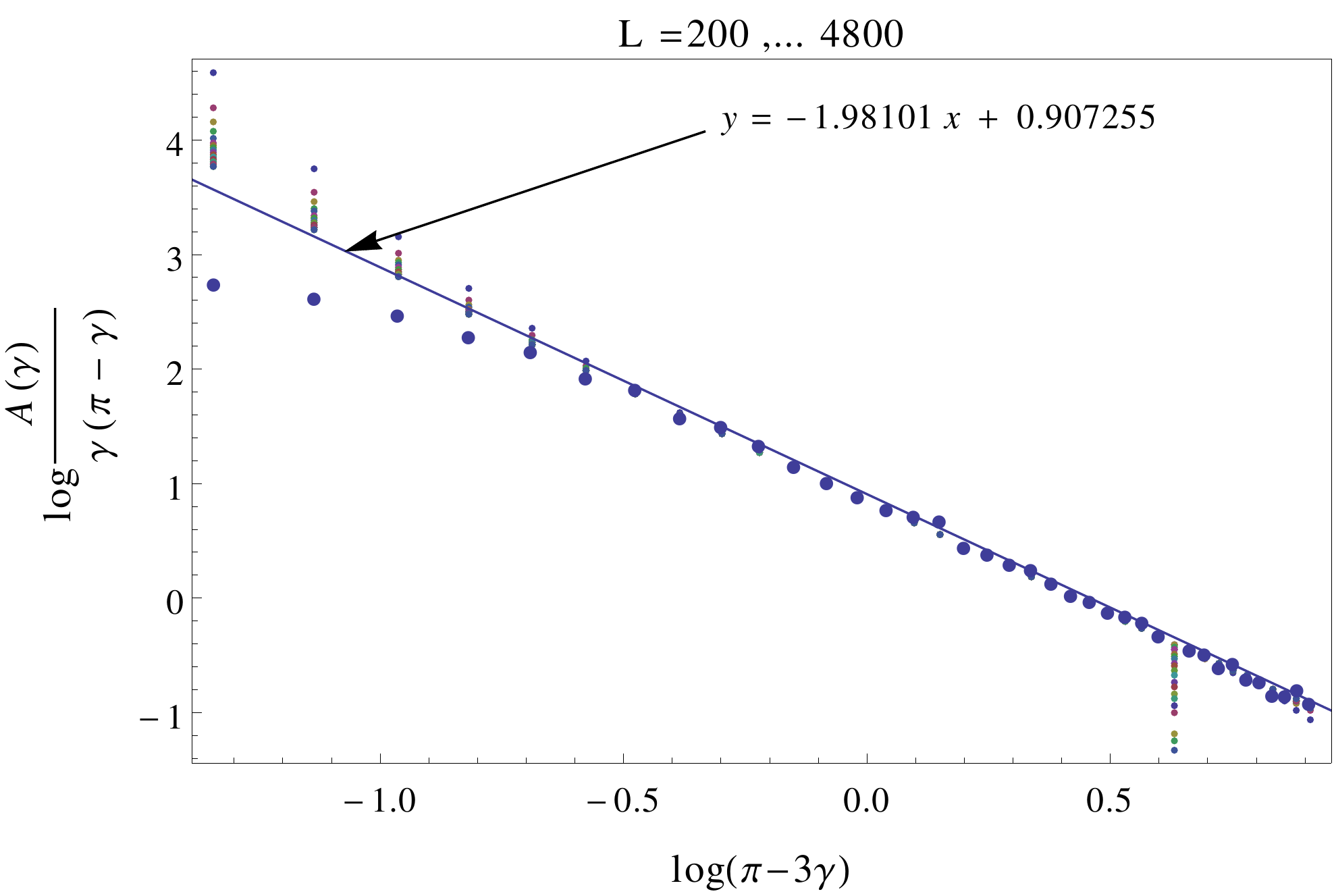}     
 \end{center}
\caption{Estimation of $\log \frac{A(\gamma)}{\gamma (\pi - \gamma)}$ for $L=200$ to $4800$, plotted as a function of $\log \left({\pi}-3\gamma\right)$.
The thick blue dots show an extrapolation to $L\to \infty$. 
The affine function plotted is a fit of the latter from values of $\gamma$ far enough from $\gamma=0$ or $\gamma={\pi \over 3}$; its parameters are given in the figure title. 
}                                 
\label{fig:LogA}                          
\end{figure}

\begin{figure}
\begin{center}
 \includegraphics[scale=0.7]{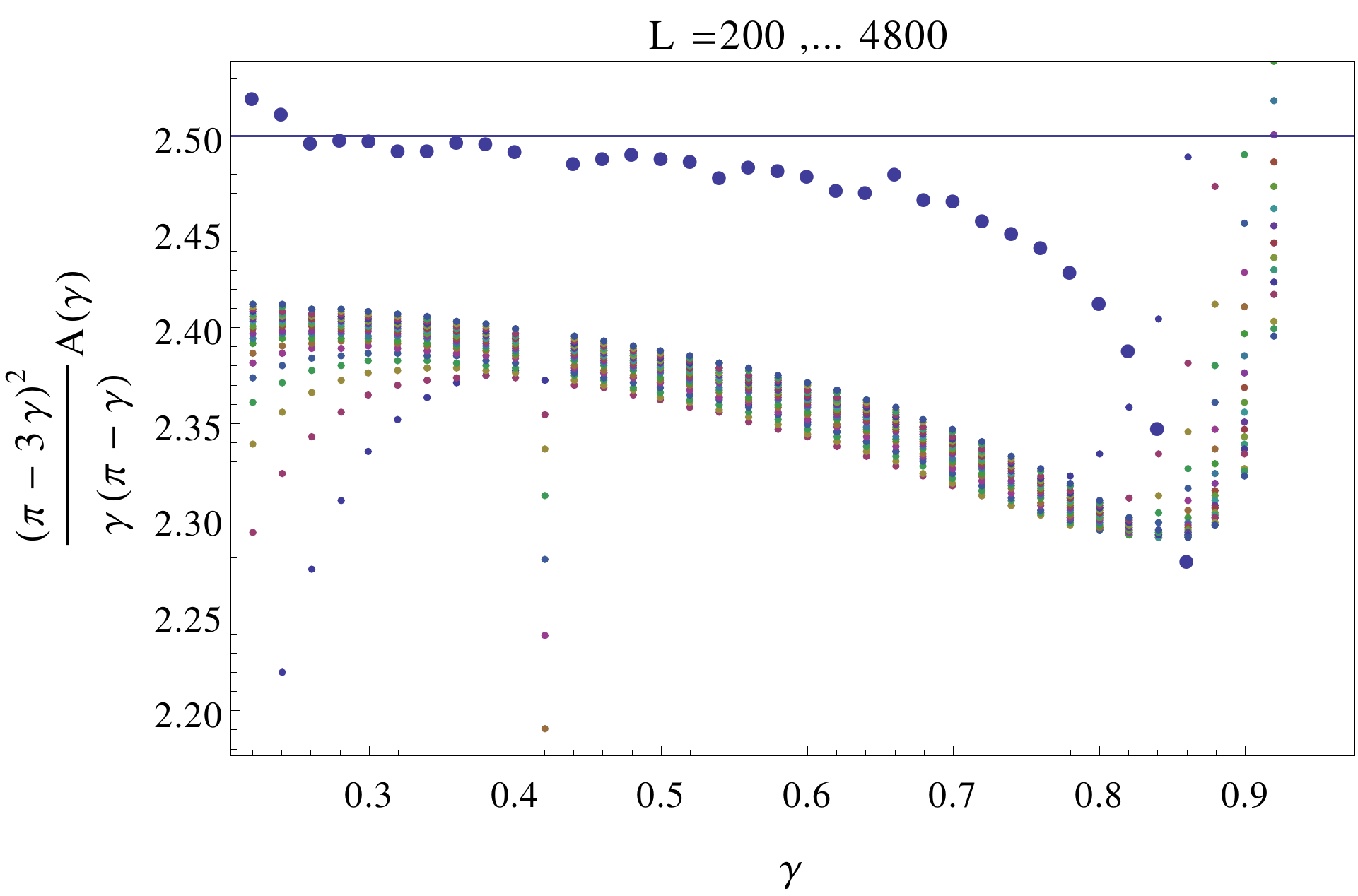}     
 \end{center}
\caption{Estimation of $ \frac{\gamma (\pi - \gamma)}{\left({\pi}-3\gamma\right)^2} A(\gamma)$ for $L=200$ to $4800$.
The thick blue dots show an extrapolation to $L\to \infty$. 
}                                 
\label{fig:Ath}                          
\end{figure}

Clearly there is a factor $\gamma$ in $A(\gamma)$. 
We also recognize a factor $\pi-\gamma$, and it looks like there is a pole at $\gamma = {\pi \over 3}$.
This is supported by the results in figure~\ref{fig:LogA}.
From there it is apparent that the pole is of order $2$. 
The remaining factor is determined with some uncertainty; see figure~\ref{fig:Ath}.

These elements lead to our final conjecture:
\begin{equation}
 A(\gamma) ={5 \over 2} \frac{\gamma (\pi - \gamma)}{\left(\pi -3\gamma\right)^2} \,.
 \label{numresforA}
\end{equation}

\subsubsection{A word on the functions $B_{(n,j)(\gamma)}$}
\label{section:Bni}

The interpretation of (\ref{fconjecture2}) is the following : the index $j$ is a discretized version of what becomes in the thermodynamic limit a continous quantum number $s \propto \frac{j}{\log L}$. In this limit, the functions $B_{(n,j)(\gamma)}$ should rather be considered as continous functions of $s$. Estimating these functions from our finite size results would imply solving the Bethe ansatz equations for large values of $j$ for large sizes $L$, which turns out to be out of our reach. 

We therefore will not focus on the determination of these functions, but leave some hope in the fact that the techniques of NLIE might help us in this direction \cite{CanduIkhlef}.

\subsection{Non zero twist and discrete states}
 
In this section we follow continously the states $(n,j)$ as the twist $\varphi$ is turned on. 
 
\begin{figure}
\begin{center}
 \includegraphics[scale=0.5]{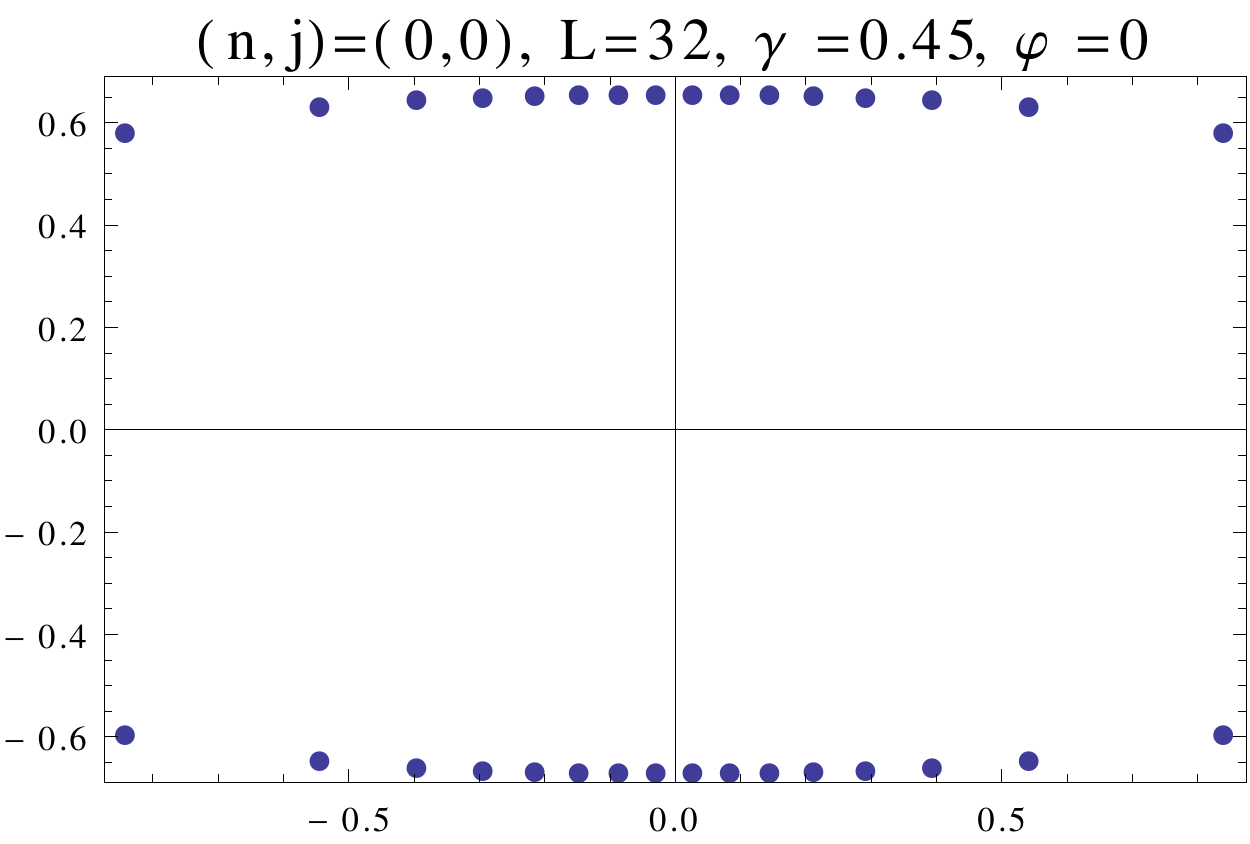}
 \hspace{1cm}
  \includegraphics[scale=0.5]{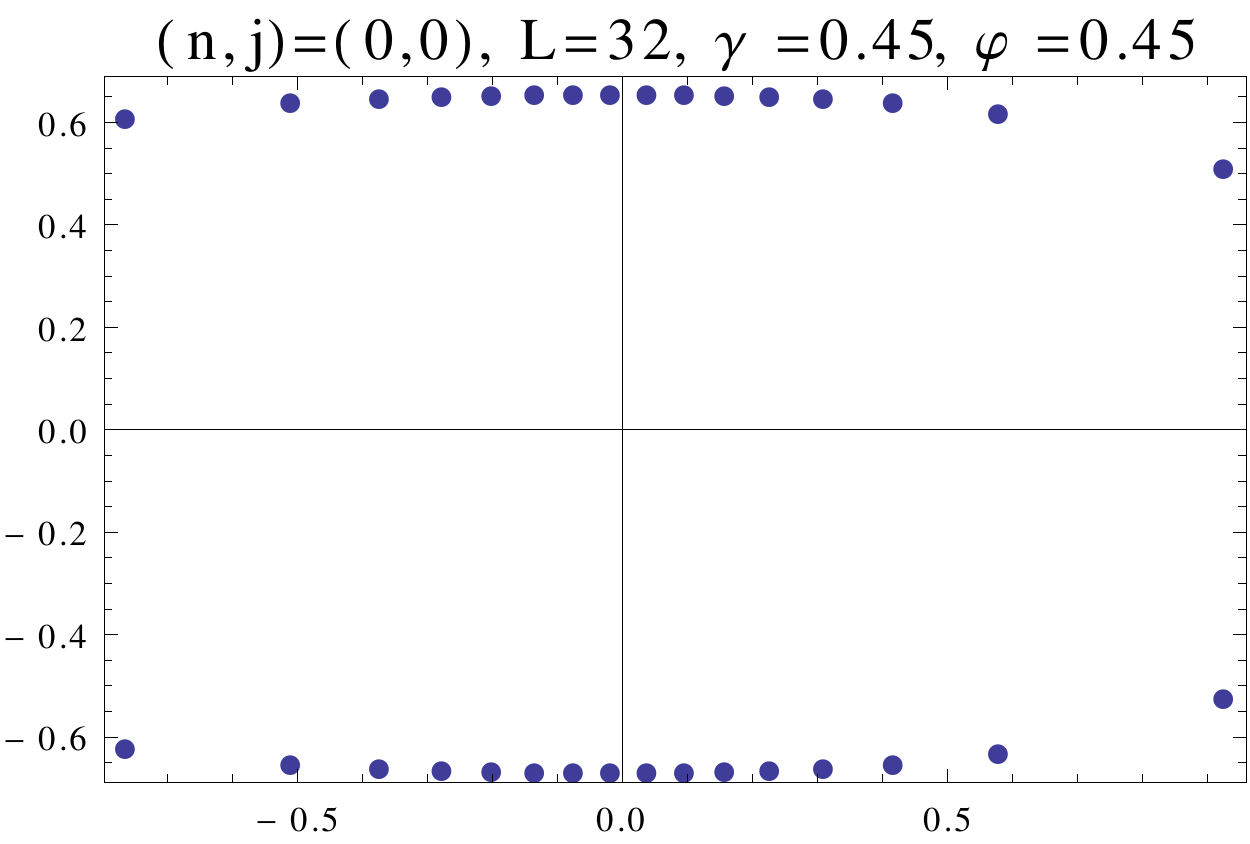}  
   \hspace{1cm}
   \includegraphics[scale=0.5]{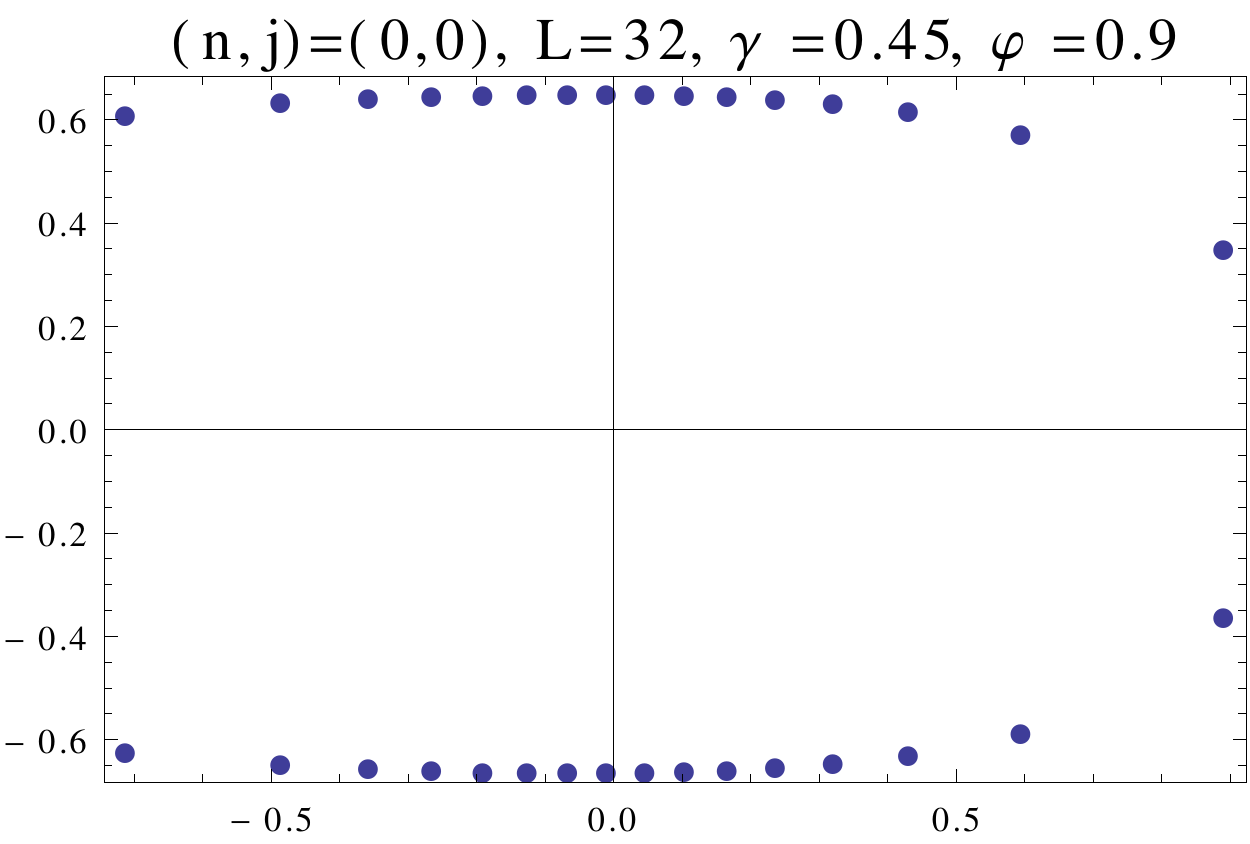}  
 \end{center}
\caption{
Roots for the ground state $(n,j)=(0,0)$ at $L=32$ and $\gamma=0.45$, for different values of the twist parameter. 
No qualtitative changes occur in the roots configurations as the twist varies. 
}                               
\label{fig:rootsE0twist}                          
\end{figure}

Our first concern is to check whether the regime change of the central charge corresponds to a qualitative change of the corresponding Bethe roots. 
We used the McCoy method at small sizes to produce the latter from exact diagonalization, and directly solved the BAE at larger sizes. 
See figure~\ref{fig:rootsE0twist} for a view of what happens at $L=32$: There seems to be no qualitative change of the roots.

\subsubsection{$n=0$ sector}
 
We now go further and describe the general pattern describing the roots configurations of the low-lying excitations as the twist value is raised. 
Let us recall the notations:  $(0,j)$ corresponds to the $j$'th lowest lying excitation in the sector of zero magnetization. 
In terms of the Bethe roots at $\varphi = 0$, it corresponds to replacing $j$ 2-strings by $j$ antistrings with $\Im \lambda_i = {\pi \over 2}$.
We looked at the first three of those excitations at $L=6$.

\paragraph{$n=0, j=0$ (ground state).}
As mentioned before, the ground state is made of ${L \over 2}$ 2-strings for all values of $\varphi$.

\begin{figure}
\begin{center}
 \includegraphics[scale=0.5]{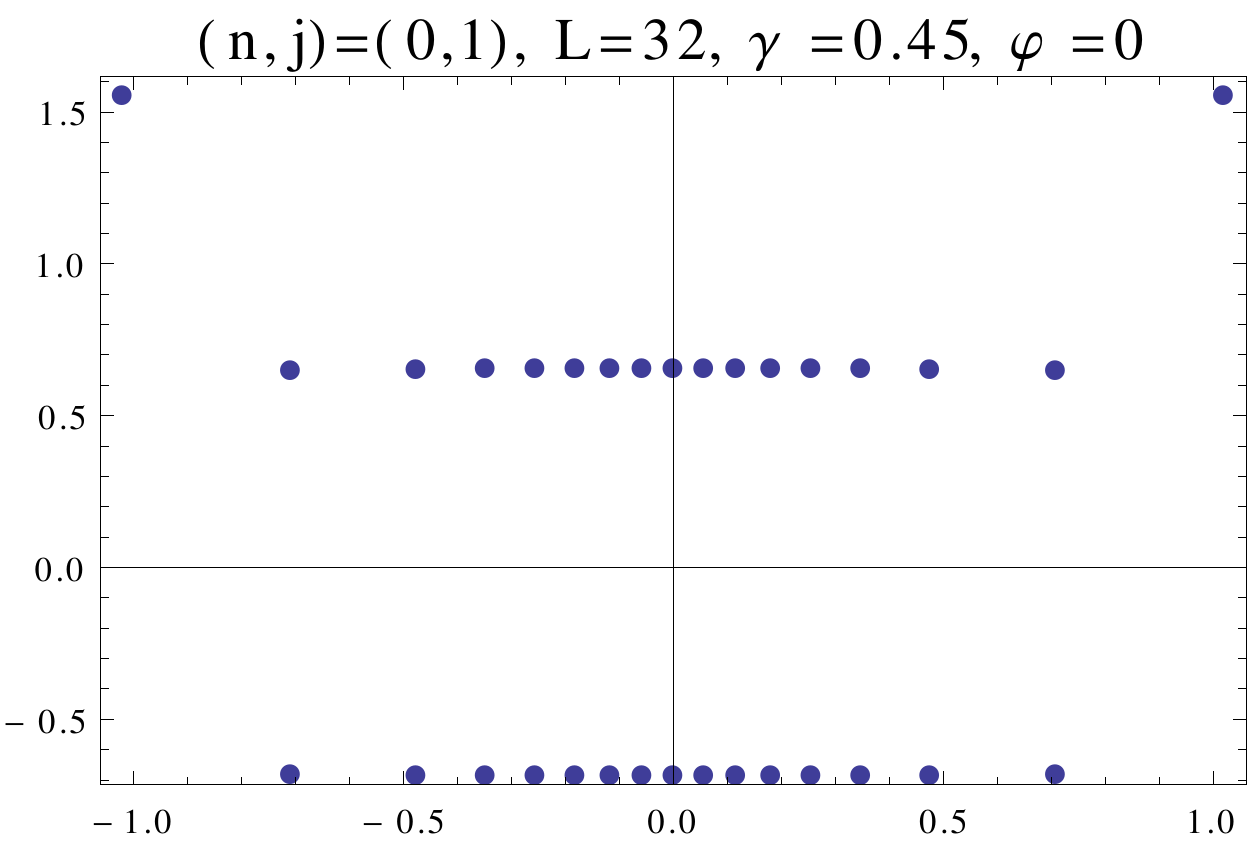}
 \vspace{1cm}  \hspace{1cm}
  \includegraphics[scale=0.5]{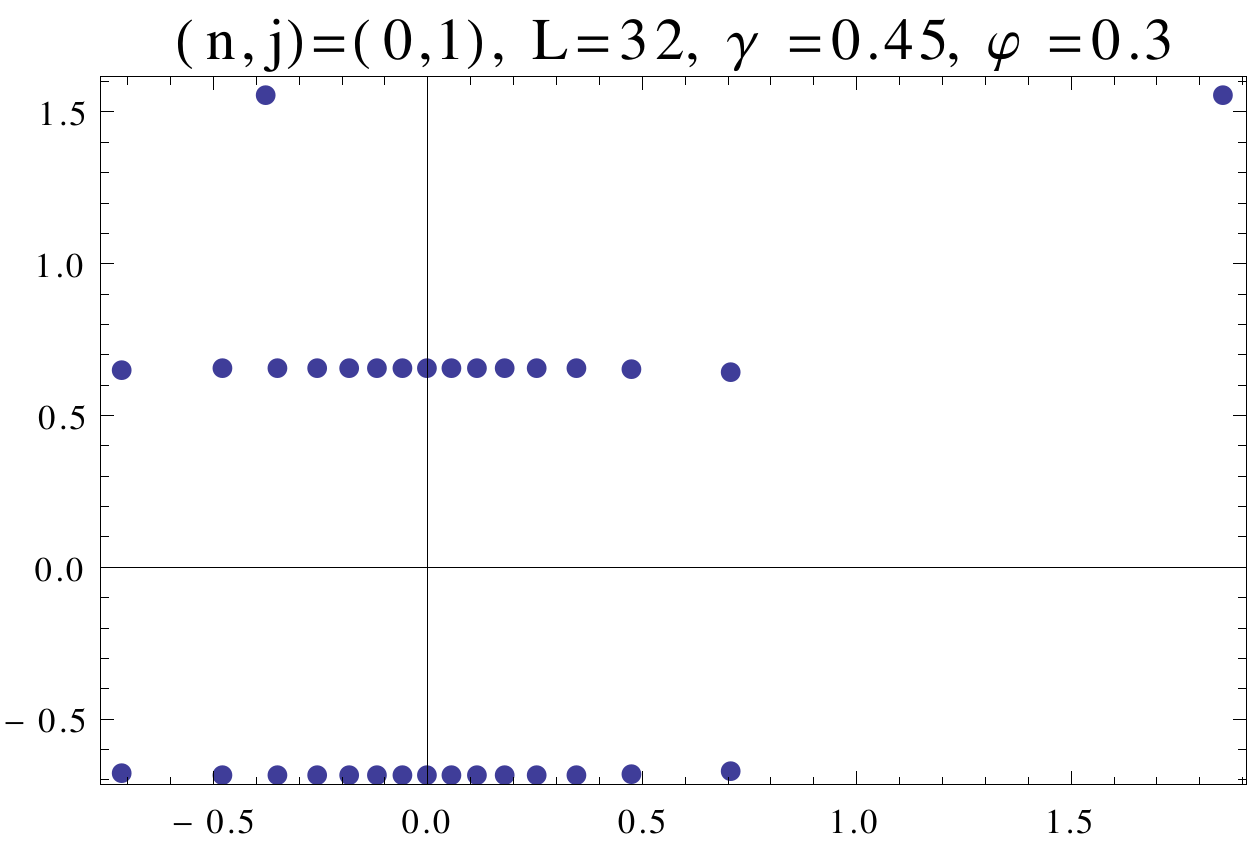}  
   \vspace{1cm}    \hspace{1cm}
   \includegraphics[scale=0.5]{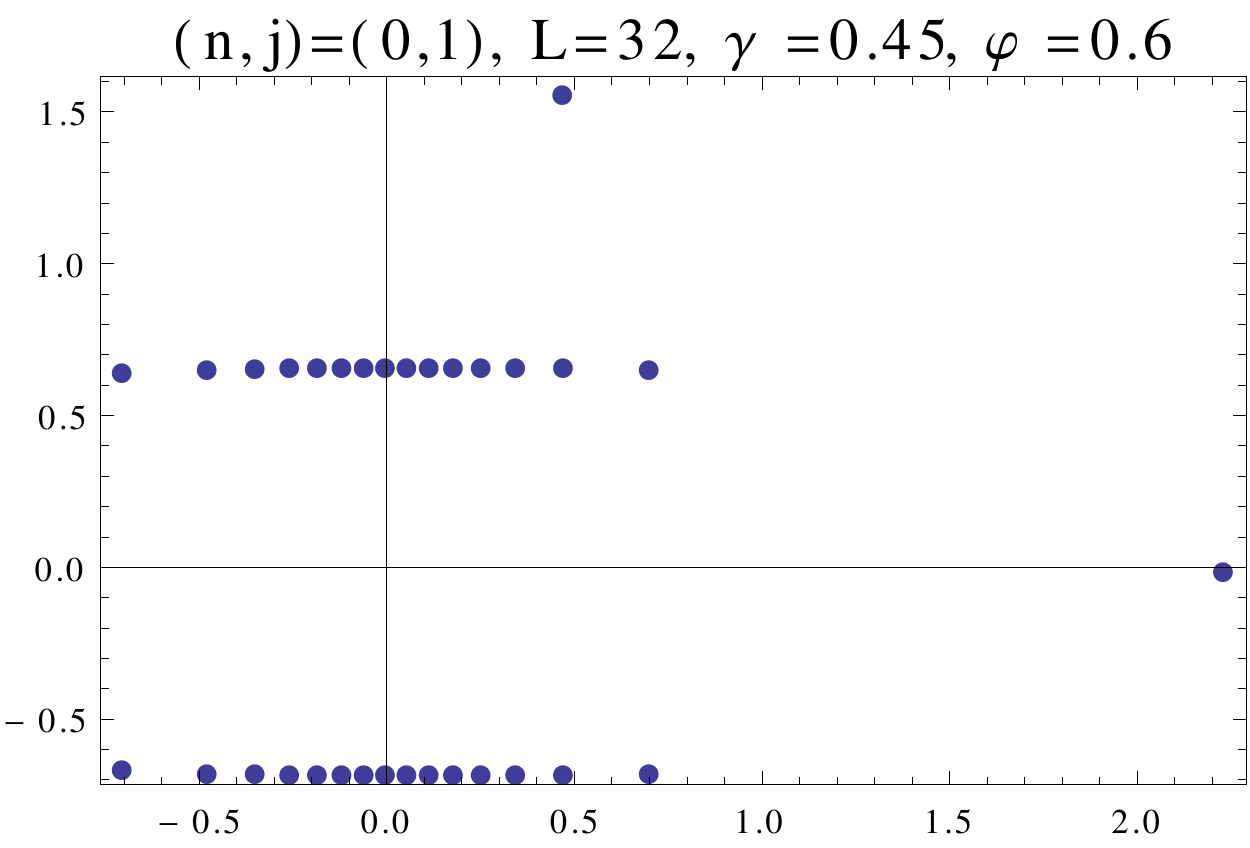}  
     \vspace{1cm}  \hspace{1cm}
   \includegraphics[scale=0.5]{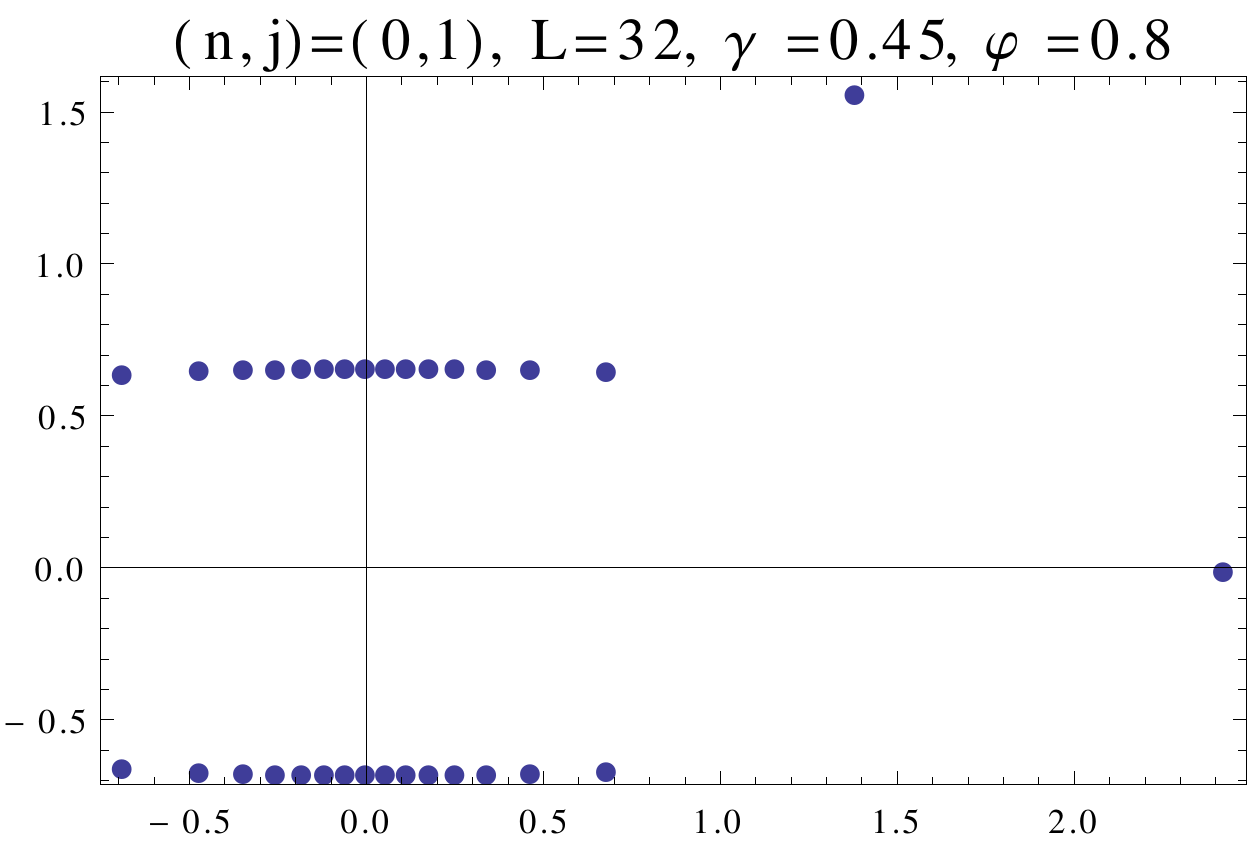}  
      \vspace{1cm}  \hspace{1cm}
   \includegraphics[scale=0.5]{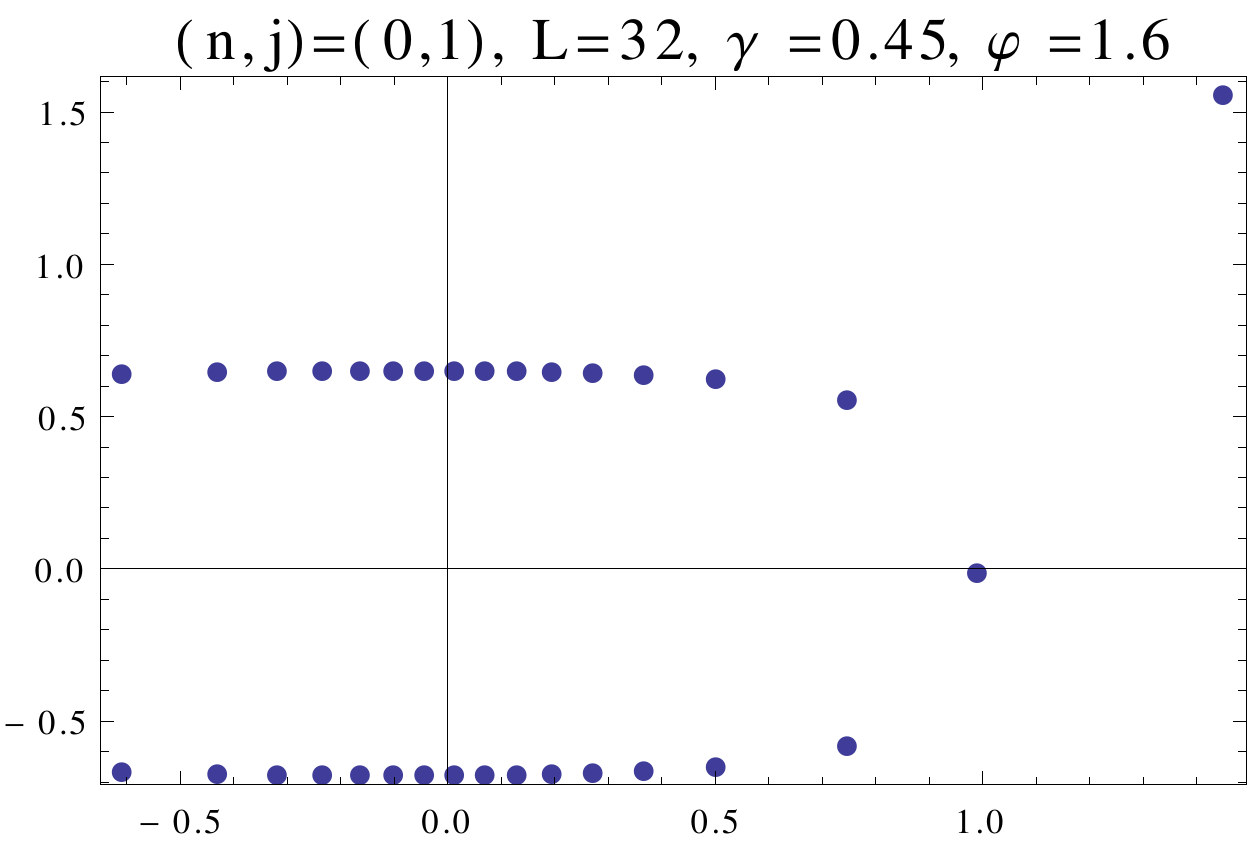}  
 \end{center}
\caption{
Roots for the state $(n,j)=(0,1)$ at $L=32$, $\gamma=0.45$, for different values of the twist parameter $\varphi$.
}                               
\label{fig:rootsE01twist}                          
\end{figure}

\paragraph{$n=0, j=1$.}
As shown in figure \ref{fig:rootsE01twist} the roots undergo several qualitative changes as $\varphi$ varies.
There are $\frac{L}{2}-1$ $2$-strings for all $\varphi$. The remaining two roots behave as follows:
\begin{itemize}
 \item $0 \leq \varphi < \gamma$ : There is 1 antistring. We label the associated roots $r_1, r_2$ by order of increasing real parts.
 \item $\varphi = \gamma$ : $r_2$ goes to $\infty + \mathrm{i} \frac{\pi}{2}$.
 \item $\gamma < \varphi < 2\gamma$ : $r_2$ has jumped to the real axis.
 \item $\varphi = 2\gamma$ : $r_1$ and $r_2$ go to $\infty$ and $\infty + \mathrm{i} \frac{\pi}{2}$ respectively.
 \item $\varphi > 2 \gamma$ : now $r_1$ is the one with the largest real part.
\end{itemize}

\paragraph{$n=0, j=2$.}
The are $\frac{L}{2}-2$ $2$-strings that do not undergo qualitative changes as $\varphi$ varies. The remaining four roots behave as follows:
\begin{itemize}
 \item $0 \leq \varphi < \gamma$ : There are 2 antistrings. We label the associated roots $r_1, r_2, r_3, r_4$ by order of increasing real parts.
 \item $\varphi <3 \gamma$ : $r_3$ and $r_4$ undergo the same process as $r_1$ and $r_2$ in the $E_{0,1}$ case. 
 For $2\gamma < \varphi < 3 \gamma$ this results in : $r_1, r_2, r_3$ have imaginary parts ${\pi \over 2}$, $r_4$ is real, and $\Re r_1 < \Re r_2 < \Re r_4 < \Re r_3$
 \item $\varphi = 3\gamma$: $r_2$ and $r_3$ get infinitely close. Their reals parts also get infinitely close to $r_4$ as the three go to $\infty$.

  \item $3\gamma < \varphi < 4 \gamma$: $r_2$ and $r_3$ are now conjugate to each other, with imagnary parts quickly decreasing from $\pi \over 2$ as $\varphi$ goes away from $3 \gamma$. As $\varphi$ is increased it seems to converge to ${\gamma \over 4}$, hence the denomination 2$^{*}$-string for the set $\{r_2,r_3\}$. 
As for $r_4$, we now have $\Im r_4 = \frac{\pi}{2}$.
 
 \item $3\gamma < \varphi < 4 \gamma$: $r_1$ and $r_3$ now form a 2-string (of imaginary part much smaller than ${\pi \over 4}-{\gamma \over 4}$, and twist-dependent), $r_2$ is back on the axis of im part $\pi \over 2$.
 \item $\varphi = 4\gamma$: $r_1$ and $r_4$ undergo the same process as that described above for $r_2$ and $r_3$, and merge into a 2-string
 \item $\varphi > 4 \gamma$ : we are left with only 2- and 2$^{*}$- strings
\end{itemize}

\paragraph{$n=0, j=3$.}
The are $\frac{L}{2}-3$ $2$-strings that do not undergo qualitative changes as $\varphi$ varies. The remaining six roots behave as follows:
\begin{itemize}
 \item $0 \leq \varphi < \gamma$ : There are 3 antistrings. We label the associated roots $r_1,\ldots,r_6$ by order of increasing real parts.
 \item $\varphi \leq 4 \gamma$ : $r_3, \ldots r_6$ undergo the same process as $r_1, \ldots, r_4$ in the $i=2$ case.
 \item $4\gamma < \varphi <6 \gamma$ : $r_1$ and $r_2$ undergo the same process as $r_1$ and $r_2$ in the $i=1$ case.
 \item $\varphi > 6 \gamma$ : we are left with ${L\over 2} -1$ 2-strings, + 1 root of imaginary part $\pi \over 2$, + 1 real root.
\end{itemize}

\paragraph{General pattern}

The general pattern is can easily be deduced from these cases.
The root configurations corresponding to the excited states $(0,j)$ undergo a series of transformations, starting at $\varphi = 0$ from a set of ${L\over 2} -n$ 2-strings and $j$ antistrings (with $\Im \lambda_i =  \frac{\pi}{2}$), and ending at $\varphi = 2 i \gamma$ with 
\begin{itemize}
 \item for $j$ even, a set of ${L\over 2}$ 2-strings
 \item for $j$ odd, a set of ${L \over 2}-2$ 2-strings, one root with imaginary part ${\pi \over 2}$, and one real root.
\end{itemize}

\begin{figure}
\begin{center}
 \includegraphics[scale=0.7]{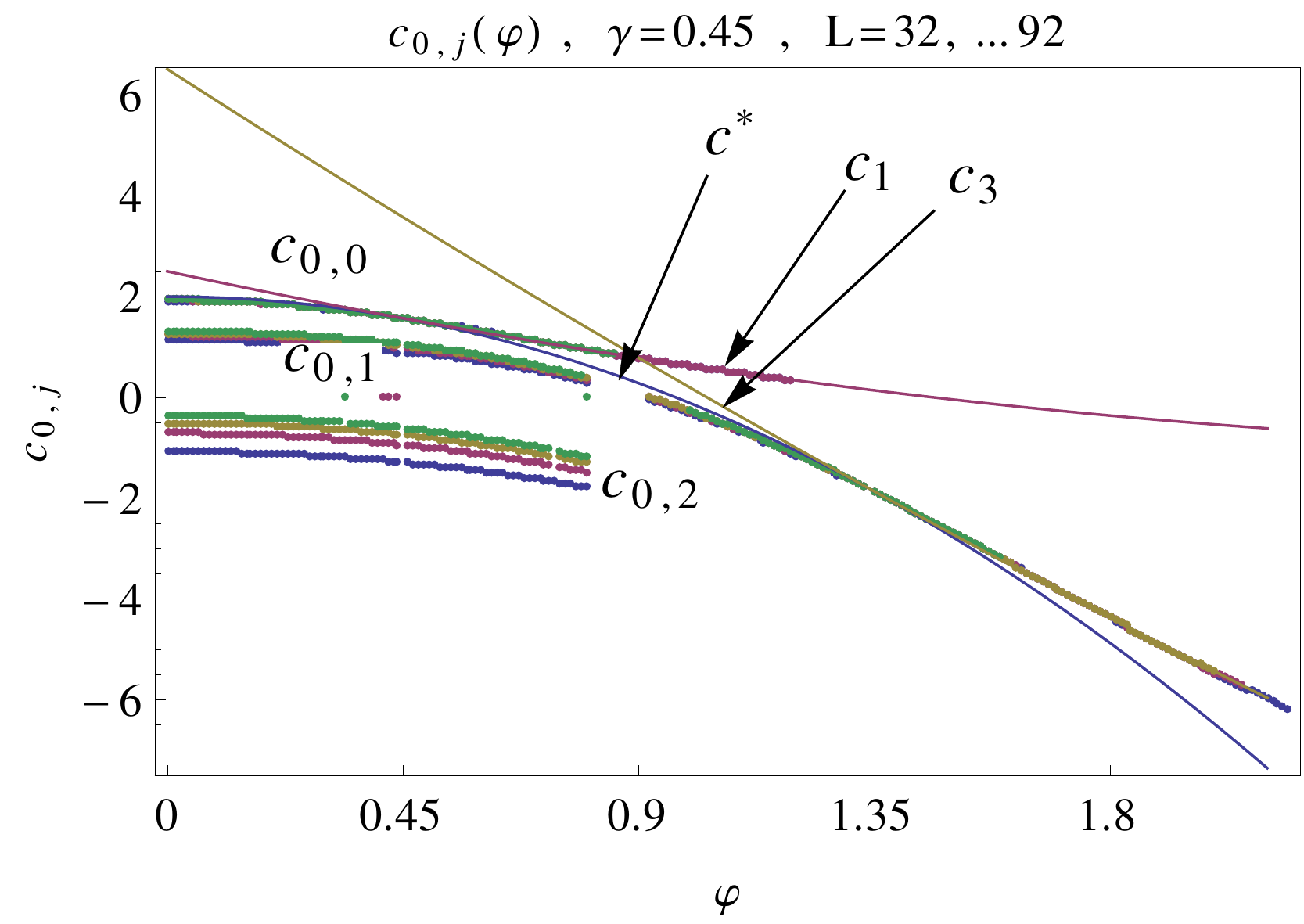}
 \end{center}
\caption{
Effective central charges measured for the first few excitations in the $S_z = 0$ sector, for $\gamma = 0.45$, as a function of the twist.
We also plotted the predicted values $c^*$, $c_1$, $c_3$.
}                               
\label{fig:cpS0}                          
\end{figure}

Note however that these transformations do not coincide with the lift of the effective central charges from the continuum, as we already saw in the $(0,0)$ case. 
Let us make this more precise. 
As in (\ref{cstar}) and (\ref{c2p+1}) we define the following effective central charges 
\begin{eqnarray}
 c^{*} &=& 2 - 3 {\varphi^2 \over \pi \gamma} \,, \\
 c_{2p+1} &=& c^{*} + \frac{3}{\gamma(\pi - \gamma)}\left[(2p+1)\gamma - \varphi\right]^2 \,.
\end{eqnarray}
$c^{*}$ is the effective central charge of the continuum states, and in particular this is the real central charge of the model for $\varphi \leq \gamma$.
The result of section~\ref{sec:twist} is that when $\varphi$ reaches $(2p+1)\gamma$ a discrete level pops out of the continuum, with effective central charge $c_{2p+1}$. 
See figure \ref{fig:cpS0}, where we plotted the effective central charges measured for $n=0, j=0, 1, 2$, at $\gamma = 0.45$, as a function of the twist.
It was established that for $\varphi$ small all the $c_{0,j}$ s are part of the continuum, namely with effective central charge $c^*$. 
What we observe here for $c_{0,0}$ and $c_{0,1}$ gives a support for the following scenario : 
\begin{itemize}
\item For $\varphi \leq (2j+1)\gamma$ the level $(0,j)$ is part of the continuum. Its effective central charge is $c^{*}$.
\item For  $\varphi > (2j+1)\gamma$ it pops out of the continuum, with an effective central charge $c_{2j+1}$.
\end{itemize}

We conclude by summing up what we found about the excitations $(0,j)$ : 
{\bf
\begin{itemize}
 \item The roots configurations undergo qualitative transformations from $\varphi =0$ to $\varphi = 2 j \gamma$.
 \item There is a change of the effective central charge from $c^{*}$ to $c_{2j+1}$ for $\varphi = (2j+1)\gamma$. 
\end{itemize}
}

\subsubsection{Excited winding modes}

At $\varphi = 2\pi-\gamma$, we expect the ground state to return to the continuum, the continuum now corresponding to the addition of an electric charge to the previous continuum. 
The same should happen with the excited states $(0,j)$ at $\varphi = 2\pi - (2 j + 1)\gamma$. 
Obviously for $j$ large enough the corresponding levels never leave the continuum, but one can see more and more discrete states by taking small enough values of $\gamma$. 
More precisely for a fixed value of $\gamma$, the discrete states that leave the continuum are those such that $(2j+1)\gamma \leq \pi$, the number of which is therefore $\left\lfloor {\pi \over 2\gamma} + {1 \over 2}\right\rfloor$.

At the level of roots configurations, similar mechanisms as those described above should take place as the states fall back into the continuum, but we refrain from going into more details about this here.

\subsubsection{Sectors of non zero magnetization}

We now consider sectors of nonzero magnetization $n \equiv S_z$. The states are labeled, in the same way, $(n,j)$. 
Remember that the effective central charges (their compact parts) of the untwisted chain were found to be (\ref{c/12conj}) that we rewrite here as
\begin{equation}
 c_{n,j} (\varphi = 0) = 2 - \frac{3 \gamma n^2}{\pi} \,.
\end{equation}

Let us describe the roots configuration for such states as $\varphi$ is varied, observed at size $L=6$.

\paragraph{$n=1, j=0$.}
\begin{itemize}
 \item $0 \leq \varphi < 2\gamma$ : ${L \over 2} -1$ 2-strings + one root with imaginary part $\pi \over 2$, labeled $r_1$.
 \item $\varphi = 2\gamma$ : $r_1$ goes to $\infty + \mathrm{i} \frac{\pi}{2}$.
 \item $2\gamma < \varphi $: $r_1$ has jumped to the real axis.
\end{itemize}

\paragraph{$n=1, j=1$.}
\begin{itemize}
 \item $0 \leq \varphi < 2\gamma$ : ${L \over 2} -2$ 2-strings + $3$ roots with imaginary part $\pi \over 2$, labeled $r_1, r_2, r_3$ by order of increasing real parts.
 \item $0 < \varphi < 3\gamma $: $r_3$ undergoes the same process as $r_1$ in the $n=1, i=0$ case.
  \item $\varphi=3\gamma$ : $r_2$ and $r_3$ go to $\infty$ and $\infty + \mathrm{i} \frac{\pi}{2}$ respectively 
  \item $3\gamma < \varphi < 4 \gamma $: $r_2$ now has a larger real part than $r_3$.
   \item $\varphi = 4\gamma$: $r_1$ and $r_2$ get infinitely close. Their real parts also get infiniely close to $r_3$, as the three go to $\infty$.
     \item $4\gamma < \varphi$: $r_1$ and $r_2$ now form a 2$^{*}$-string, $r_3$ is back on the axis of imaginary part $\pi \over 2$.
   \end{itemize}

\paragraph{$n=2, j=0$.}   
${L \over 2} -1$ 2-strings, no qualitative change when $\varphi$ is varied.

\paragraph{$n=2, j=1$.}
\begin{itemize}
 \item $0 \leq \varphi < 3\gamma$ : 1 antistring. We label the associated roots $r_1, r_2$ by order of increasing real parts.
 \item $\varphi = 3\gamma$ : $r_2$ goes to $\infty + \mathrm{i} \frac{\pi}{2}$.
 \item $3\gamma < \varphi < 4\gamma$ : $r_2$ has jumped to the real axis.
 \item $\varphi = 4\gamma$ : $r_1$ and $r_2$ go to $\infty$ and $\infty + \mathrm{i} \frac{\pi}{2}$ respectively.
 \item $\varphi > 4 \gamma$ : now $r_1$ is the one with larger real part.
\end{itemize}

\paragraph{General pattern}

The general pattern is quite easily understandable from what we observe here and from the conclusions in the $n=0$ case. 
Trying to be quite general, we can say that the roots of the $(n,j)$ excitations undergo the same transformations as those of the $(0,j)$ excitations, but the values of the twist at which these happen is shifted by $|n| \gamma$.

\begin{figure}
\begin{center}
 \includegraphics[scale=0.7]{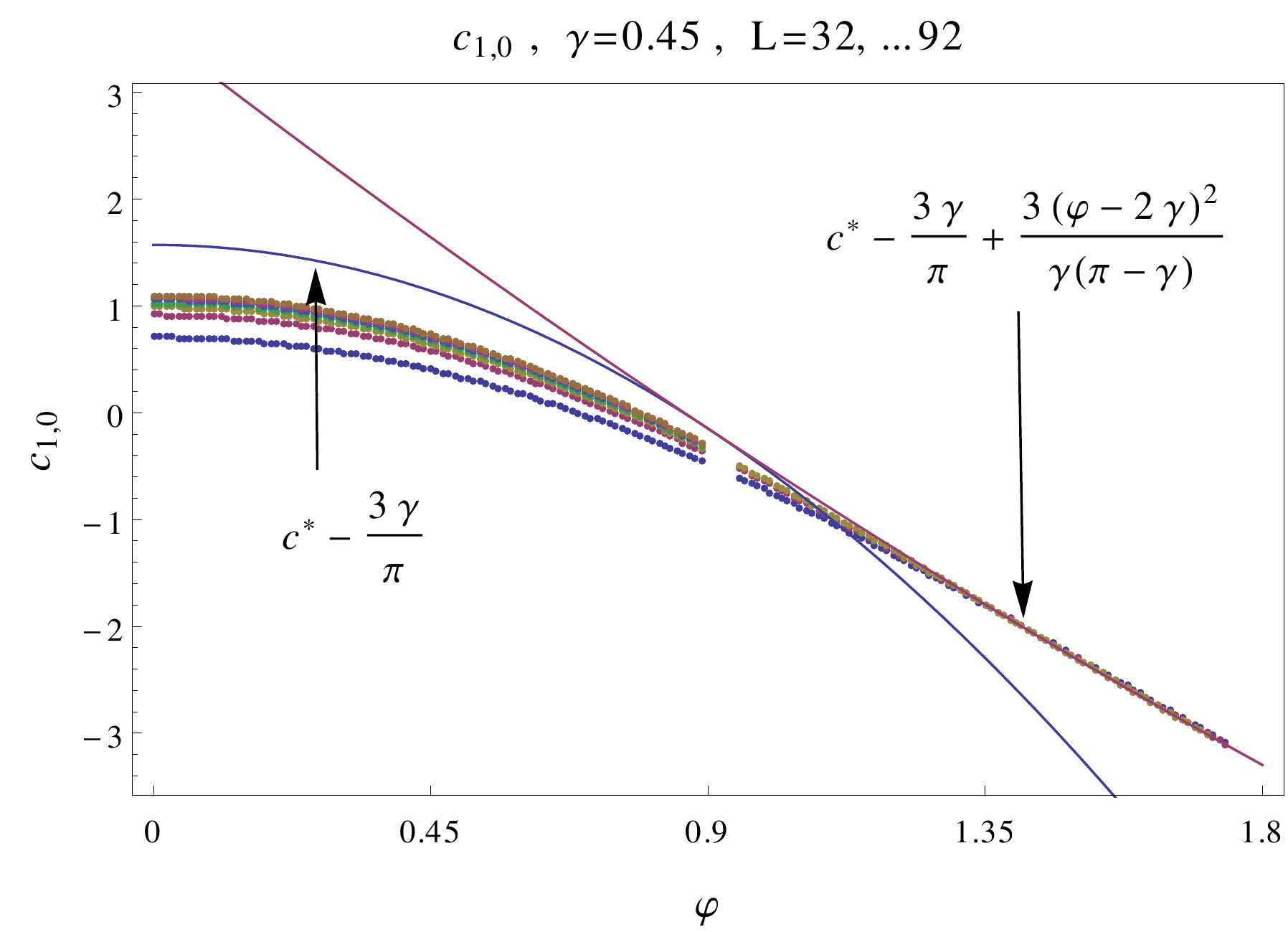}
 \end{center}
\caption{
Effective central charge measured for the ground state in the sector $S_z = 1$, for $\gamma = 0.45$.
The two plotted curves correspond to the prediction (\ref{eq:cm}) 
}                               
\label{fig:c0s1}                          
\end{figure}

Now we turn to the corresponding central charges. 
We checked numerically the following,
\begin{eqnarray}
 c_{n,0} = \begin{cases}  c_n^{*} \equiv c^* -\frac{3 \gamma n^2}{\pi}    & \mbox{for } \varphi\leq\ (|n|+1) \gamma 
\\  c_n^{*} + {3 \over \gamma(\pi - \gamma)}\left(\varphi - (1+|n|)\gamma\right)^2 & \mbox{for } \varphi \geq (|n|+1) \gamma \,,
\end{cases}
\label{eq:cm}
\end{eqnarray}
 see figure \ref{fig:c0s1} for a display of the results at $n=1$. 

\begin{figure}
\begin{center}
 \includegraphics[scale=0.7]{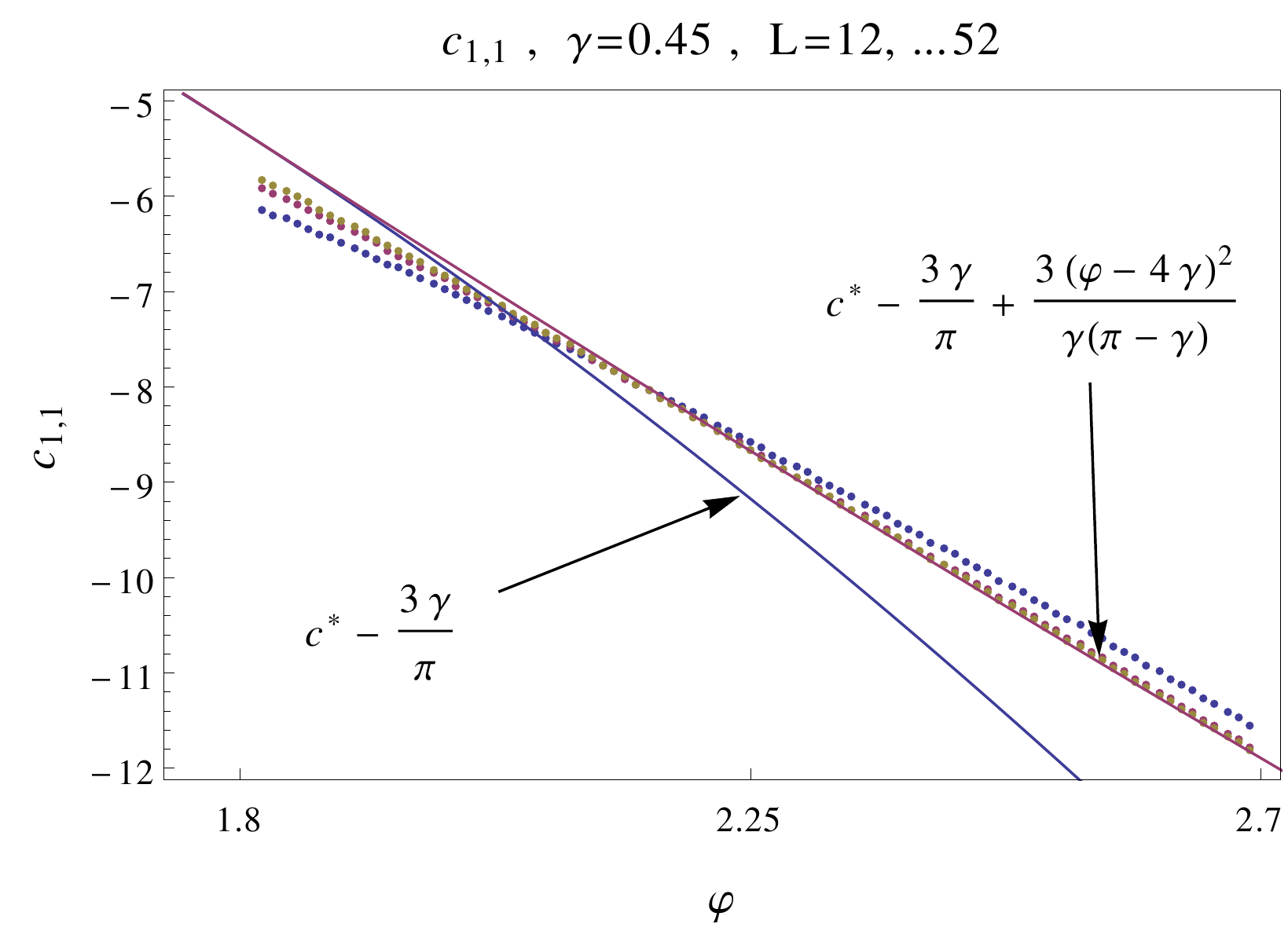}
 \end{center}
\caption{
Effective central charge $c_{1,1}$ measured for $\gamma = 0.45$ in the region $\varphi \geq 4 \gamma$.
}                               
\label{fig:c1s1}                          
\end{figure}

Now consider the first excitation in the $n=1$ sector, $(1,1)$. 
For $\varphi$ small it is part of the continuum, namely $c_{1,1}=c_m^*$. 
With our experience of the $n=0$ case, we now expect it to leave the continuum at $\varphi =4 \gamma$. 
See figure \ref{fig:c1s1}, where the corresponding effective central charge is plotted for $\varphi > \gamma$. 
We indeed that at $\varphi =4 \gamma$ it lifts off the continuum, and becomes 
\begin{equation}
 c_1^{*} + {3 \over \gamma(\pi - \gamma)}\left(\varphi - 4\gamma\right)^2 \,.
\end{equation}
This observation gives some confidence in the following conjecture:
the effective central charge of the excitation $(n,j)$ is given (as long as we do not intersect the excited winding modes) by 
\begin{equation}
  c_{n,j} = \begin{cases}  c_n^{*}  -\frac{3 \gamma n^2}{\pi}  = 2 - 3 {\varphi^2 \over \pi \gamma} -\frac{3 \gamma n^2}{\pi}  & \mbox{for } \varphi\leq\ (|n|+2 j+1) \gamma 
\\  c_n^{*} -\frac{3 \gamma n^2}{\pi} + {3 \over \gamma(\pi - \gamma)}\left[\varphi - (|n|+2 j+1)\gamma\right]^2 & \mbox{for } \varphi \geq (|n|+2 j+1)  \gamma \,.
\end{cases}
\label{eq:cmappendix}
\end{equation}

\end{document}